\documentclass{aastex63}
\usepackage{amsmath}

\newcommand\gaia{\textit{Gaia}}
\newcommand\gdr[1]{\gaia~DR#1}

\newcommand\gbp{\ensuremath{G_\mathrm{BP}}}
\newcommand\grp{\ensuremath{G_\mathrm{RP}}}

\newcommand{\domain}{\mathcal{D}}

\defcitealias{abergel2014planck}{Planck14}
\newcommand\finalgpqnum{\ensuremath{{{160,946}}}} 

\graphicspath{{./}{figures/}}

\received{10 Sep, 2020}
\revised{30 Jan, 2021}
\accepted{18 Feb, 2021}
\submitjournal{ApJS}

\shorttitle{Quasars behind the Galactic plane}
\shortauthors{Fu et al.}

\begin{document}

\title{Finding Quasars behind the Galactic Plane. I. Candidate Selections with Transfer Learning}

\email{fuym@pku.edu.cn; wuxb@pku.edu.cn}

\author[0000-0002-0759-0504]{Yuming Fu}
\affil{Department of Astronomy, School of Physics, Peking University, Beijing 100871, China}
\affil{Kavli Institute for Astronomy and Astrophysics, Peking University, Beijing 100871, China}

\author[0000-0002-7350-6913]{Xue-Bing Wu}
\affil{Department of Astronomy, School of Physics, Peking University, Beijing 100871, China}
\affil{Kavli Institute for Astronomy and Astrophysics, Peking University, Beijing 100871, China}

\author[0000-0002-6893-3742]{Qian Yang}
\affil{Department of Astronomy, University of Illinois at Urbana-Champaign, Urbana, IL 61801, USA}

\author[0000-0002-7419-9679]{Anthony G.~A.~Brown}
\affil{Leiden Observatory, Leiden University, Niels Bohrweg 2, 2333 CA, Leiden, The Netherlands}

\author{Xiaotong Feng}
\affil{Department of Astronomy, School of Physics, Peking University, Beijing 100871, China}
\affil{Kavli Institute for Astronomy and Astrophysics, Peking University, Beijing 100871, China}

\author{Qinchun Ma}
\affil{Department of Astronomy, School of Physics, Peking University, Beijing 100871, China}
\affil{Kavli Institute for Astronomy and Astrophysics, Peking University, Beijing 100871, China}

\author{Shuyan Li}
\affil{Department of Astronomy, School of Physics, Peking University, Beijing 100871, China}



\begin{abstract}
    Quasars behind the Galactic plane (GPQs) are important astrometric references and useful probes of Milky Way gas. However, the search for GPQs is difficult due to large extinctions and high source densities in the Galactic plane. Existing selection methods for quasars developed using high Galactic latitude (high-$b$) data cannot be applied to the Galactic plane directly because the photometric data obtained from high-$b$ regions and the Galactic plane follow different probability distributions. To alleviate this dataset shift problem for quasar candidate selection, we adopt a Transfer Learning Framework at both data and algorithm levels. At the data level, to make a training set in which dataset shift is modeled, we synthesize quasars and galaxies behind the Galactic plane based on SDSS sources and Galactic dust map. At the algorithm level, to reduce the effect of class imbalance, we transform the three-class classification problem for stars, galaxies, and quasars to two binary classification tasks. We apply XGBoost algorithm on Pan-STARRS1 (PS1) and AllWISE photometry for classification, and additional cut on \gaia\ proper motion to remove stellar contaminants. We obtain a reliable GPQ candidate catalog with \finalgpqnum\ sources located at $|b|\leq 20^{\circ}$ in PS1-AllWISE footprint. Photometric redshifts of GPQ candidates achieved with XGBoost regression algorithm show that our selection method can identify quasars in a wide redshift range ($0<z\lesssim5$). This study extends the systematic searches for quasars to the dense stellar fields and shows the feasibility of using astronomical knowledge to improve data mining under complex conditions in the Big Data era.
     
\end{abstract}


\section{Introduction}
\label{sec:intro}

The Galactic plane has long been the ``zone of avoidance'' for extragalactic astronomy, including quasar surveys. The Half Million Quasar \citep[HMQ;][]{Flesch2015} catalog contains a total of 510,764 objects, but only 35,105 located at $b\leq |30^{\circ}|$ (half of the whole sky area), 3,730 at $b\leq |20^{\circ}|$, and 255 at $b\leq |10^{\circ}|$. Although it is difficult to search for Quasars behind the Galactic Plane (hereafter GPQs), such quasars are important references for astrometry and useful probes of Milky Way gas.

Quasars are used as astrometric references due to their small parallaxes and proper motions. GPQs enable the accurate measurement of positions, distances, and proper motions of stars in the Galactic disk, which is key to understanding our own Galaxy. The high-precision astrometry provided by the \gaia\ mission defines a celestial reference frame through the positions of 556,869 candidate quasars, however only a tiny fraction of these quasars are located at $|b|\leq 15^{\circ}$ \citep{mignard2018gaia}. A large sample of GPQs will help build a better reference frame in the optical, through direct coverage of the sky in the Galactic plane, and will help to better understand the systematic astrometry errors of Gaia in the Galactic plane region \citep{Arenou2018}.

Line-of-sight absorption towards quasars can probe gas structures of the Milky Way. While quasars at high Galactic latitude have been useful in studying the Milky Way halo gas \citep[e.g.][]{savage1993hubble,savage2000hubble,2008A&A...487..583B,2012A&A...542A.110B}, GPQs allow absorption line studies on gaseous structures in the Galactic plane \citep[e.g. Anti-Center Shell, H Complex; see][]{2018MNRAS.474..289W}. Moreover, a high density sample of GPQs can map the gas distribution with a higher angular resolution than that is possible with the 21 cm surveys.

Another application of GPQs is adaptive-optics observation on quasar host galaxies, which is achieved by their proximity to nearby bright stars as natural guide stars \citep{Im2007,2019ApJ...875..102F}. For adaptive optics, natural guide stars should be located within a few arcseconds of the science target, which rarely occurs outside of the Galactic plane but is more common in the plane.

The difficulty of finding quasars behind the Galactic plane is caused by several challenges, including:
\begin{itemize}
    \item In comparison to objects at high Galactic latitude (high-$b$), sources in the Galactic plane suffer from higher extinction and reddening. As a result, many sources (especially extragalactic sources) can not be detected within the survey detection limit. For other detectable sources, their colors are different from those at high Galactic latitude. 
    \item The source density in the Galactic plane is high. The quality of photometry can be worse in dense regions, because sources can be easily contaminated by visible or unseen neighbors.
    \item A lot of ``unusual'' stars are located within the Galactic plane, including some white dwarfs, M/L/T dwarfs, and Young Stellar Objects (YSOs), that share many similar observational properties with quasars. These sources can be contaminants for quasars at different redshifts \citep[e.g.][]{1997AJ....113.1421K,2002MNRAS.335..673V,2006AJ....131.2722C,2009ApJ...701..508K}. 
\end{itemize}

Since the first identification of quasar \citep[3C 273;][]{schmidt19633c273}, many methods for quasar candidate selection have been developed, including ultraviolet excess \citep[e.g.][]{1965ApJ...141.1560S,1986ApJS...61..305G}, radio sources \citep[e.g.][]{1996AJ....112..407G,2000ApJS..126..133W,2001ApJS..135..227B}, X-ray sources \citep[e.g.][]{1979RSPSA.366..375P,2000AJ....119.2540G}, optical/near-infrared (near-IR) colors \citep[e.g.][]{richards2002spectroscopic,2001AJ....122.2833F,2010MNRAS.406.1583W}, mid-IR colors \citep[e.g.][]{2004ApJS..154..166L,2005ApJ...631..163S,2012ApJ...753...30S,2012MNRAS.426.3271M,2012AJ....144...49W,yan2013characterizing}, and quasar variability \citep[e.g.][]{2003AJ....125.1330D,2011A&A...530A.122P}. In addition, tools based on statistical machine learning \citep[e.g.][]{2004ApJS..155..257R,2011ApJ...729..141B} and deep learning \citep[e.g.][]{2010A&A...523A..14Y,2018A&A...611A..97P} have also been established to find quasars with various data that are available. 

A few studies have focused on finding quasars/AGNs behind dense stellar fields such as the Galactic plane, Magellanic Clouds, and M31 and M33 galaxies. Most of these studies used infrared selection methods to efficiently find quasars. For example, \citet{Im2007} discovered 40 bright quasars at $|b| \leq 20^{\circ}$ by applying the combination of a near-IR color cut of $J-K>1.4$ on Two Micron All Sky Survey \citep[2MASS;][]{skrutskie2006two} and detection of a radio counterpart from the NRAO VLA Sky Survey \citep[NVSS;][]{condon1998nrao}.
\citet{2009ApJ...701..508K} identified 5,000 AGNs behind the Magellanic Clouds with mid-IR color cuts modified from method of \citet{2005ApJ...631..163S}.
\citet{huo2010new,huo2013lamost,huo2015lamost} discovered 1,870 new quasars around the Andromeda (M31) and Triangulum (M33) galaxies, with the Large Sky Area Multi-Object Fiber Spectroscopic Telescope (LAMOST) from 2009 to 2013.

Recently, searches for quasars have been focused on large datasets with big data volumes and large sky coverage. \citet{Secrest2015} obtained an all-sky AGN candidates catalog with $\sim 1.4$ million sources using two-color infrared photometric selection criteria from the Wide-field Infrared Survey Explorer final catalog release (AllWISE) \citep[][]{wright2010wide,mainzer2011preliminary}. \citet{Assef2018} built two catalogs of AGN candidates also based on AllWISE photometry, while excluding regions around Galactic center and Galactic plane. \citet{2019MNRAS.485.4539J} selected quasar candidates with machine learning method using Pan-STARRS1 \citep[PS1;][]{Chambers2016} and AllWISE data. \citet{2019MNRAS.490.5615B} classified objects in \textit{Gaia} Data Release 2 \citep[{\gdr{2}};][]{prusti2016gaia,brown2018gaia} as stars, quasars, and galaxies with Gaussian Mixture Model and addressed the problem of \textit{class imbalance} in \gdr{2}. 

However, the studies listed above either treated sources in the Galactic plane and high Galactic latitude as the same, or removed the Galactic plane from consideration. Selection methods for quasars at high Galactic latitude are not generic and can not be applied to the Galactic plane directly, because data (e.g. PS1 and AllWISE photometry) obtained from high-$b$ and low-$b$ follow different probability distributions. For example, apparent colors of quasars (stars) vary from high-$b$ to low-$b$ regions, and so do the source density of quasars (stars). Such behavior of data is a kind of non-stationarity called \textit{dataset shift} \citep{quionero2009dataset}, which leads to significant estimation bias of supervised machine learning algorithms. The color cuts for quasar selection can also be regarded as simple decision tree models in machine learning regime. Previous color cuts obtained from high Galactic latitude regions fail in the Galactic plane due to the \textit{dataset shift}.

To deal with these \textit{dataset shift} problems, \textit{transfer learning} \citep{pan2009survey} has been proposed and studied extensively by data scientists. The idea of \textit{transfer learning} is to use knowledge gained in one problem and apply it to a different but related problem. Although spectroscopically identified (i.e. ``labeled'') samples of extragalactic objects are inadequate in the Galactic plane, such labeled samples are available at high Galactic latitude. The labeled data make it possible to build a good selection method for GPQs, once the knowledge transfer from high Galactic latitude to low Galactic latitude is successful.

This paper is the first one of this series for finding GPQs. In this paper we present a transfer learning method for quasar selection, as well as a GPQ candidate catalog with \finalgpqnum\ sources. In Section \ref{sec:data}, we introduce the archival data used for this study. In Section \ref{sec:TLF}, we describe the algorithm design for GPQ selection. In Section \ref{sec:mock}, we synthesize quasars and galaxies behind the Galactic plane with extragalactic objects at high Galactic latitude from the Sloan Digital Sky Survey \citep[SDSS;][]{2000AJ....120.1579Y}, to make a training set in which \textit{dataset shift} is modeled. In Section \ref{sec:class}, we transform the three-class classification problem for stars, galaxies and quasars to two binary classification tasks: stars versus extragalactic objects, and quasars versus galaxies to reduce the class imbalance and \textit{class-balance change}. In Section \ref{sec:photoz}, we calculate the photometric redshifts for GPQ candidates. In Section \ref{sec:results}, we present the GPQ candidate catalog and some statistical properties of the sample. We summarize the results in Section \ref{sec:summary}. Throughout this paper, we use AB magnitude for PS1 photometry and Vega magnitude for AllWISE photometry unless mentioned.


\section{Data} 
\label{sec:data}

We make use of optical and infrared photometric data from PS1 and AllWISE, and astrometric data from \gdr{2}. We also retrieve samples of spectroscopically identified objects from SDSS and LAMOST.

\subsection{PS1 DR1 photometry}
\label{sec:data-ps1}

Pan-STARRS1 \citep[PS1;][]{Chambers2016} has carried out a set of synoptic imaging sky surveys including the 3$\pi$ Steradian Survey and the Medium Deep Survey in 5 bands $(grizy_{P1})$. The mean 5$\sigma$ point source limiting sensitivities in the stacked 3$\pi$ Steradian Survey in $(grizy_{P1})$ are (23.3, 23.2, 23.1, 22.3, 21.4) and the single epoch 5$\sigma$ depths in $(grizy_{P1})$ are (22.0, 21.8, 21.5, 20.9, 19.7). For better astrometry in the crowded Galactic plane field, we use mean coordinates from the PS1 MeanObject table. Mean PSF magnitudes are used for all bands $(grizy_{P1})$, and mean Kron magnitudes \citep{1980ApJS...43..305K} are used for $i_{P1}$ and $z_{P1}$ bands. The Galactic extinction coefficients for $(grizy_{P1})$ are $R_{g},R_{r},R_{i},R_{z},R_{y}=3.5805,2.6133,1.9468,1.5097,1.2245$. These coefficients are calculated using $R_{\lambda}=A_{\lambda}/A_{V} \times R_{V}$, where $A_{\lambda}/A_{V}$ is the relative extinction value for band $\lambda$ given by a new optical to mid-IR extinction law \citep{wang2019optical}, and $R_{V}=3.1$.

We set a few constraints on the PS1 data to ensure the data quality. All sources should be: (i) detected in all PS1 bands ($grizy_{P1} > 0$) and significantly detected in $i_{P1}$ (error in PSF mag of $i_{P1}$ band $i\_err<0.2171$, equivalent to $i_{P1}$-band SNR larger than 5); (ii) not too bright in $i_{P1}$ to avoid possible saturation ($i>14$); (iii) measured with Kron magnitude \citep{1980ApJS...43..305K} in the $i_{P1}$ and $z_{P1}$ bands ($i_{\mathrm{Kron}}>0\ \&\ z_{\mathrm{Kron}}>0$). For simplification, we use ($g,r,i,z,y$) to represent the PSF magnitudes of PS1 bands ($grizy_{P1}$) in color indexes (e.g. $g-r$, $g-W1$) and derived quantities $i-i_{\mathrm{Kron}}$ and $z-z_{\mathrm{Kron}}$. The $z_{P1}$ PSF magnitude does not appear alone and will not be confused with the redshift symbol $z$.

\subsection{AllWISE photometry for point-like sources}
\label{sec:data-wise}

The AllWISE catalog is built upon the work of the Wide-field Infrared Survey Explorer mission \citep[WISE;][]{wright2010wide} by combining data from the WISE cryogenic and NEOWISE \citep{mainzer2011preliminary} post-cryogenic survey. WISE has 4 bands at 3.4, 4.6, 12, and 22 $\mu$m (W1, W2, W3, and W4). The 5$\sigma$ limiting magnitudes of the AllWISE catalog in W1, W2, W3, and W4 bands are 19.6, 19.3, 16.7, and 14.6 mag. The Galactic extinction coefficients for W1, W2, W3 used in this study are $R_{W1},R_{W2},R_{W3}=0.1209,0.0806,0.124$. These coefficients are also calculated with relative extinction $A_{\lambda}/A_{V}$ values from \citet{wang2019optical}.

We cross-match the PS1 sources with AllWISE using a radius of $1''$ to avoid source confusion in the dense fields of the Galactic plane. We also set a few constraints on the AllWISE data. All sources should be: (i) AllWISE point sources ($ext\_flg = 0$); (ii) not too bright to avoid possible saturation ($W1>8\ \&\ W2>7$); (iii) significantly detected in $W1$ and $W2$ bands ($W1snr>5\ \&\ W2snr>5$); (iv) unaffected by prioritized image artifacts in each band ($cc\_flags=$``0000''); (v) unblended with nearby detections, so that only one component is used in each profile-fitting for each source ($nb=1$). 

\subsection{\gdr{2} astrometry}

\gdr{2} \citep{prusti2016gaia,brown2018gaia} contains celestial positions and the apparent brightness in $G$ band for approximately $1.7$ billion sources. For $1.3$ billion of those sources, parallaxes and proper motions are available. Broad-band photometry in the {\gbp} (330--680~nm) and {\grp} (630--1050~nm) bands are available for $1.4$ billion sources. We use the proper motions and their uncertainties from the \gaia\ DR2 catalog (columns \texttt{pmra}, \texttt{pmra\_error}, \texttt{pmdec}, and \texttt{pmdec\_error}) to find quasars.

\subsection{SDSS Quasar Catalog: the fourteenth data release}

\begin{figure}[htb!]
    \centering
    \includegraphics[width=0.5
    \textwidth]{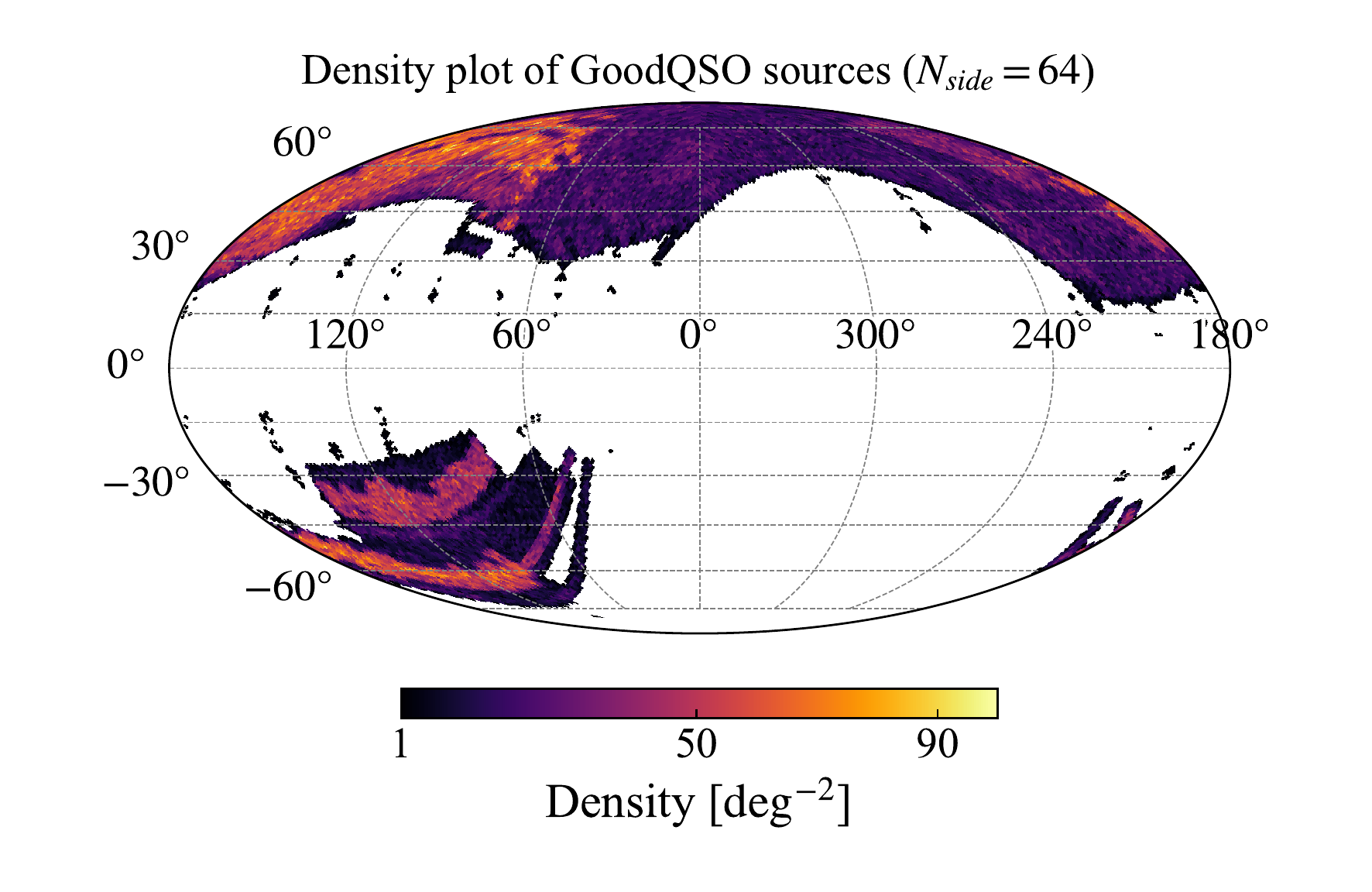}
    \caption{HEALPix density map of $GoodQSO$ sources from SDSS DR14Q (in Galactic coordinate system) with a median density of 20.3 $\mathrm{deg}^{-2}$. The HEALPix parameter $N_{side}=64$ and the sky area per pixel is 0.839 $\mathrm{deg}^2$.}
    \label{fig:densgoodqso}
\end{figure}

SDSS \citep[][]{2000AJ....120.1579Y} has mapped the high Galactic latitude northern sky and obtained imaging as well as spectroscopy data for millions of objects including stars, galaxies, and quasars. The 14th data release of the SDSS Quasar Catalog \citep[SDSS DR14Q;][]{paris2018sloan} contains 526,356 quasars. We cross-match the DR14Q catalog with PS1 and AllWISE both with a radius of $1''$. To ensure the data quality, we use the same constraints in Section \ref{sec:data-ps1} and Section \ref{sec:data-wise} to retrieve a subset of DR14Q. This subset has 289,271 sources and is denoted as $GoodQSO$ hereafter. As can be seen from the HEALPix \citep{2005ApJ...622..759G} density map of $GoodQSO$ (Figure \ref{fig:densgoodqso}), very few sources of $GoodQSO$ are located at $|b|\leq20^{\circ}$.

\subsection{SDSS spectroscopically identified stars and galaxies}

In order to compare high-$b$ sources with Galactic plane sources that we use, a sample of stars and a sample of galaxies are extracted from SpecPhotoAll table of SDSS Data Release 15 \citep{2017AJ....154...28B,2019ApJS..240...23A}. We cross-match both the star and galaxy sample with PS1 and AllWISE with a radius of $1''$. The SDSS star sample has 23,693 sources. We also apply quality constraints in Section \ref{sec:data-ps1} and Section \ref{sec:data-wise} to select galaxy subset with good photometry for later use. The resulted subset of galaxy (denoted as $GoodGal$ hereafter) has 1,635,053 sources. Most SDSS stars and galaxies are located at high Galactic latitude ($|b|>20^{\circ}$).

\subsection{Stars from LAMOST general catalog} \label{sec:data-lamost}
The Large Sky Area Multi-Object Fiber Spectroscopic Telescope (LAMOST, also called the Guoshoujing Telescope) is a special reflecting Schmidt telescope, the design of which allows both a large effective aperture of 3.6 m--4.9 m and a wide field of view of $5^{\circ}$ \citep{wang1996special,2004ChJAA...4....1S,2012RAA....12.1197C}. The LAMOST spectral survey \citep[][]{2012RAA....12..723Z,2012RAA....12.1243L,2015RAA....15.1095L} consists of two major components, i.e. the LAMOST Experiment for Galactic Understanding and Exploration \textbf{\citep[LEGUE;][]{2012RAA....12..735D}}, and the LAMOST ExtraGAlactic Survey (LEGAS). The LEGUE observes stars in different sky regions with different magnitude ranges, including the Galactic halo with $r<16.8$ mag at $|b|>30^{\circ}$, the Galactic anti-center with $14.0<r<17.8$ mag at $150^{\circ}\leq l \leq 210^{\circ}$ and $|b|<30^{\circ}$ \citep{yuan2015lamost}, as well as the Galactic disk with $r\lesssim16$ mag at $|b|\leq20^{\circ}$ with uniform coverage along Galactic longitude. The LEGAS mainly identifies galaxies and quasars that are within the SDSS footprint but complementary to the SDSS spectroscopic samples \textbf{\citep[e.g.][]{2016RAA....16...43S,2019ApJS..240....6Y}}. Nevertheless, extragalactic objects in the LEGUE plates are also targets of the LEGAS.
The LAMOST spectral survey has obtained the largest stellar spectra sample to date. We retrieve star sample from LAMOST general catalog from DR1 to DR7v0. A total number of 3,940,076 LAMOST stars meet the the same constraints in Section \ref{sec:data-ps1} and Section \ref{sec:data-wise}. From this LAMOST star sample, we select 1,334,577 Galactic plane stars with $|b| \leq 20^{\circ}$ (denoted as $T_{Star}$ hereafter). Most $T_{Star}$ sources are from the LEGUE survey and are brighter than 18 mag in $i_{P1}$ band.

\subsection{The Million Quasars (Milliquas) Catalog}
\label{sec:data-mlq}

The Million Quasars (Milliquas) Catalog \citep{2019arXiv191205614F} is a compilation of quasars and quasar candidates from the literature. The Milliquas v6.4c update includes 758,908 type-I QSOs and AGN up to 31 December 2019. We use this catalog to extract extant GPQ sample within PS1 footprint. There are 4,344 quasars (with ``Q'' label in the ``Descrip'' column) located in $|b|\leq~20^{\circ}$ in Milliquas v6.4c. Cross-matching these 4,344 known GPQs with PS1 and AllWISE both with a radius of $1^{\prime\prime}$ gives 2,757 sources. After applying same constraints as in Section \ref{sec:data-ps1} and Section \ref{sec:data-wise}, we get a subset of 1,853 sources. This Galactic plane subset of Milliquas quasars, denoted as $MLQSUB$, will be used later for candidate validation.

\section{Design of the Transfer Learning Framework} \label{sec:TLF}

\subsection{Dataset shift problem in the Galactic plane} 

The task of quasar selections can be described by classification problems in machine learning. Here we look into the three-class classification for stars, galaxies, and quasars with photometric data. The learning process requires two independent datasets for model training and model validation respectively. Training and validation sets can be two nonoverlapping subsets from a common parent sample with both features (colors and/or magnitues) and class labels (star, galaxy, and quasar). Usually the class labels are given by spectroscopic identifications. The classification algorithm learns a mapping relation from features to class labels with the training set. Often, the trained classification model (classifier) is applied to another dataset without class labels (i.e. no spectroscopic identifications), which is called application set or test set. The classifier takes features from the test set as inputs $X$ (a.k.a. covariates) and gives class labels as outputs $Y$.

A basic assumption for traditional machine learning is that training and test data follow the same probability distribution \citep{bishop2006pattern,hastie2009elements,vapnik2013nature}. However, this assumption no longer holds if we use high-$b$ data for model training and low-$b$ data for application, because the joint distribution of inputs and outputs $P(X,Y)$ differs between training and test data \citep[i.e. \textit{dataset shift};][]{quionero2009dataset}. 

For our GPQ selections, the dataset shift includes changes in both source colors and prior probabilities of different classes. Sources in the Galactic plane become fainter and redder than those at high Galactic latitude due to greater reddening, which changes the distribution of input features and the conditional probability of the output labels given the inputs $P(Y|X)$ \citep[i.e. \textit{covariate shift};][]{shimodaira2000improving,sugiyama2012machine}. Prior probabilities of stars are much higher than those of quasars (and galaxies) in the Galactic plane, which means the marginal probability ${P(Y)}$ differs from that at high Galactic latitude \citep[i.e. \textit{class-balance change};][]{saerens2002adjusting,du2014semi}. Moreover, class ratio between extragalactic objects and stars may vary significantly from one place to another in the Galactic plane, which we refer to as ``internal'' class-balance change of the test data.

\textit{Transfer learning} can be applied to improve the learning performance under dataset shift from a source domain to the target domain \citep[see a review in][]{pan2009survey}, where \textit{domain} is a set $\domain$ that consists of a feature space $\mathcal{X}$ and a marginal probability distribution $P(X)$, $\domain=\{\mathcal{X},P(X)\}$. For our classification task, the source domain data are from high Galactic latitude ($|b|>20^{\circ}$) and the target domain data are from the Galactic plane ($|b|\leq 20^{\circ}$). In this study we only care about areas at $\delta > -30^{\circ}$ due to the limit of PS1 survey coverage. Comparison of some properties of the source and target domains are listed in Table \ref{table:domains}. 

\begin{deluxetable*}{ccccc}[htb!]

    \tablecaption{Comparison of the two domains of learning\label{table:domains}}
    \tablenum{1}
    \tablehead{\colhead{Domains of learning} & \colhead{Location} & \colhead{Labels of stars} & \colhead{Labels of quasars/galaxies} & \colhead{Internal class-balance change}} 
    
    \startdata
    Source domain & $|b| > 20^{\circ}$ & Available & Available & Moderate \\
    Target domain & $|b| \leq 20^{\circ}$ & Available & Unavailable & Severe \\
    \enddata
\end{deluxetable*}

As large numbers of stars, quasars and galaxies have been spectroscopically identified at $|b|>20^{\circ}$, labels for these three classes are available in the source domain. Since spectroscopically identified samples of quasars and galaxies are significantly lacked at $|b|\leq 20^{\circ}$, labels for these two classes are unavailable in the target domain. Nevertheless, labels of many stars in the target domain are available with the help of LAMOST spectroscopic survey.

According to the classification scheme for different settings of transfer learning by \citet{pan2009survey}, the set-up of classification in the Galactic plane can be categorized into \textit{Transductive Transfer Learning}, where source domain labels are available and target domain labels are unavailable. A popular approach to \textit{Transductive Transfer Learning} is \textit{Feature-based Transfer} \citep[e.g.][]{blitzer2006domain,argyriou2006multi}, which reduces the difference between the source and target domain through feature transformation in either one or both of the domains.

To solve the dataset shift problem of classification in the Galactic plane, we borrow the idea of \textit{Feature-based Transfer Learning}. Using the mapping relation between the features of high-$b$ and low-$b$ objects, we can generate mock samples of quasars and galaxies in the Galactic plane to simulate the covariate change of their colors and magnitudes. The LAMOST Galactic plane stars also contribute to a more accurate probability distribution of data in the target domain. To reduce the effect of class-balance change, we manually go through two binary classification steps rather than running a three-class classification algorithm only once.

\subsection{Modelling covariate change with mock samples}

As data of LAMOST Galactic plane stars are available, we only focus on reducing the differences in features of extragalactic objects between training and test data. For our classification problem, all features will be constructed with photometric data from PS1 and AllWISE. We assume that the differences in photometric properties between extragalactic objects in the Galactic plane and those off the plane are only caused by different extinctions/reddening along their sight-lines. In this way, we can simply generate mock extragalactic objects behind the Galactic plane with data obtained at high Galactic latitude, using the mapping relation determined by the Galactic extinction law and Galactic dust map. 

The covariate change can then be shown as color change of extragalactic objects on a set of color-color diagrams. Our classification will perform better by adding mock samples of quasars and galaxies behind the Galactic plane into the training set. 

\subsection{Dealing with class imbalance and class-balance change in machine learning} \label{sec:TLFsub3}

With the data-level improvements above, the covariate change can be reduced. Efforts on the algorithm level are required to handle the class imbalance and class-balance change. During the GPQ selections, instead of performing a star-quasar binary classification, we additionally take galaxies into account and perform a three-class classification. 

Many machine learning software packages support multi-class classification jobs, by transforming the task into multiple binary classification problems. However, the built-in treatment is often inflexible and sometimes destructive when dealing with class imbalance problems. For example, in the scenario of using \textit{one-vs-rest} (also known as \textit{one-vs-all}) strategy for multi-class classification, at some stages, samples of one class are regarded as the positive samples while all samples of other classes are regarded as negative samples. Even if all the classes in the training set have a same sample size, the binary classification situation is imbalanced as the positive class (the ``one'') has less samples than the negative class (the ``rest''). In our case, the GPQ set (in both training and test set) has significantly less samples than the sets of galaxies and stars, thus severe class imbalance will happen.

To reduce the disadvantage of \textit{one-vs-rest} strategy which is commonly used in machine learning algorithms, we convert this three-class classification problem into two binary classification problems manually. In the first step, the Galactic plane sources are classified into two classes: stars and extragalactic objects. Extragalactic objects are then classified into quasars and galaxies in the second step. By combining the two minority classes of quasar and galaxy into one, we would expect the class imbalance to be better controlled in the first step. The physical basis for merging the quasar and galaxy classes is that quasars are a special type of galaxies. For the second step, we expect that the quasar-to-galaxy ratio to be nearly constant across different locations in the Galactic plane. Thus the variable quasar-to-star or galaxy-to-star ratio is avoided and the internal class-balance change is lessened in the learning process.

\section{Mock catalogs for quasars and galaxies behind the Galactic plane} \label{sec:mock}

In order to construct training samples for extragalactic objects, as well as understand the covariate shift of them from high Galactic latitude to the Galactic plane, we synthesize quasars and galaxies behind the Galactic plane using $GoodQSO$ and $GoodGal$ samples. The synthesis is plausible if we assume the distribution of quasars on the celestial sphere is homogeneous and isotropic on large scale, just as the Cosmological principle has suggested. We not only observe the changes in colors of quasars and galaxies as they are placed in low Galactic latitudes in this modeling process, but also get a rough estimation on the sky distributions of the sources that could be detected by a certain sky survey. 

\subsection{Synthesizing procedures} \label{sec:synt_proc}

Let $E$ be a set of extragalactic objects ($E$ can be $GoodQSO$ or $GoodGal$). The synthesis process consists of following steps.

\textit{1. Correcting for extinctions.} Extinctions of objects in set $E$ are corrected according to a two-dimensional dust map provided by \citet[][hereafter \citetalias{abergel2014planck}]{abergel2014planck}, and the optical to mid-IR extinction law from \citet{wang2019optical} with $R_{V}=3.1$. The $E(B-V)$ values are retrieved using a Python module, dustmaps \citep{2018JOSS....3..695M}.

\textit{2. Assigning new locations.} We generate a random sample of points that are uniformly distributed on the sky with $|b|\leq 20^{\circ}$. The number of these random points is equal to the sample size of $E$. Coordinates of these points are randomly assigned to objects of $E$ as their new locations. Now we get a new set $E_{m}$ ($MockGPQ$, $MockGal$) without line-of-sight extinctions.

\textit{3. Adding new extinctions.} We add extinctions to the $E_{m}$ sample using the \citetalias{abergel2014planck} dust map based on their new (mock) locations.

\textit{4. Setting limiting magnitudes.} We obtain a subset of $E_{m}$ by choosing sources brighter than the PS1 single epoch 5$\sigma$ depths in all PS1 passbands: $(grizy_{P1}) < (22.0,\ 21.8,\ 21.5,\ 20.9,\ 19.7)$. This subset, denoted as $E_{gm}$ ($GoodMockGPQ$, $GoodMockGal$), represents ``good'' mock sample that can be detected by PS1 survey in all bands. However, we don't apply similar constraints to AllWISE bands as the magnitude which corresponds to a 5$\sigma$ sensitivity varies with location. Also, this extinction-selection effect relies more on the optical survey depth than the IR survey depth. Factors such as observation strategies and source confusions in dense fields are not taken into consideration in this step. Therefore we may overestimate the detection rate of GPQs (and galaxies) through this synthesis. We select sources within the PS1 footprint (i.e. $\delta \geq -30^{\circ}$) and obtain set $E_{gm\textnormal{-}\mathrm{PS1}}$ ($GoodMockGPQ\textnormal{-PS1}$, $GoodMockGal\textnormal{-PS1}$).

\textit{5. Constructing training sets with mock and real data.} For mock quasars that are not included in $GoodMockGPQ$-PS1, their original counterparts (high-$b$ quasars in the input set $GoodQSO$; denoted as $C_{QSO}$) are also added to the training and validation sets along with $GoodMockGPQ$-PS1. For mock galaxies that are not included in$GoodMockGal$-PS1, 25\% of their original counterparts (high-$b$ galaxies in the input set $GoodGal$; denoted as $C_{Gal}$) are added to the training and validation sets. The resulted quasar and galaxy samples for training and validation are denoted as $T_{QSO}$ and $T_{Gal}$, respectively. $T_{QSO}$, $T_{Gal}$, and the LAMOST Galactic plane star sample $T_{Star}$ form the training and validation sets for machine learning classification.

By adding good mock samples and real data ($C_{QSO}$ and $C_{Gal}$) together instead of using only good mock samples as training data for quasars or galaxies, we increase the data diversity as well as sample size of the training set. This \textit{data diversification} ensures that the training set can provide more discriminative information for the machine learning model \citep{gong2019diversity}. In addition, more training data can help reduce overfitting.

\begin{figure}[htb!]
    \centering
    \includegraphics[width=0.5\textwidth]{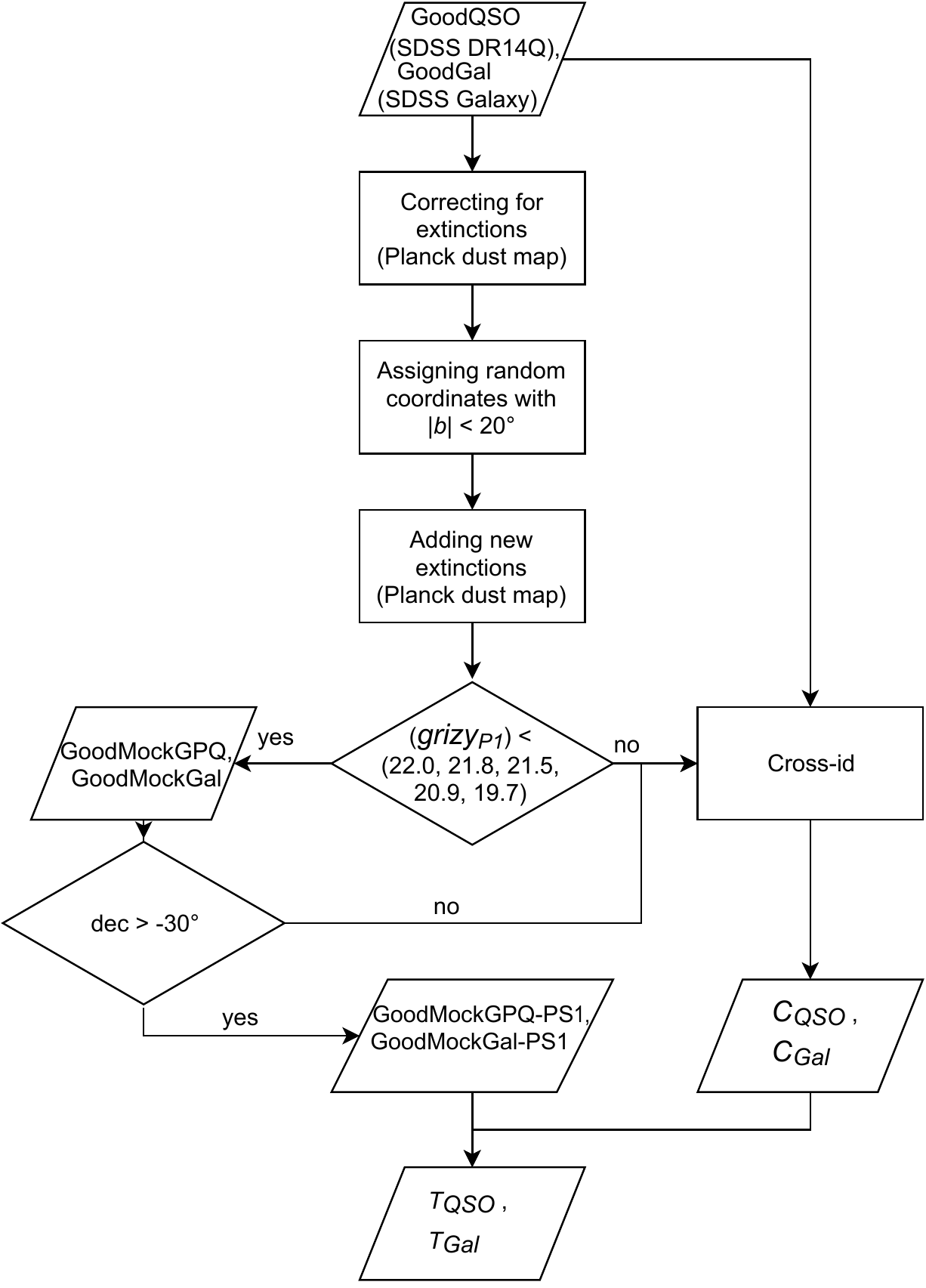}
    \caption{Flowchart of synthesizing procedures for the mock catalogs.}
    \label{fig:flow_mock}
\end{figure}

The flowchart of the synthesizing procedures is displayed in Figure \ref{fig:flow_mock}.

In the synthesizing process, we adopt the \citetalias{abergel2014planck} dust map because it detects dust at a greater depth and better estimates the 2D extinctions in the Galactic plane than do the dust maps constructed with stellar photometry \citep[e.g.][]{2018MNRAS.478..651G,2019ApJ...887...93G}. We assume a uniform $R_{V}=3.1$ although $R_{V}$ varies slightly in the Galaxy with a dispersion of about 0.18 \citep{2016ApJ...821...78S}. Such minor variations in $R_{V}$ can lead to small uncertainties of magnitudes and colors of individual mock quasars (galaxies), but have limited impacts on the statistical properties of the training sample because mock sources with large extinctions (and thus large uncertainties caused by $R_{V}$ variations) are removed by the magnitude limits, as we shall see in Section \ref{sec:synt_res}.

\subsection{Synthesizing results and \textit{dataset shift}} \label{sec:synt_res}

\begin{figure*}[htb!]
    \centering
    \includegraphics[width=1\textwidth]{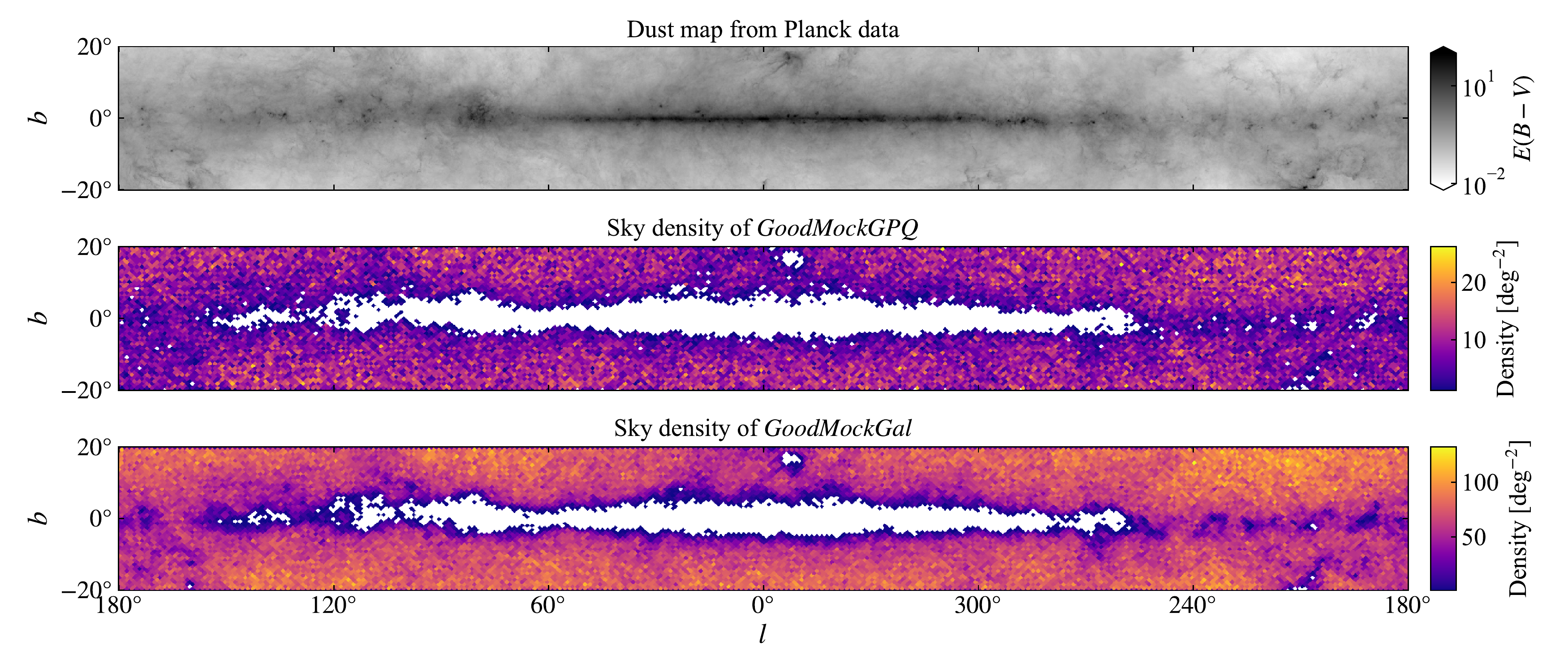}
    \caption{Dust extinction map along the Galactic plane retrived from \citetalias{abergel2014planck} (top panel), sky density of $GoodMockGPQ$ (middle panel) and $GoodMockGal$ (bottom panel).}
    \label{fig:ebv_goodmock}
\end{figure*}

\begin{figure*}[htb!]
    \centering
    \gridline{\fig{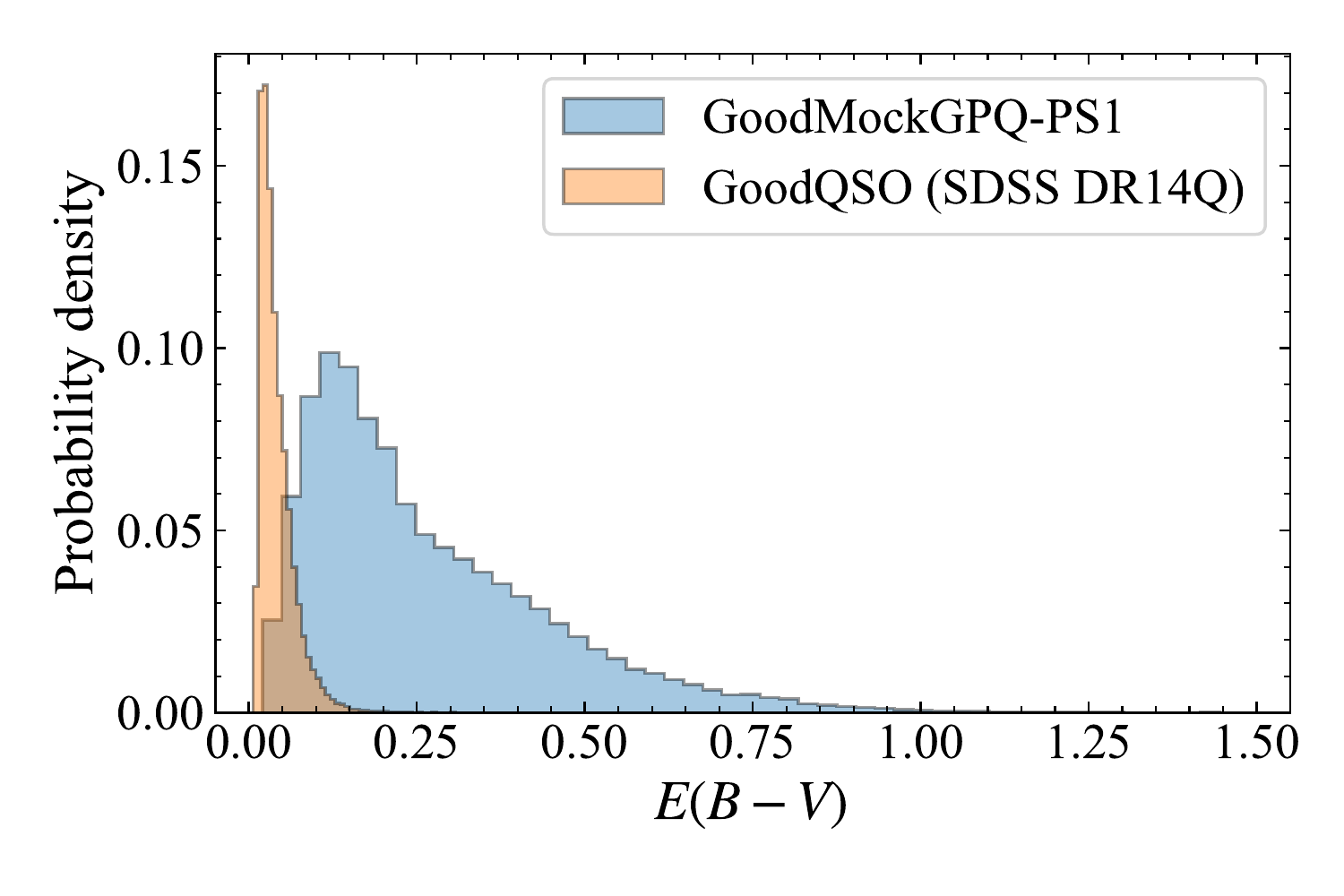}{0.49\textwidth}{(a)}
              \fig{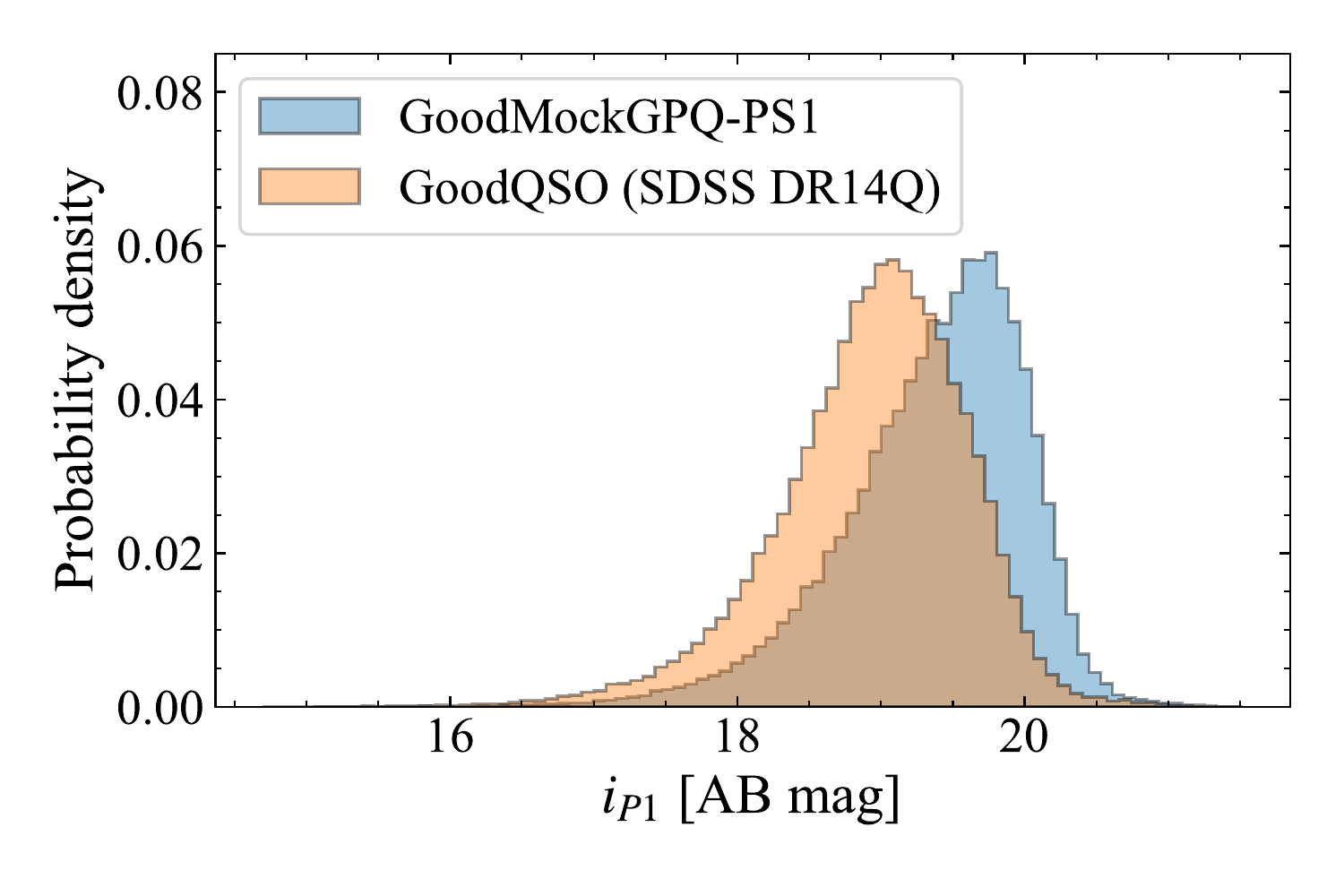}{0.49\textwidth}{(b)}}
    \caption{Histograms of (a) line-of-sight $E(B-V)$ and (b) $i_{P1}$ band magnitudes of $GoodMockGPQ$-PS1 and $GoodQSO$. The $i_{P1}$ band magnitudes are not corrected for extinction.}
    \label{fig:hist_ebv}
\end{figure*}

We define the extinction-based selection rate in the Galactic plane as $R = |E_{gm}| / |E|$, where $|E|$ is the cardinality, i.e. number of elements/sources of set $E$. The source numbers of the input samples are $|GoodQSO|=289,271$ and $|GoodGal|=1,635,053$; the source numbers of the output samples are $|GoodMockGPQ|=101,482$ and $|GoodMockGal|=771,392$. Therefore the selection rates for GPQs and galaxies are $R_{GPQ} = |GoodMockGPQ|/|GoodQSO| = 0.35$ and $R_{Gal} = |GoodMockGal|/|GoodGal|= 0.47$, respectively. The selection rate of galaxies is higher than that of GPQs because the input galaxies are on average brighter than the input quasars. With Step 2 in Section \ref{sec:synt_proc}, sources of $MockGPQ$ and $MockGal$ are randomly and evenly distributed in the Galactic plane ($|b|<20^{\circ}$). But after Step 4, the densities of remaining sources ($GoodMockGPQ$ and $GoodMockGal$) are inversely related to the dust map (Figure \ref{fig:ebv_goodmock}): more extragalactic sources remain detectable in regions with smaller $E(B-V)$, and voids of detection present at regions with large $E(B-V)$. A sky survey deeper than PS1 might help make up some fraction of the gap in the middle of the Galactic plane. The $GoodMockGPQ$ sample is sparser compared to $GoodMockGal$, simply because the input quasars are fewer than galaxies. Most sources of $GoodMockGPQ$ and $GoodMockGal$ have line-of-sight color excess of $E(B-V)<1.5$, which corresponds to extinction of $A_{V}<4.65$ with $R_{V}=3.1$. The medians of line-of-sight $E(B-V)$ of $GoodMockGPQ$-PS1 and $GoodQSO$ are 0.21 and 0.03, respectively. In general, $GoodMockGPQ$-PS1 sample has significantly larger $E(B-V)$ compared to $GoodQSO$ (see Figure \ref{fig:hist_ebv} (a)). Therefore covariate change for color indexes from high-$b$ to low-$b$ regions cannot be ignored. In addtion, $GoodMockGPQ$-PS1 sources are fainter than $GoodQSO$ sources (Figure \ref{fig:hist_ebv} (b)).

\begin{figure*}[htb!]
    \gridline{\fig{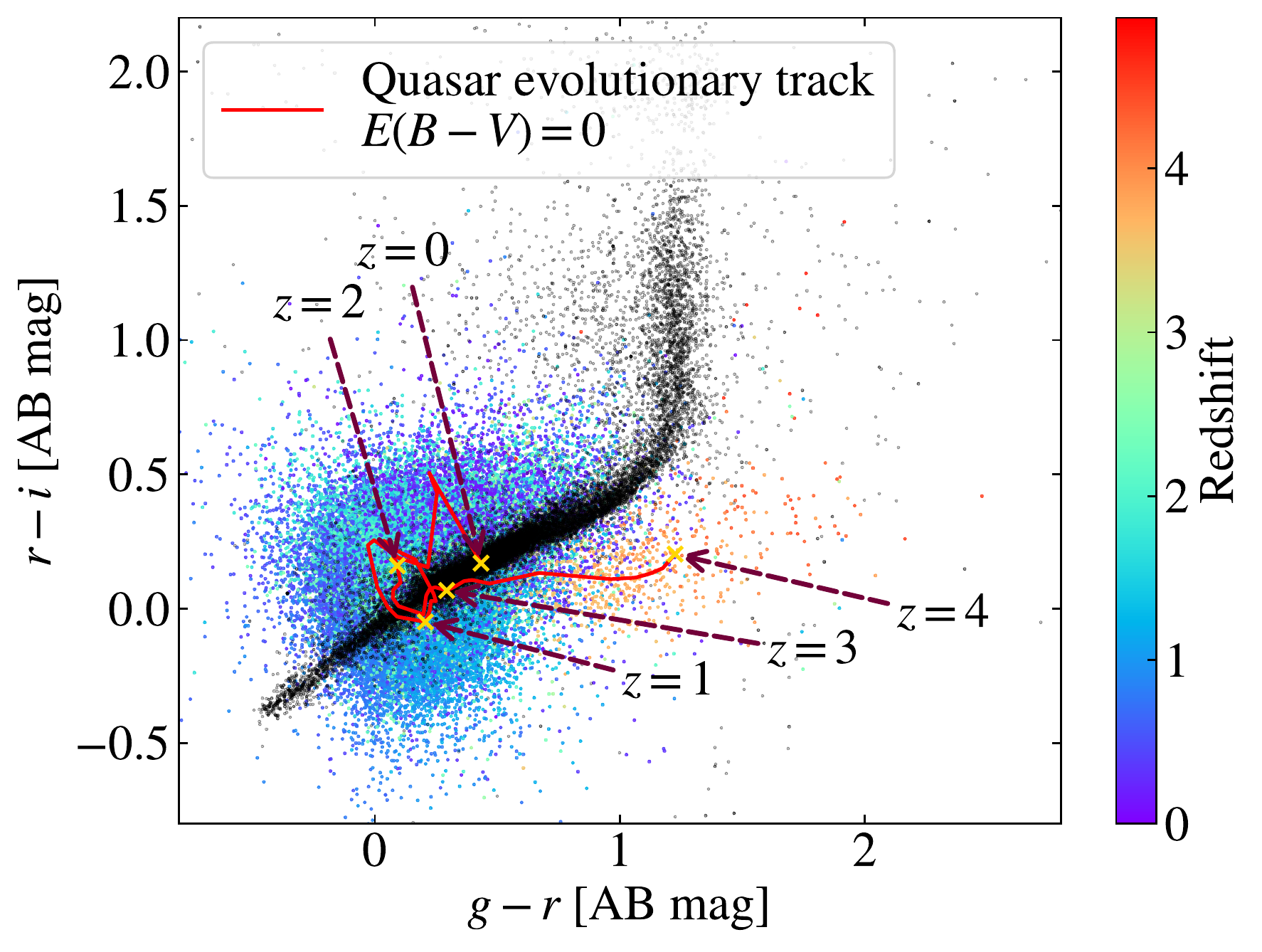}{0.34\textwidth}{(1a)}
              \fig{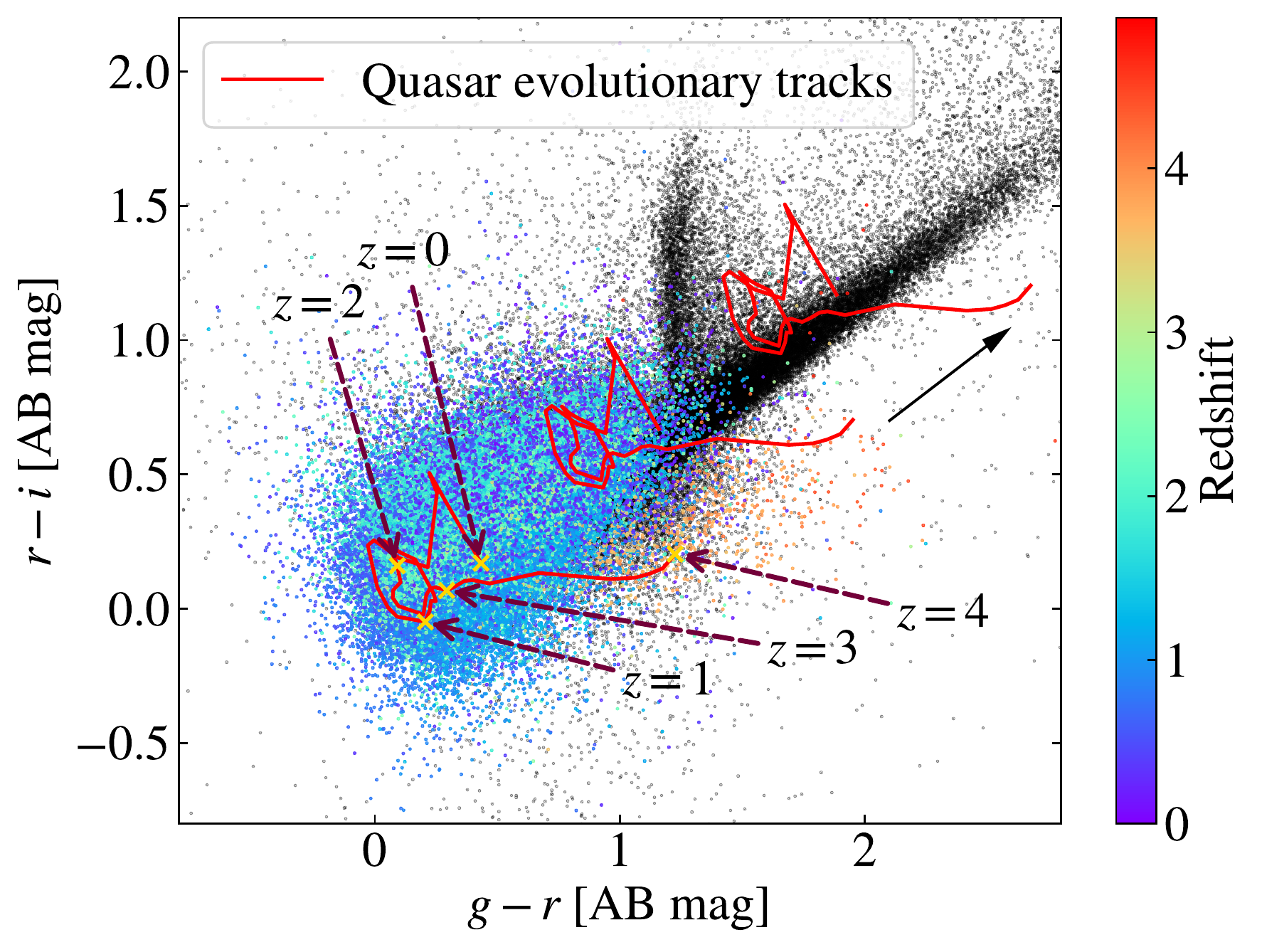}{0.34\textwidth}{(1b)}
              \fig{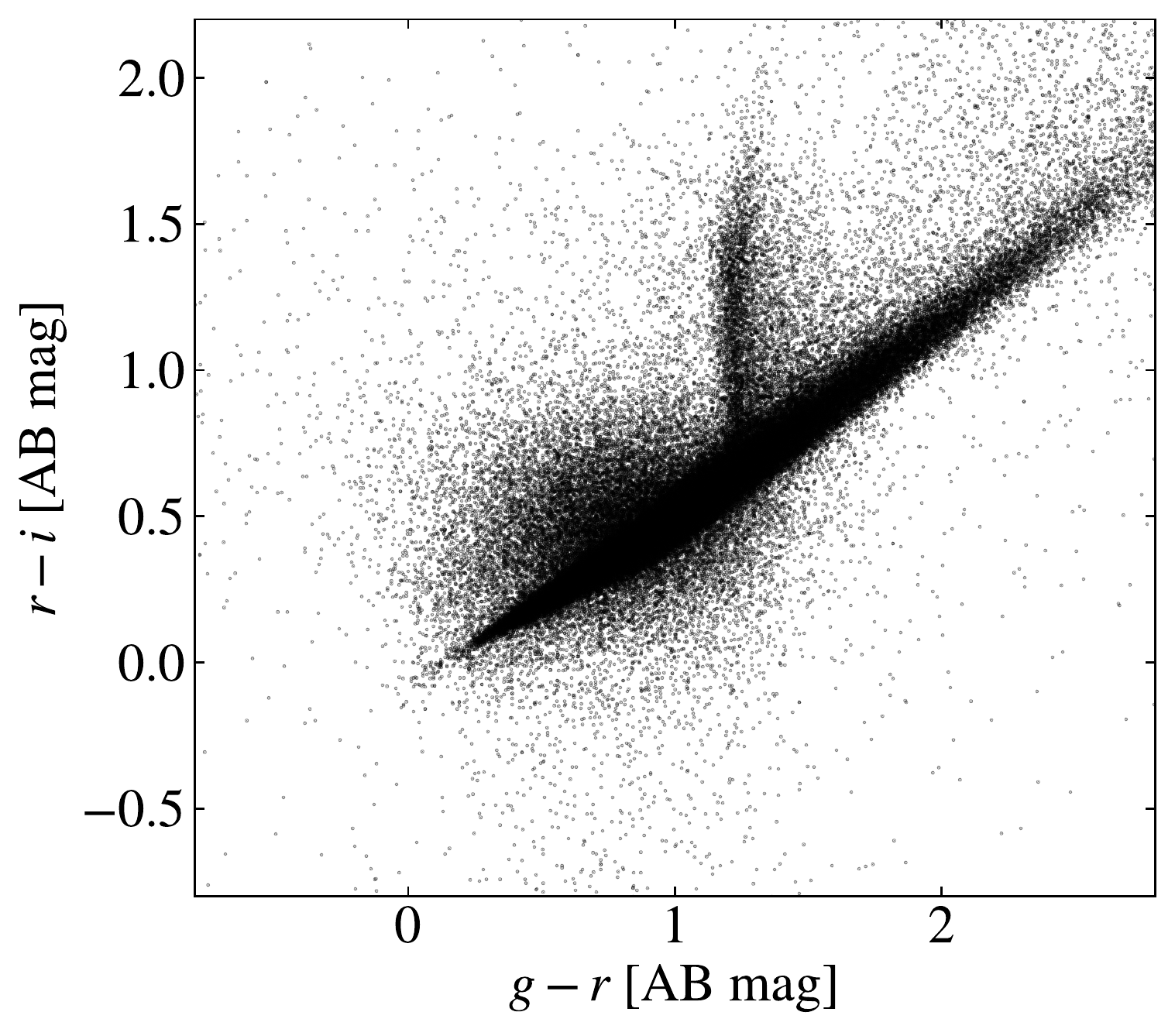}{0.29\textwidth}{(1c)}}
    \gridline{\fig{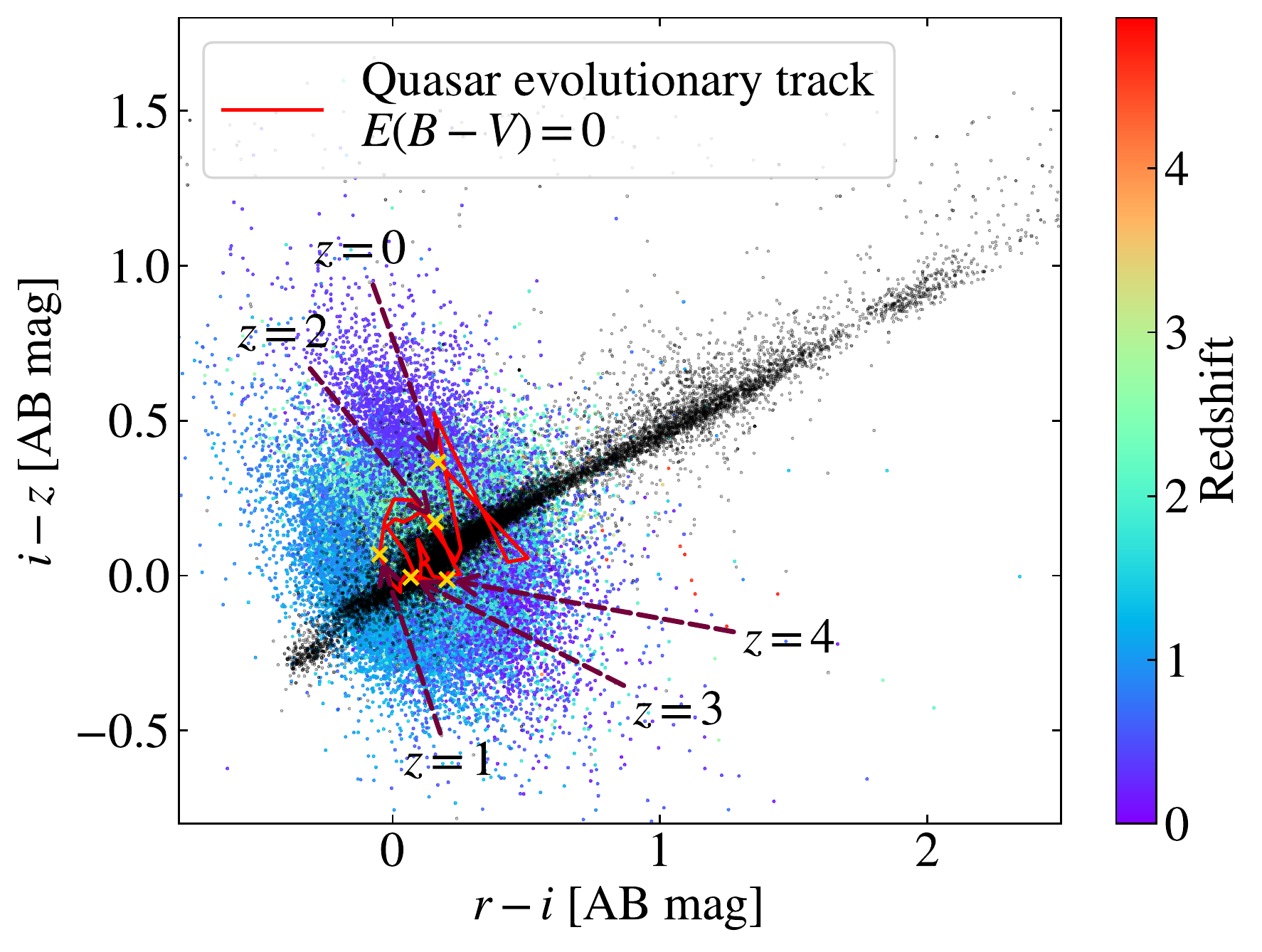}{0.34\textwidth}{(2a)}
              \fig{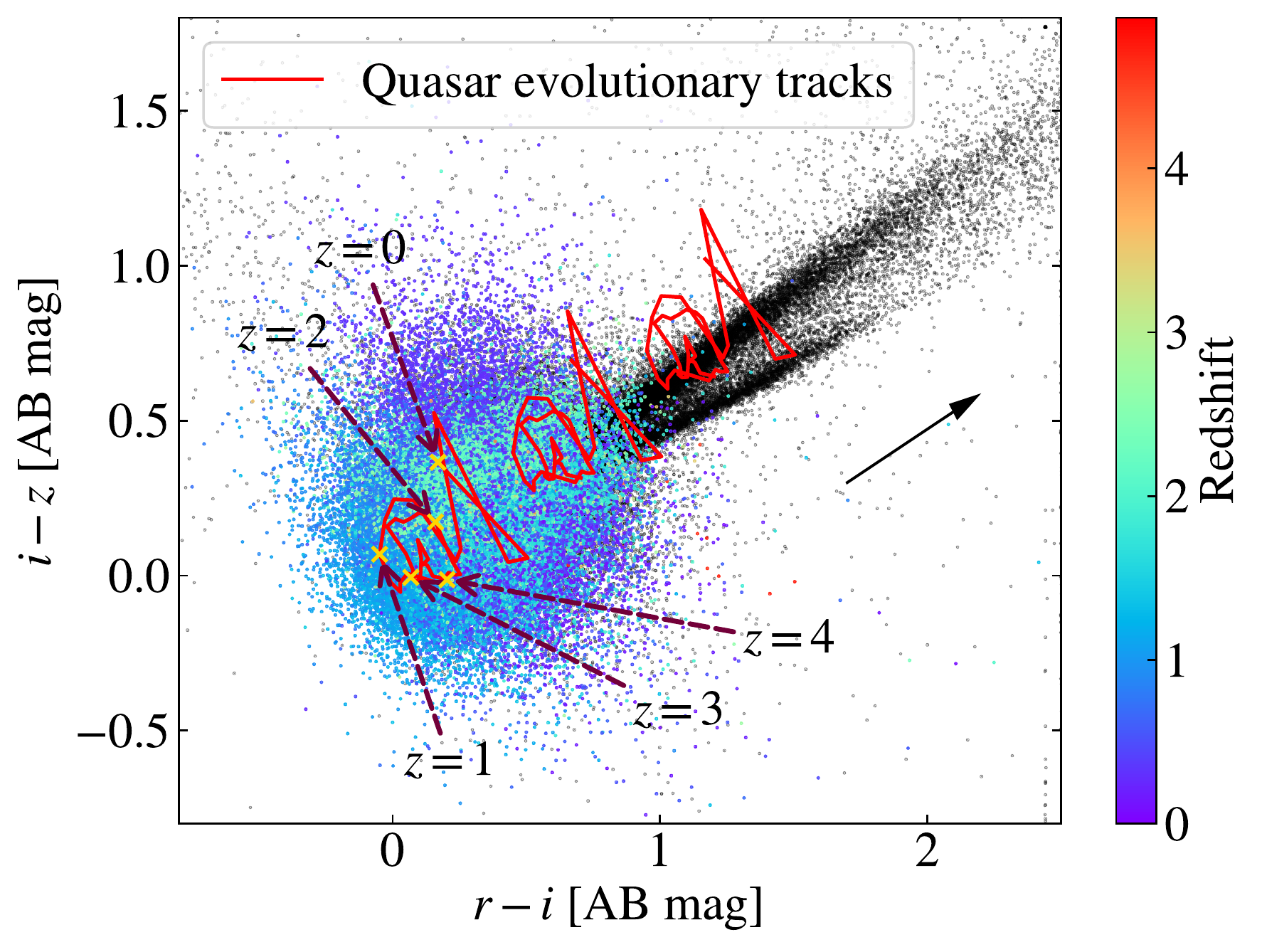}{0.34\textwidth}{(2b)}
              \fig{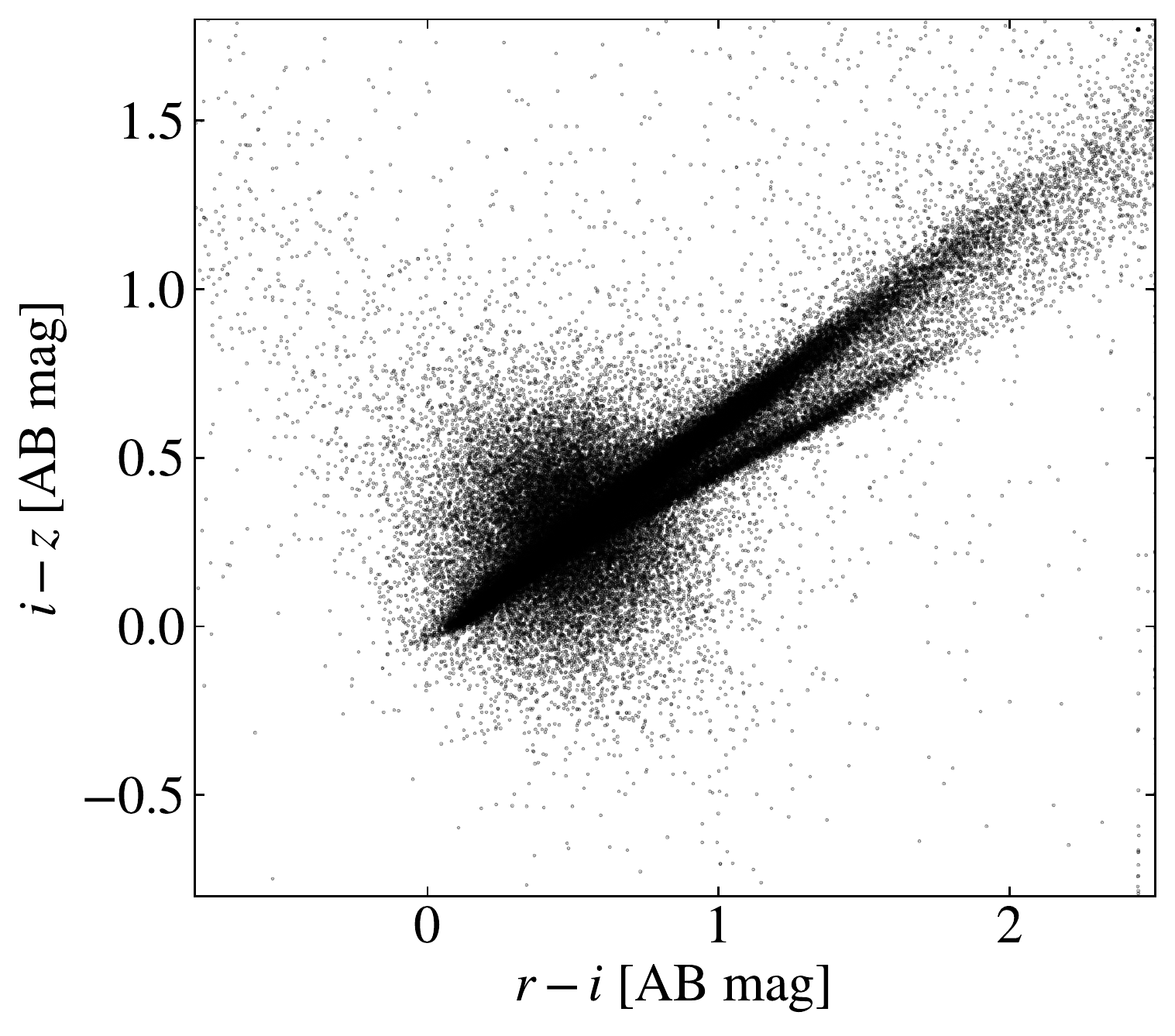}{0.29\textwidth}{(2c)}}
    \gridline{\fig{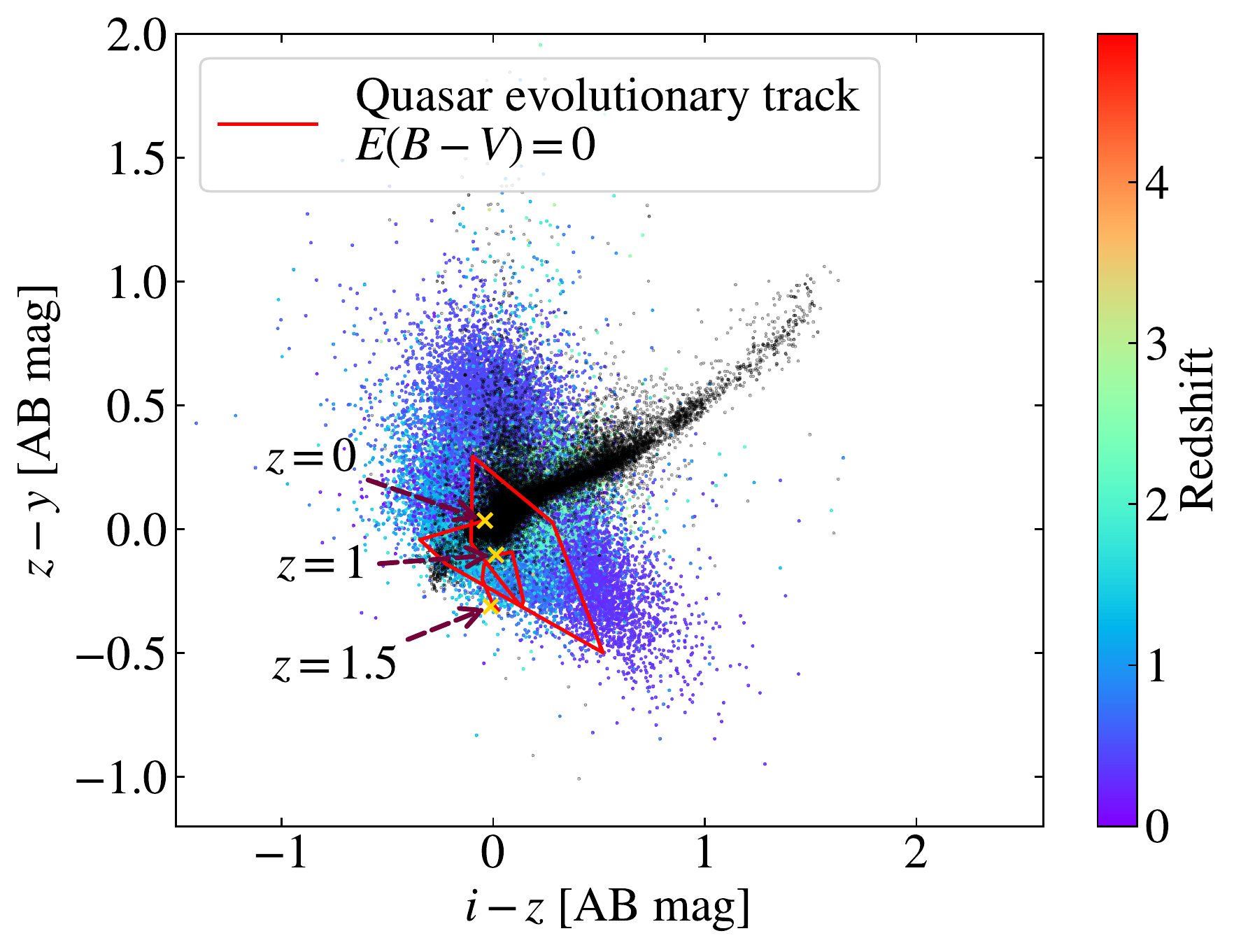}{0.34\textwidth}{(3a)}
              \fig{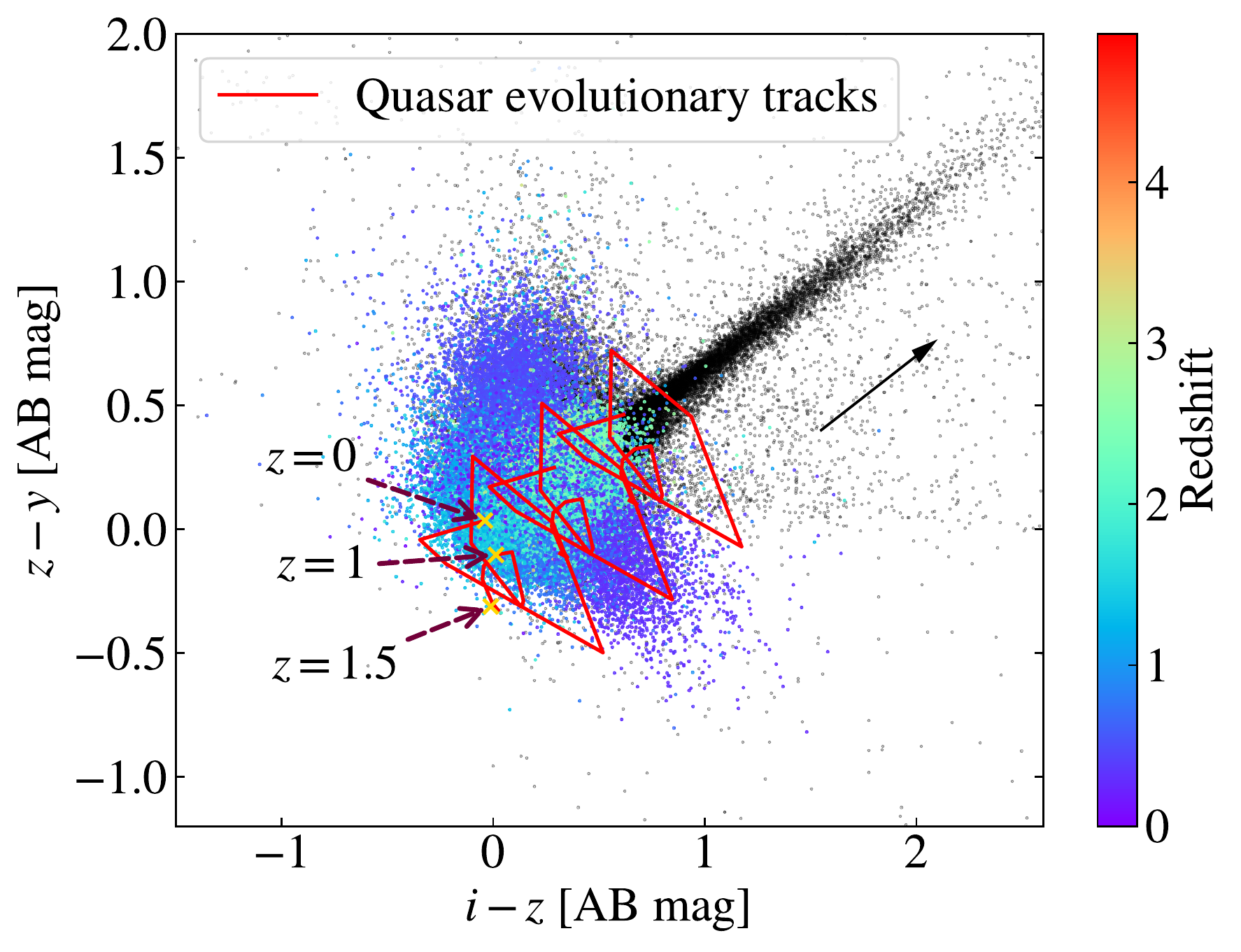}{0.34\textwidth}{(3b)}
              \fig{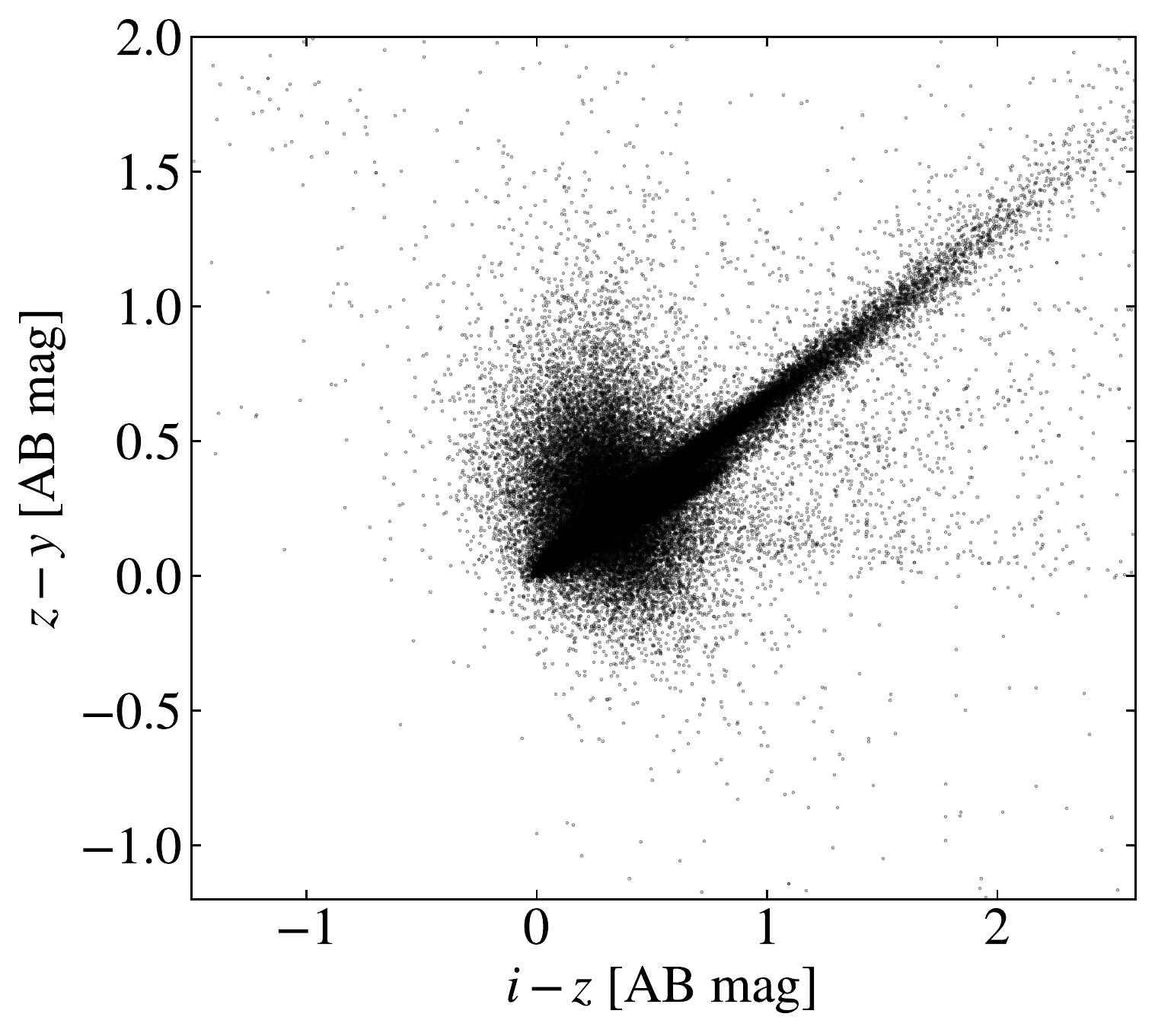}{0.29\textwidth}{(3c)}}
    \caption{Color-color diagrams of (1a, 2a, 3a) reddening-corrected $GoodQSO$ (color coded dots) and SDSS stars (black dots); (1b, 2b, 3b) $GoodMockGPQ$-PS1 (color coded dots) and a random sample of PS1-AllWISE point sources (black dots) in Galactic plane ($|b|\leq 20^{\circ}$); and (1c, 2c, 3c) the same sample of PS1-AllWISE point sources in Galactic plane. For panel on the left (1a, 2a, 3a), quasar evolutionary tracks from redshift 0 to 4 without Galactic reddening ($E(B-V)=0$) are shown in red. While for panel in the middle (1b, 2b, 3b), quasar evolutionary tracks from redshift 0 to 4 with $E(B-V)=0,0.75,1.5$ are displayed. The black arrows (reddening vectors) indicate evolution directions of source colors with increasing $E(B-V)$. Yellow crosses denote points on the quasar evolutionary tracks without Galactic reddening with $z = 0,1,2,3,4$ in (1b, 2b) and $z = 0,1,1.5$ in (3b). The quasar evolutionary tracks are calculated using the optical composite quasar spectrum from \citet{berk2001composite} and near-IR composite quasar spectrum from \citet{2006ApJ...640..579G}.}
    \label{fig:ccds}
\end{figure*}

\begin{figure*}[htb!]

    \gridline{\fig{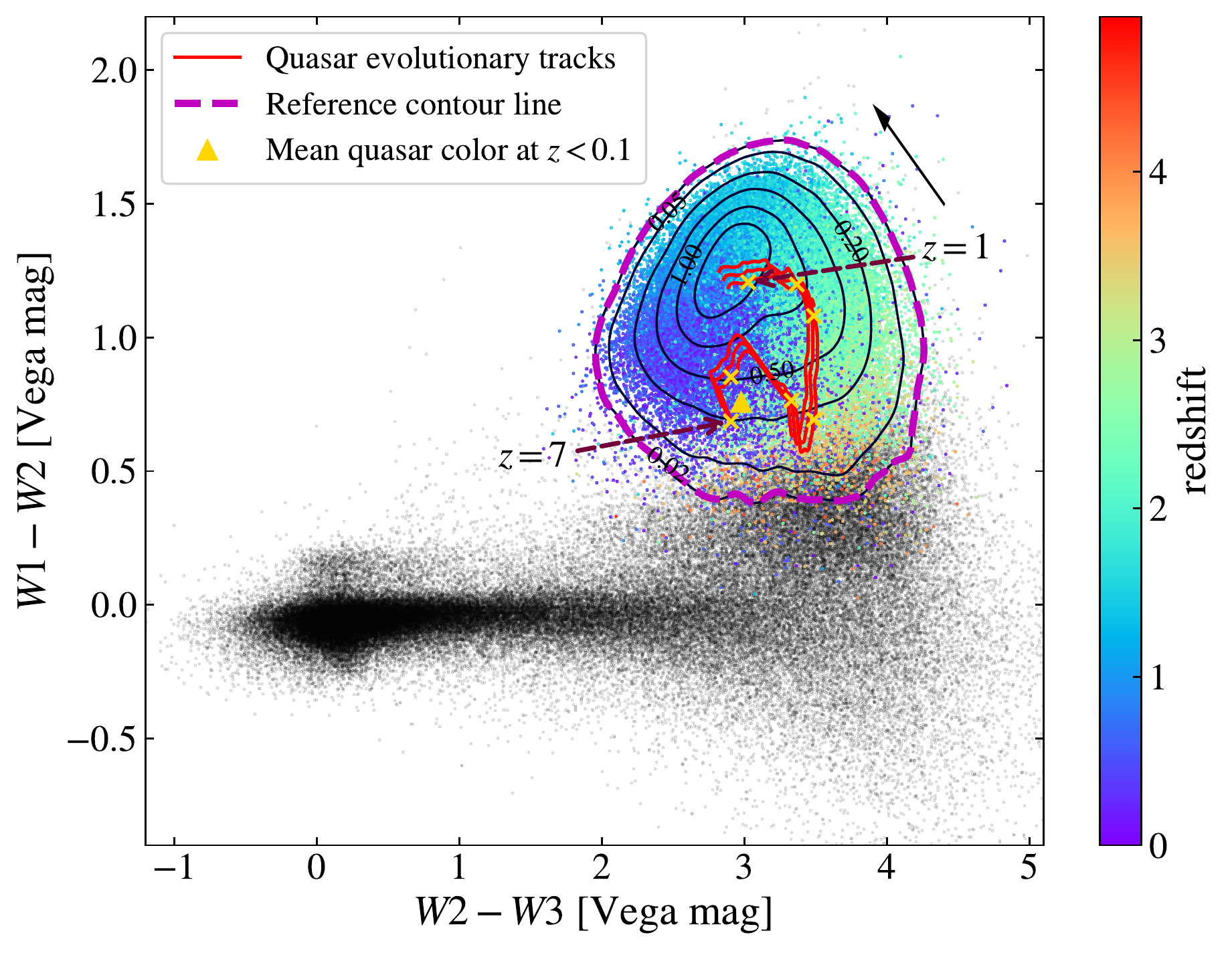}{0.49\textwidth}{(a)}
              \fig{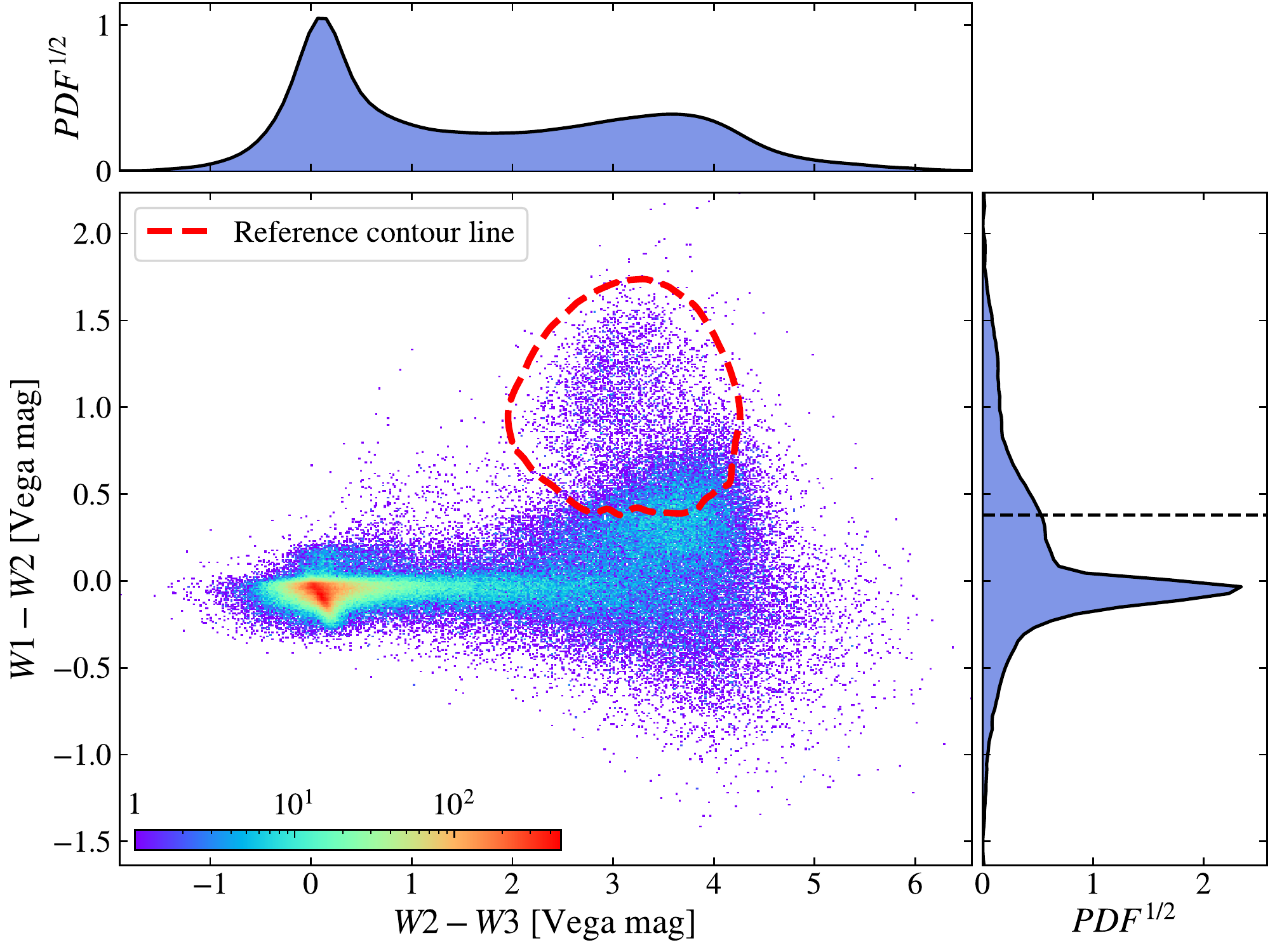}{0.49\textwidth}{(b)}}
    \caption{(a) The $W1-W2$ versus $W2-W3$ color-color diagram of  $GoodMockGPQ$-PS1 (color coded dots) and PS1-AllWISE point sources (black dots) in Galactic plane ($|b|\leq 20^{\circ}$), and (b) The $W1-W2$ versus $W2-W3$ color-color diagram of PS1-AllWISE point sources in Galactic plane (color coded according to density) and marginal probability distribution plots in $W1-W2$ and $W2-W3$ axes. For panel (a), contour lines from 2d Kernal Density Estimation (KDE) for $GoodMockGPQ$-PS1 sample is plotted, and a reference contour line (in magenta) with density of 0.02 is specified. Quasar evolutionary tracks that begin at $z=0.6$ (due to the template coverage) and end at $z=7$ with $E(B-V)=0,0.75,1.5$ are displayed. The black arrow (reddening vector) indicates evolution direction of source colors with increasing $E(B-V)$. Gold cross marks denotes points on the quasar evolutionary tracks without Galactic reddening with $z = 1,\cdots,7$. The mean color of quasars with $z<0.1$ is marked with gold triangle. For panel (b), the same reference contour line of $GoodMockGPQ$-PS1 sample from panel (a) is shown in red dashed line; the lowest $W1-W2$ value of the reference contour line is plotted as black dashed line over the $W1-W2$ marginal distribution plot. The quasar evolutionary tracks are calculated based on the template from \citet{hernan2016near}.}
    \label{fig:ccdwise}
\end{figure*}

A series of color-color diagrams for $GoodQSO$ and $GoodMockGPQ$-PS1 along with SDSS stars and Galactic plane point sources are shown in Figure \ref{fig:ccds} and Figure \ref{fig:ccdwise}. In Figure \ref{fig:ccds}, from the left to the middle panels, the covariate change of colors of quasars from high Galactic laititude to the Galactic plane can be directly observed. The Galactic reddening makes the cluster of GPQs in a color-color plane extend towards redder colors (to the upper right along the reddening vector) and scatter more than high-$b$ quasars. The scattering is greater in color indexes of bluer bands, while less at redder bands. This trend is also observable in the quasar evolutionary tracks with $E(B-V)=0,0.75,1.5$. From the top to the bottom panels in Figure \ref{fig:ccds}, the distance between two quasar evolutionary tracks with different reddening decreases.

The covariate change of stellar colors is also evident from the color-color diagrams. The stellar loci are simple and clear for high-$b$ (SDSS) stars (see Figure \ref{fig:ccds} (1a, 2a, 3a)). However, additional spikes along the direction of increasing $E(B-V)$ appear in the stellar loci of Galactic plane stars due to reddening, as can be seen from Figure \ref{fig:ccds} (1b, 1c, 2b, 2c). Therefore we expect to better distinguish stars from other sources using Galactic plane stars from LAMOST instead of high-$b$ SDSS stars in the training set. 

Since mid-IR bands are less sensitive to extinction and reddening, the covariate change of AllWISE colors are less obvious than that of PS1 colors. For instance, in Figure \ref{fig:ccdwise} (a), the quasar evolutionary tracks with $E(B-V)=0,0.75,1.5$ stay very close to each other. Since the AllWISE colors of quasars do not change much from high Galactic latitude to the Galactic plane, we can use the $W1-W2$ versus $W2-W3$ color-color diagram to examine the purity of the final quasar candidates, by comparing probability distributions of the candidates and $GoodMockGPQ$-PS1 sources.

The AllWISE color-color diagram also gives ``hardness'' information on the classification problems that separating quasars from galaxies is harder than separating quasars from stars. In general, quasars have redder $W1-W2$ and $W2-W3$ colors than stars and galaxies due to the power-law SEDs and hot dust of quasars. From Figure \ref{fig:ccdwise} (b), we can recognize stellar locus ($W1-W2\approx 0$; on the lower-left) and galaxy locus ($W1-W2\approx 0.5,\ W2-W3\approx 3.5$; on the middle-right) from the density plot. The quasar reference contour line marks a ``quasar region'' where most quasars are located in the $W1-W2$ versus $W2-W3$ diagram. Most stars are away from the quasar region, while a large number of galaxies enter the quasar region, indicating that such galaxies can contaminate the mid-IR quasar selection.   

\begin{figure*}[htb!]
    \gridline{\fig{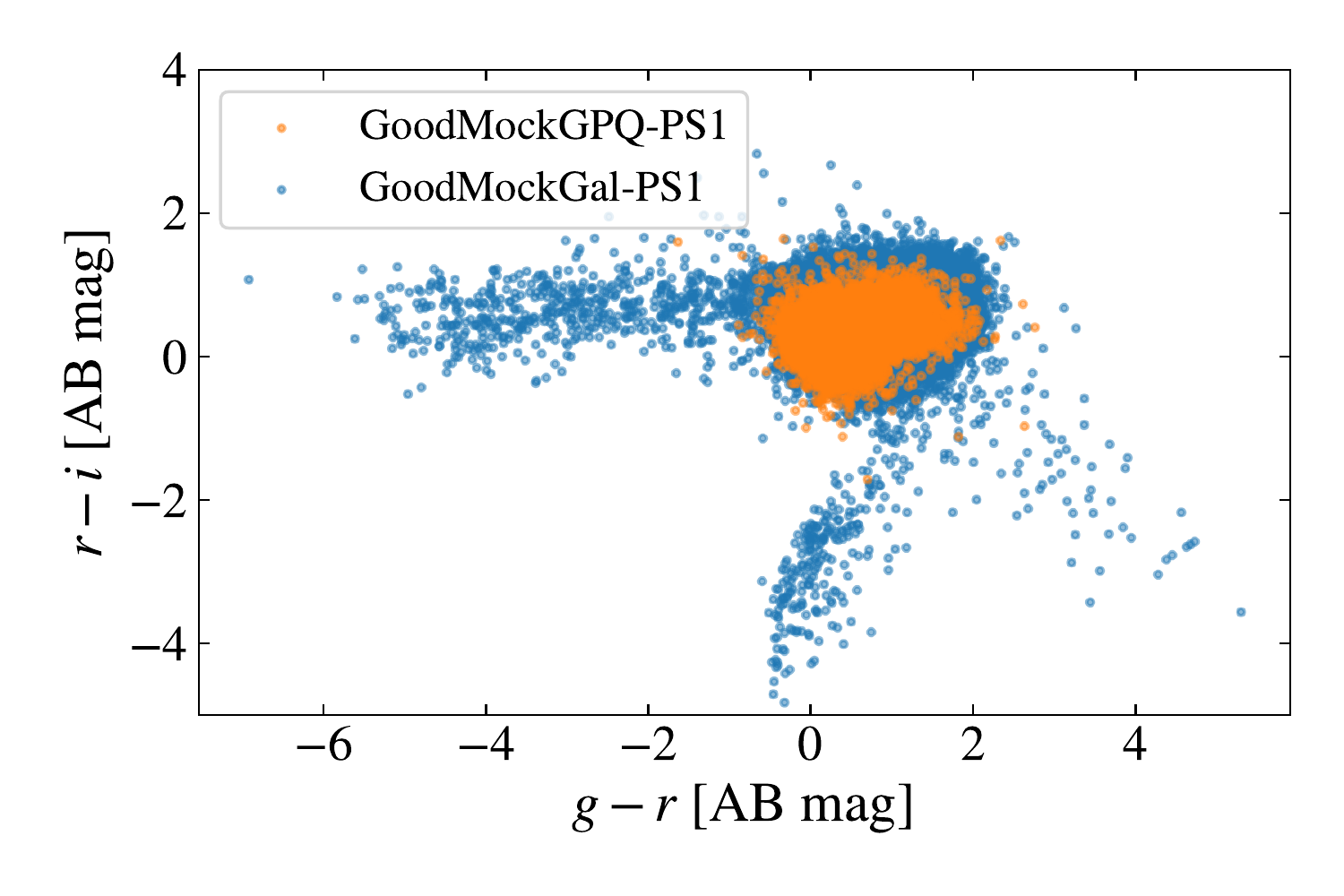}{0.33\textwidth}{(a)}
              \fig{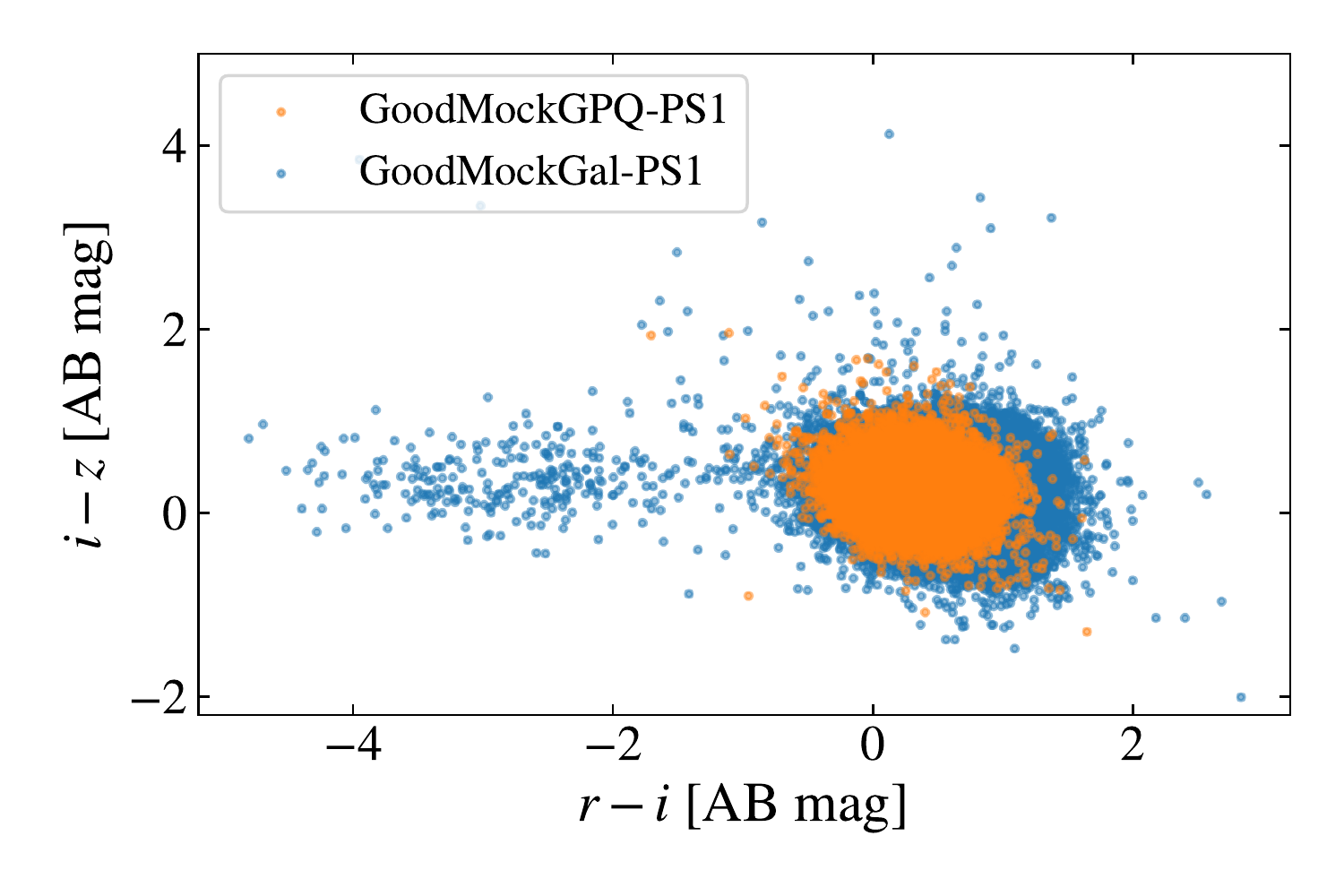}{0.33\textwidth}{(b)}
              \fig{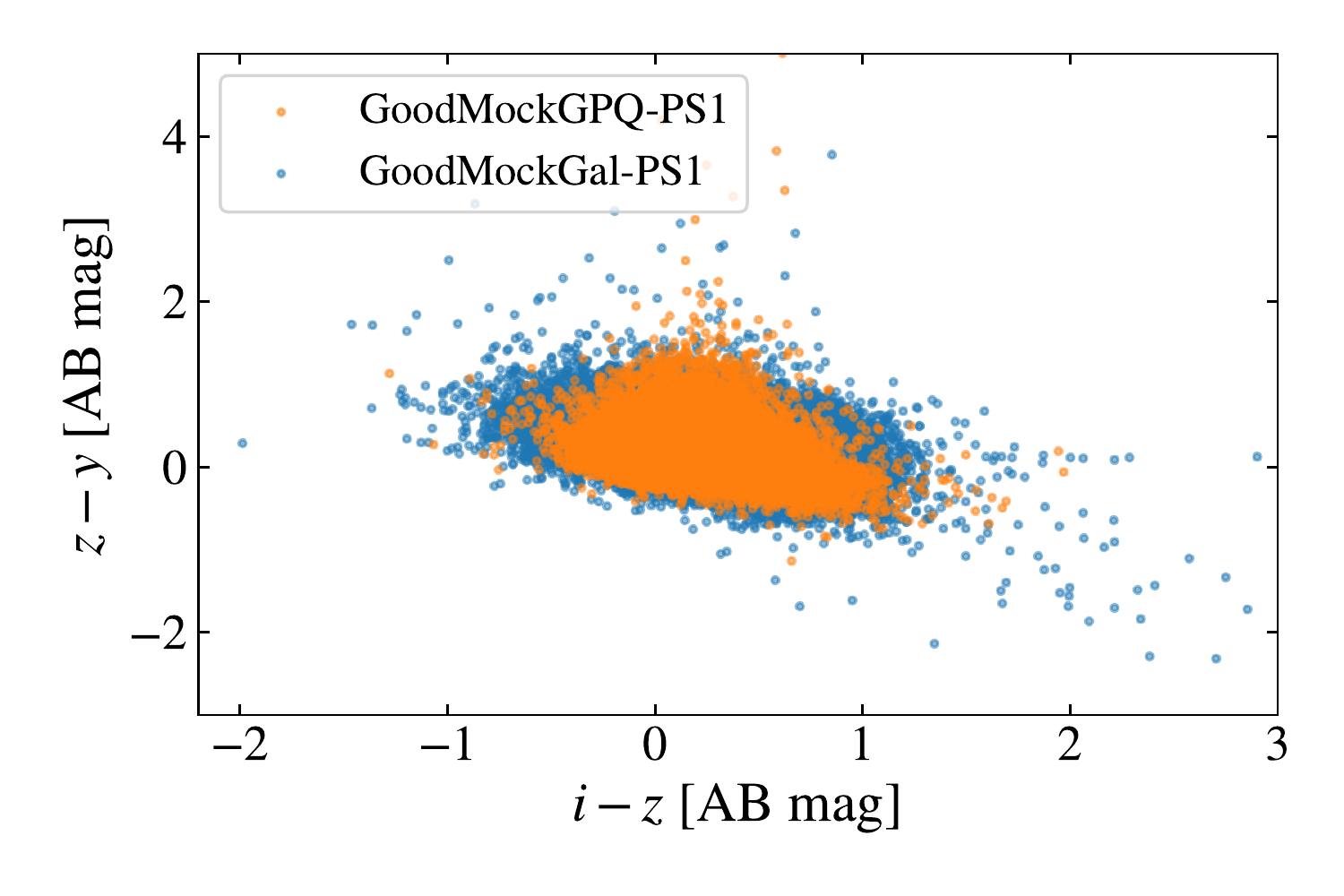}{0.33\textwidth}{(c)}}
    \gridline{\fig{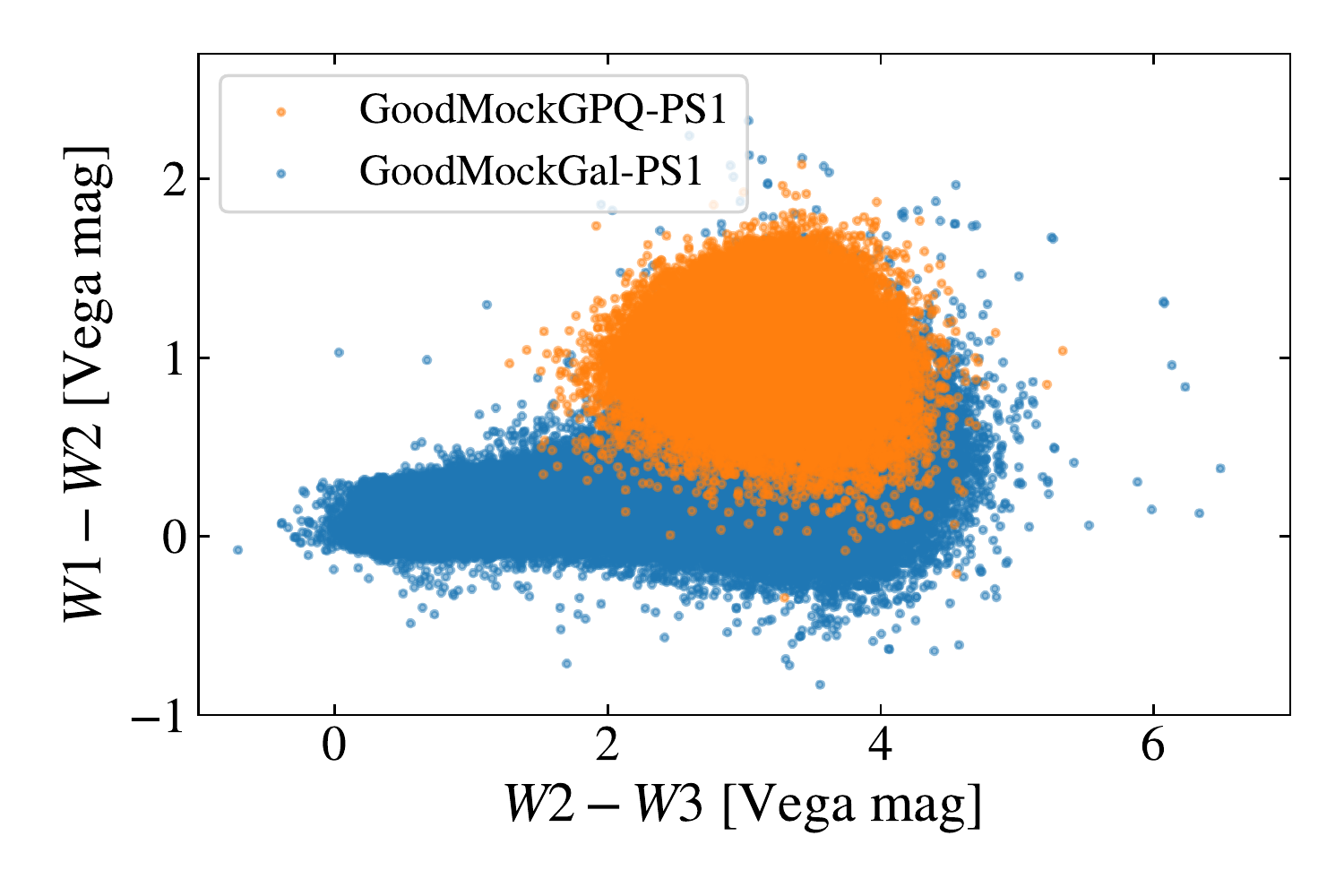}{0.33\textwidth}{(d)}
              \fig{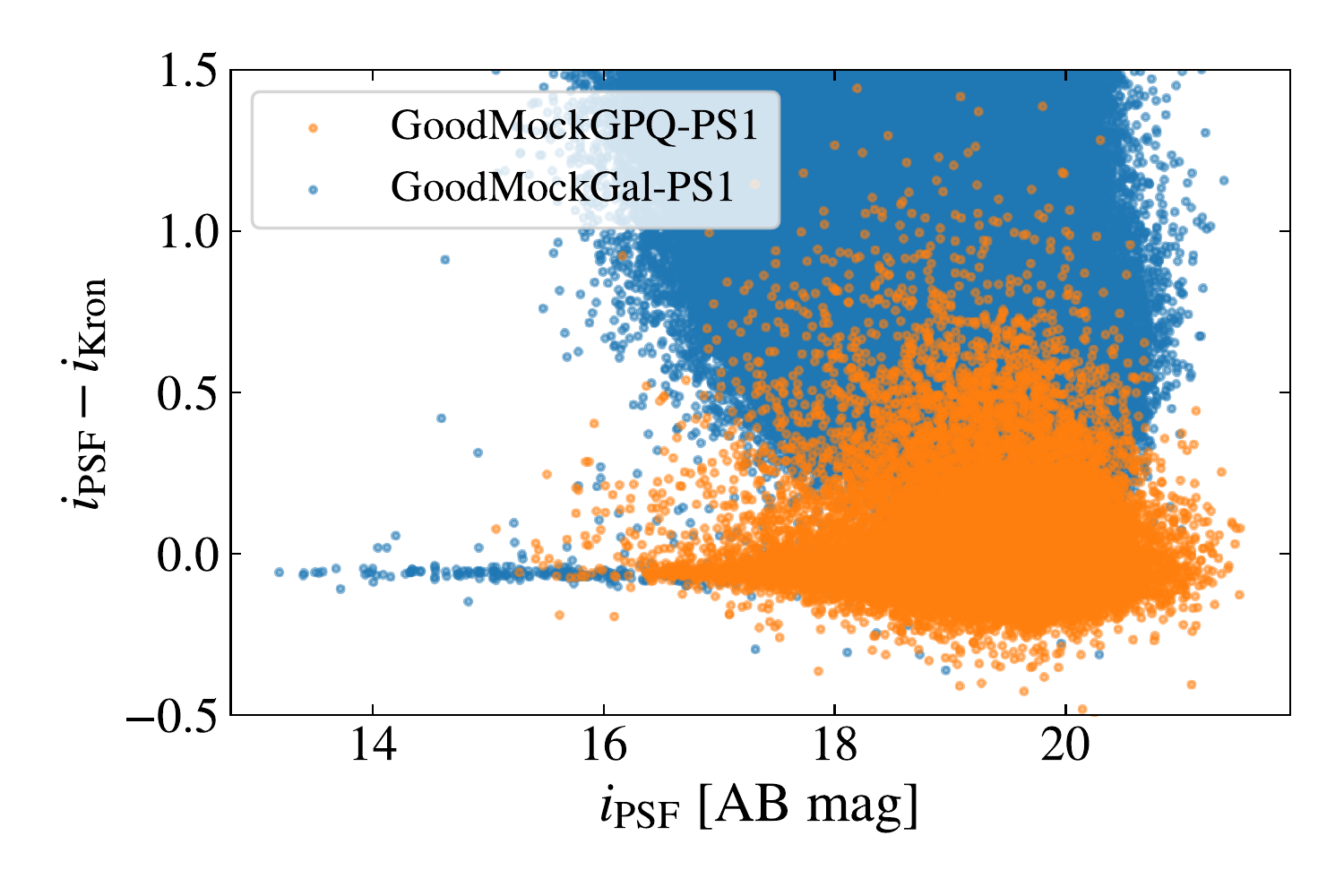}{0.33\textwidth}{(e)}}
    \caption{Color-color diagrams of $GoodMockGPQ$-PS1 and $GoodMockGal$-PS1 (a, b, c, d), and $i_{\mathrm{PSF}}-i_{\mathrm{Kron}}$ versus $i_{\mathrm{PSF}}$ plot for the two samples (e). Orange circles represent $GoodMockGPQ$-PS1 sources while blue circles represent $GoodMockGal$-PS1 sources.}
    \label{fig:ccdqg}
\end{figure*}

Some comparisons between mock quasars ($GoodMockGPQ$-PS1) and mock galaxies ($GoodMockGal$-PS1) behind the Galactic plane are also shown in Figure \ref{fig:ccdqg}. In the color-color diagrams of PS1 bands, galaxies largely overlap with quasars (Figure \ref{fig:ccdqg} (a, b, c)); while on the $W1-W2$ versus $W2-W3$ plane, these two classes are slightly more separable (Figure \ref{fig:ccdqg} (d)). Except for the colors, the difference between PSF magnitude and \citet{1980ApJS...43..305K} magnitude is often used as a morphological separator \citep{2002AJ....124.1810S,2014MNRAS.437..748F} for galaxies and point sources including quasars. However, separating quasars from galaxies becomes harder with $i_{\mathrm{PSF}}-i_{\mathrm{Kron}}$ at the faint end (Figure \ref{fig:ccdqg} (e)), as has been pointed out by \citet{yang2017quasar}. Among all the $\sim1.6$ million $GoodGal$ sources, $\sim200$ are point sources with $i_{\mathrm{PSF}}<18$ and $i_{\mathrm{PSF}}-i_{\mathrm{Kron}}<0$, which can also been seen from Figure \ref{fig:ccdqg} (e). These point sources include few quasars with ``galaxy'' labels and may also include some stars that are misclassified as galaxies. We do not pay attention to these point sources because they only contribute to a tiny fraction of the whole galaxy sample.

To sum up, we examine the properties of $GoodMockGPQ$, $GoodMockGal$ and PS1-AllWISE point-like sources in the color-color spaces. For quasar candidates selection,  contamination from both stars and galaxies should be taken care of. Simple PS1 color cuts are only capable of selecting quasars that are away from the stellar loci. Using a series of PS1 colors in high-dimensional space might help reduce the overlap between the stellar loci and clusters of quasar and galaxy. Moreover, with AllWISE colors, quasars can be better separated from stars and galaxies. Therefore we expect that the combination of PS1 and AllWISE data will make quasar selection more efficient. 

\subsection{A rough estimation on the lower limit to the sky density of GPQs} \label{sec:est_dens}

An estimation on the sky density of GPQs will be useful for evaluating the final GPQ candidate sample and the selection method. However, the $GoodMockGPQ$ sky distribution in the middle panel of Figure \ref{fig:ebv_goodmock} does not reflect the true density of GPQs due to two reasons: (i) the synthesizing process does not consider the source crowdedness and its effects on the photometric data quality; (ii) the source number of $GoodMockGPQ$ only depends on the size of input $GoodQSO$ sample when the dust extinction map is fixed.

\begin{figure*}[htb!]
    \centering
    \includegraphics[width=1\textwidth]{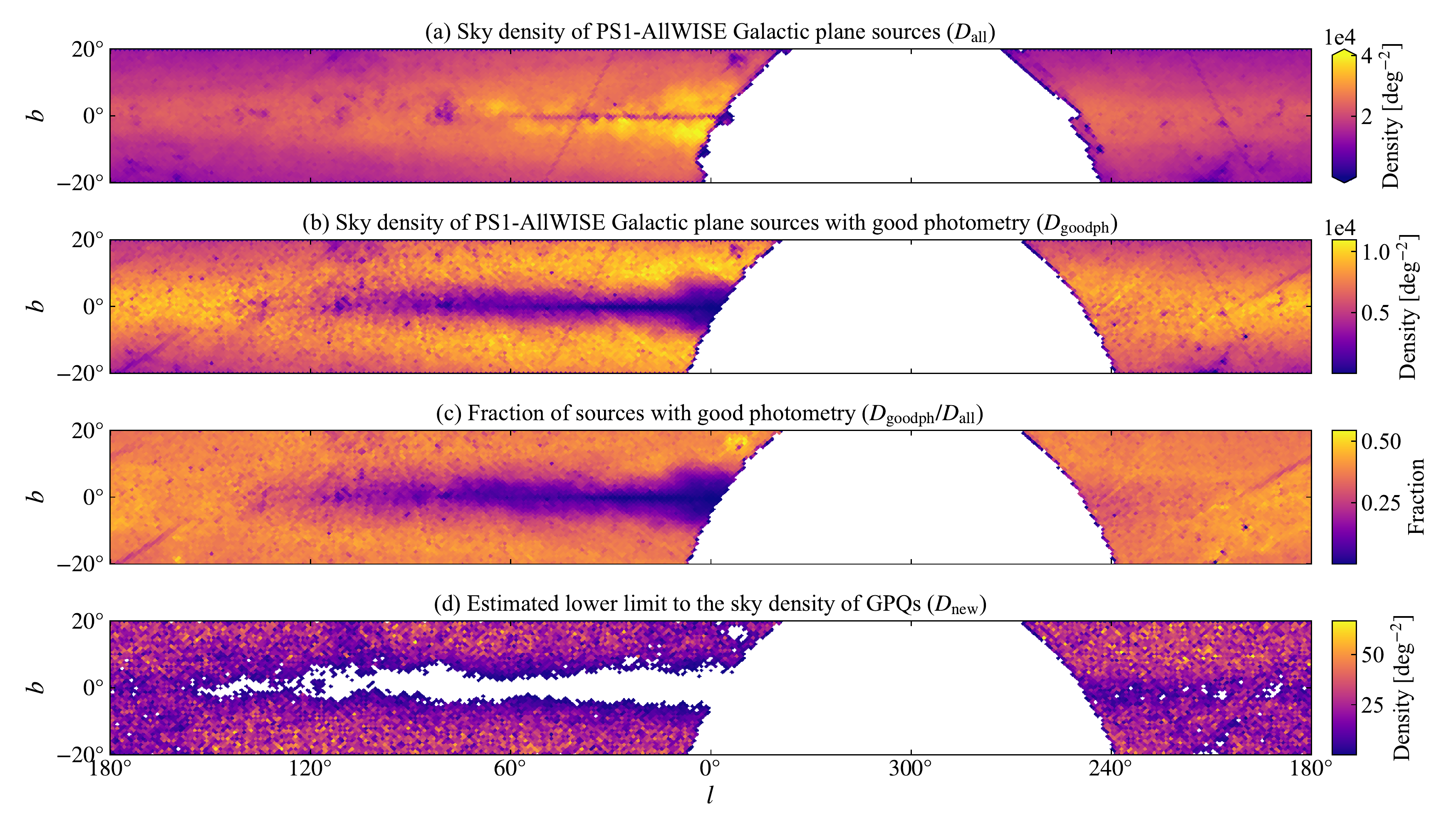}
    \caption{Sky density of all PS1-AllWISE Galactic plane sources (a) and its subset with good photometry (b), fraction of sources with good photometry in the PS1-AllWISE sample (c), and estimated lower limit to the sky density of GPQs (d).
    }
    \label{fig:est_dens_gpq}
\end{figure*}

Let the density of $GoodMockGPQ$ be $D_{\mathrm{old}}$, then the relative density of quasars with good photometry across the Galactic plane is: 

\begin{equation}
    D^{\prime}_{\mathrm{new}}=D_{\mathrm{old}}\times\frac{D_{\mathrm{goodph}}}{D_{\mathrm{all}}}
\end{equation}

\noindent where $D_{\mathrm{all}}$ is the sky density of all PS1-AllWISE sources in the Galactic plane, and $D_{\mathrm{goodph}}$ is the sky density of sources with good photometry as defined in Section \ref{sec:data-ps1} and Section \ref{sec:data-wise}. The fraction $D_{\mathrm{goodph}}/D_{\mathrm{all}}$ roughly quantifies the effects of source crowdedness on the photometric quality.
We expect that the median sky density of GPQs is no higher than that of $GoodQSO$ ($\mathrm{Median}(D_{\mathrm{new}})\leq\mathrm{Median}(D_{GoodQSO})$), therefore  the lower limit ``absolute'' sky density of GPQs can be computed as: 

\begin{equation}
    D_{\mathrm{new}}\geq D^{\prime}_{\mathrm{new}}\times \frac{\mathrm{Median}(D_{GoodQSO})}{\mathrm{Median}(D^{\prime}_{\mathrm{new}})}
\end{equation}

\noindent where $\mathrm{Median}(D_{GoodQSO})=20.3~\mathrm{deg}^{-2}$, and $\mathrm{Median}(D^{\prime}_{\mathrm{new}})=2.9~\mathrm{deg}^{-2}$. Figure \ref{fig:est_dens_gpq} shows the sky distribution of $D_{\mathrm{all}}$, $D_{\mathrm{goodph}}$, $D_{\mathrm{goodph}}/D_{\mathrm{all}}$ and $D_{\mathrm{new}}$. The estimated $D_{\mathrm{new}}$ has a median of $20.3~\mathrm{deg}^{-2}$ and a maximum of $66.7~\mathrm{deg}^{-2}$.

The predicted marginal probability of GPQs to the PS1-AllWISE sample with good photometry is $D_{\mathrm{new}}/D_{\mathrm{goodph}}$, which ranges from $2\times10^{-4}$ to 0.17 with a median of $3\times10^{-3}$. The maximum value of 0.17 is not reliable because it locates at edges of the HEALPix map ($\delta\sim -30^{\circ}$), where source count in a pixel does not correspond to the true number in the sky region.

\section{GPQ candidate selections with XGBoost} \label{sec:class}

We use XGBoost \citep{chen2016xgboost}, a scalable tree boosting system, to perform machine learning classification for GPQ selection. XGBoost is an implementation of the original gradient boosting framework \citep{friedman2000additive,friedman2001greedy}, known for high efficiency and outstanding performance in machine learning competitions \citep{chen2016xgboost}. Compared to traditional Gradient Boosting Machines (GBM), XGBoost has made a few improvements in the algorithm level. For example, XGBoost includes regularization terms in the objective function to control the model complexity, therefore can reduce overfitting and improve the model generalization; XGBoost is optimized for sparse input data, i.e. data with missing values; Other than \textit{greedy algorithm} by \citet{friedman2001greedy}, XGBoost supports a weighted quantile sketch algorithm that can more effectively find the optimal split points. Moreover, system enhancements for parallelization, tree pruning, and cache optimization have been integrated into XGBoost. Recently, XGBoost has been applied to astronomy and showed its capabilities of handling astronomical problems, including identifying Galactic candidates among unassociated sources from the Third Fermi Large Area Telescope (LAT) catalog \citep[3FGL;][]{2015ApJS..218...23A} \citep[e.g.][]{2016ApJ...825...69M}, distinguishing M giants from M dwarfs for spectral surveys \citep[e.g.][]{2019ApJ...887..241Y}, and selecting quasar candidates with photometric data \citep[e.g.][]{2019MNRAS.485.4539J}.

\begin{figure}[htb!]
    \centering
    \includegraphics[width=0.5\textwidth]{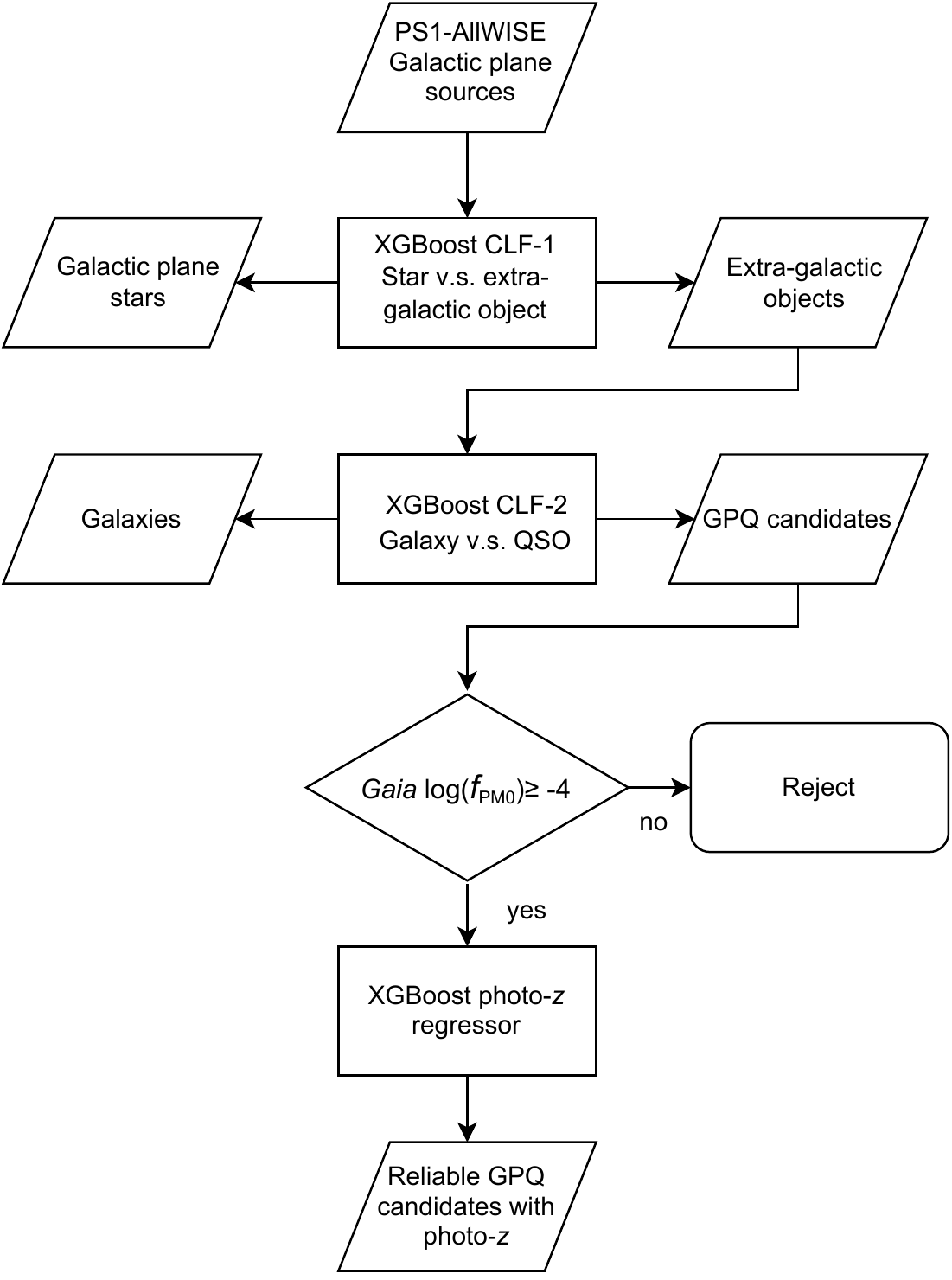}
    \caption{Flowchart of GPQ selection and photometric redshift calculation.}
    \label{fig:flow_clf}
\end{figure}

In order to obtain the optimal models, we use optuna \citep{akiba2019optuna}, a hyperparameter optimization framework to tune the learning hyperparameters. As has been mentioned in Section \ref{sec:TLFsub3}, we transform the three-class classification problem into two binary classification problems (stars versus extragalactic objects, and galaxies versus quasars). Under this setting, hyperparameters can be fine-tuned separately for the two classification steps. After classifying the Galactic plane sources with the two classifiers, we may use necessary additional criteria to ensure the purity of GPQ candidates. The classification scheme is shown in Figure \ref{fig:flow_clf}.

A few evaluation metrics are used in the machine learning process: $Accuracy$, $Precision$, $Recall$, $F_{1}$, $MCC$ (Matthews correlation coefficient) and $AUCPR$ (Area Under the Precision-Recall Curve). With true positive denoted as $TP$, true negative as $TN$, false positive as $FP$ and false negative as $FN$, the first five metrics are defined as: 

\begin{gather}
    Accuracy = \frac{TP+TN}{TP+TN+FP+FN} \\
    Precision = \frac{TP}{TP+FP}\\
    Recall = \frac{TP}{TP+FN}\\
    F_{1} = 2\times \frac { Precision \times Recall}{Precision + Recall} \\
    MCC = \frac{TP \times TN - FP \times FN}{\sqrt{(TP+FP)(TP+FN)(TN+FP)(TN+FN)}}.
\end{gather}

The Precision-Recall (PR) curve can be constructed by plotting precision-recall pairs (operating points) that are obtained using different thresholds on a probabilistic or other continuous-output classifier \citep{boyd2013area}. The $AUCPR$ can then be calculated with numerical integration methods. 

Among the six metrics, $Accuracy$, $Precision$, $Recall$ and $F_{1}$ are commonly used. However, the $Accuracy$ and $F_{1}$ metrics fail to measure the classification performance correctly under class-imbalanced situations, because they will be heavily biased towards the majority class. For example, given a sample with 95 from the negative class and 5 from the positive class, simply classifying all instances as negative produces $Accuracy=0.95$ and $F_{1}=0.9744$. These two scores of metrics are misleading because all the positive instances are wrongly classified while the $Accuracy$ and $F_{1}$ are high. The last two metrics, $MCC$ and $AUCPR$ are considered better evaluation measures in class-imbalanced cases. The $MCC$ takes the four confusion matrix categories ($TP$, $TN$, $FP$, $FN$) into account, and it is high only if the classifier makes good predictions on both positive and negative classes, independently of their ratios in the overall dataset \citep{chicco2020advantages}. It is also suggested by studies that the PR curve is more informative than the more famous Receiver Operator Characteristic (ROC) curve \citep[first recommended by][]{provost1998case}, especially on imbalanced datasets \citep{davis2006relationship,saito2015precision}. $AUCPR$ is useful as a measure of the overall performance of the model.

A total of 13 features are chosen for the two classification steps and the later photometric redshift regression, including 11 colors: $g-r$, $r-i$, $i-z$, $z-y$, $g-W1$, $r-W1$, $i-W1$, $z-W1$, $y-W1$, $W1-W2$, and $W2-W3$; and two morphological features: $i-i_{\mathrm{Kron}}$ and $z-z_{\mathrm{Kron}}$. As has been discussed in Section \ref{sec:synt_res}, using a set of PS1 colors ($g-r$, $r-i$, $i-z$, and $z-y$) can help reduce the overlap between clusters of quasars and stellar loci on two-dimensional diagrams. Quasars have redder $W1-W2$ and $W2-W3$ colors than stars and galaxies, which makes these two colors good features for quasars selection. \citet{2019MNRAS.485.4539J} showed that three PS1-AllWISE colors (i.e. $i-W1$, $y-W1$, and $z-W2$) can be used to efficiently distinguish quasars from stars and improve the performance of XGBoost classification. We construct similar colors as features by combining all PS1 bands and $W1$ (i.e. $g-W1$, $r-W1$, $i-W1$, $z-W1$ and $y-W1$), because $W1$ is the most sensitive one of AllWISE bands. These five new colors provide rough optical Spectral Energy Distributions (SEDs) for the objects and can characterize different objects with broader wavelength ranges than other optical colors (e.g. $g-r$) do. The difference between PSF magnitude and Kron magnitude in $i_{P1}$ and $z_{P1}$ bands ($i-i_{\mathrm{Kron}}$ and $z-z_{\mathrm{Kron}}$) are used as morphological features to separate point sources (stars and quasars) from extended sources (galaxies). We convert Vega magnitude to AB magnitude for AllWISE data when constructing all the features. As we don't set constraints on $W3$ magnitude or $W3snr$, some sources may have poor or missing $W3$ (and $W2-W3$) data. Nevertheless, the use of $W2-W3$ will not be a problem, because XGBoost can handle the missing values, and data with lower $SNR$ are more informative than missing values. 
 
\subsection{Binary classification for stars and extragalactic objects}

In the first classification step, the input data for training and validation consist of synthetic quasar sample $T_{QSO}$, synthetic galaxy sample $T_{Gal}$ (see Section \ref{sec:synt_proc}), and LAMOST Galactic plane star sample $T_{Star}$ (Section \ref{sec:data-lamost}). The input data have more than 3 million rows. For binarization, we assign the label \texttt{EXT} (extragalactic object) to all $T_{QSO}$ and $T_{Gal}$ instances, and keep the label for $T_{Star}$ as \texttt{STAR}. Here we regard extragalactic objects as the positive class and stars as the negative class.

We first apply five-fold cross validations with optuna to find the optimal setting of hyperparameters that minimizes the log loss among 500 trials. For a binary classification problem with a true label $y \in \{0,1\}$ and a probability estimate $p=\mathrm{Pr}(y=1)$, the log loss per sample is the negative log-likelihood of the classifier given the true label:

\begin{align}
    log\_loss(y,p) & = -\log\mathrm{Pr}(y|p) \\
                   & = -(y\log(p)+(1-y)\log(1-p)) .
\end{align}

\noindent Then we randomly split the whole input data into training set and validation set according to a $4:1$ ratio and calculate scores of the six metrics with the validation set. This $4:1$ split ratio is consistent with that of the five-fold cross validations. The large sample size of input data also ensures both training and validation sets have enough samples. 

Some fixed parameters in our programs are: \texttt{objective}=\texttt{binary:logistic}; \texttt{booster}=\texttt{gbtree}; \texttt{tree\_method}=\texttt{hist}. For hyperparameters that are tuned, the default values, optimal values found by the cross validations, and corresponding metric scores of these parameters are listed in Table \ref{tab:clf1par}. The number of boosting rounds (\texttt{num\_boost\_round}, a.k.a. \texttt{n\_estimators} in scikit-learn API of XGBoost) is fixed to 100 and not tuned together with \texttt{eta} (a.k.a. \texttt{learning\_rate}), because the effects of increasing \texttt{num\_boost\_round} can cancel that of decreasing \texttt{eta}, and vice versa. In the training process, we need to lower the learning rate \texttt{eta} and increase the \texttt{num\_boost\_round} to reduce the generalization error. The Classifier No.1 (CLF-1) is trained using $\mathtt{eta}=0.02$, $\mathtt{num\_boost\_round}=1,200$ with other optimal parameters in Table \ref{tab:clf1par}.

\begin{deluxetable}{lcc}[htb!]
    \tablecaption{Default and optimal hyperparameter settings for CLF-1 (star versus extragalactic object classification) \label{tab:clf1par}}
    \tablenum{2}
    \tablehead{\colhead{Hyperparameter} & \colhead{Default} & \colhead{Optimal}}
    
    \startdata
    \texttt{eta\ (learning\_rate)} & 0.3 & 0.3 \\
    \texttt{lambda\ (reg\_lambda)} & 1 & 2.32 \\
    \texttt{alpha\ (reg\_alpha)} & 0 & 1.13 \\
    \texttt{max\_depth} & 6 & 9 \\
    \texttt{gamma\ (min\_split\_loss)} & 0 & 0.60 \\
    \texttt{grow\_policy} & \texttt{depthwise} & \texttt{depthwise} \\
    \texttt{min\_child\_weight} & 1 & 1 \\
    \texttt{subsample} & 1 & 0.96 \\
    \texttt{colsample\_bytree} & 1 & 0.92 \\
    \texttt{max\_delta\_step} & 0 & 3 \\
     &  &  \\
    $Accuracy$ & 0.9993 & 0.9995 \\
    $Precision^{+}$ & 0.9993 & 0.9995 \\
    $Recall^{+}$ & 0.9995 & 0.9996 \\
    $F_{1}$ & 0.9994 & 0.9996 \\
    $MCC$ & 0.9985 & 0.9990 \\
    $AUCPR$ & 0.9992 & 0.9995 \\
    \enddata
    
    
    
\end{deluxetable}

We then classify the PS1-AllWISE point-like sources with CLF-1. To exclude as many stars as possible, we adopt a high threshold on $p_{\mathrm{EXT}}$ (model-predicted probabilities of sources for being extragalactic) to select extragalactic candidates. Sources with $p_{\mathrm{EXT}}>0.99$ are labeled as \texttt{EXT}, and the others are labeled as \texttt{STAR} and removed.

\subsection{Binary classification for galaxies and quasars}

We use $T_{QSO}$ and $T_{Gal}$ samples as input data for training and validation in the second classification step. Here we regard quasars as the positive class and galaxies as the negative class.

The same processes of parameter tuning and training as those of CLF-1 are applied to build CLF-2. We keep some parameters unchanged as: \texttt{objective}=\texttt{binary:logistic}; \texttt{booster}=\texttt{gbtree}; \texttt{tree\_method}=\texttt{hist}. For hyperparameters that are tuned, the default values, optimal values found by the cross validations, and corresponding metric scores of these parameters are listed in Table \ref{tab:clf2par}. The CLF-2 is trained using $\mathtt{eta}=0.02$, $\mathtt{num\_boost\_round}=1,500$ with other optimal parameters in Table \ref{tab:clf2par}. 

The optimal scores of the six metrics in Table \ref{tab:clf2par} are all lower than those in Table \ref{tab:clf1par}, indicating that the quasar--galaxy classification ``hardness'' is higher than that of star--extragalactic problem. Here we also use a high threshold of probability to select sources of our target class. We classify the sources labeled as \texttt{EXT} with CLF-2. Sources with $p_{\mathrm{QSO}}>0.95$ are kept as GPQ candidates, where $p_{\mathrm{QSO}}$ is the probability of a source for being a quasar predicted by the XGBoost model.

\begin{deluxetable}{lcc}[htb!]
    \tablecaption{Default and optimal hyperparameter settings for CLF-2 (quasar versus galaxy classification) \label{tab:clf2par}}
    \tablenum{3}
    \tablehead{\colhead{Hyperparameter} & \colhead{Default} & \colhead{Optimal}}
    
    \startdata
    \texttt{eta\ (learning\_rate)} & 0.3 & 0.2 \\
    \texttt{lambda\ (reg\_lambda)} & 1 & 2.32 \\
    \texttt{alpha\ (reg\_alpha)} & 0 & 1.10 \\
    \texttt{max\_depth} & 6 & 9 \\
    \texttt{gamma\ (min\_split\_loss)} & 0 & 0.81 \\
    \texttt{grow\_policy} & \texttt{depthwise} & \texttt{depthwise} \\
    \texttt{min\_child\_weight} & 1 & 2 \\
    \texttt{subsample} & 1 & 0.95 \\
    \texttt{colsample\_bytree} & 1 & 0.86 \\
    \texttt{max\_delta\_step} & 0 & 6 \\
     &  &  \\
    $Accuracy$ & 0.9969 & 0.9974 \\
    $Precision^{+}$ & 0.9894 & 0.9905 \\
    $Recall^{+}$ & 0.9894 & 0.9918 \\
    $F_{1}$ & 0.9894 & 0.9912 \\
    $MCC$ & 0.9875 & 0.9896 \\
    $AUCPR$ & 0.9938 & 0.9951 \\
    \enddata
    
    
    
\end{deluxetable}

\subsection{Additional cut based on \gaia\ proper motion to remove stellar contaminants}

In the first classification process, we classify all PS1-AllWISE point-like sources to stars and extragalactic objects. We ignore stars in the second classification step. Although the metrics of CLF-1 are high (Table \ref{tab:clf1par}), some stars can be misclassified as extragalactic objects, and then be classified either as quasars or galaxies. Faint stars are more likely to be misclassified than bright stars because stars in the training sample ($T_{Star}$) are biased towards the bright end. When using optical and near-IR colors for candidate selection, white dwarfs are major contaminants for low redshift quasars, and M/L/T dwarfs are typical contaminants for high redshift quasars \citep[e.g.][]{1997AJ....113.1421K,2002MNRAS.335..673V,2006AJ....131.2722C}. In the mid-IR regime, potential stellar contaminants for quasars are Young Stellar Objects (YSO), Asymptotic Giant Branch (AGB) stars, and Planetary Nebulae (PNe) \citep{2009ApJ...701..508K,2014ApJ...791..131K,Assef2018}. 

The YSOs are stars at the early stages of evolution, and are often divided into four subclasses \citep{lada1987star}: Class I, Class II, Flat spectrum, and Class III. Among them, Class II and Flat spectrum YSOs are the most-likely contaminants since they have optical and mid-IR SEDs similar to those of quasars. Since we require both optical and mid-IR detections for classification, optically faint Class I YSOs are eliminated in the first place. As has been studied by \citet{2014ApJ...791..131K}, Class III YSOs are clustered around $W2-W3=0$ and $W1-W2=0$, while Class I, II, and Flat spectrum YSOs occupy the region approximately with $W1-W2>0.25$ and $1.0<W2-W3<4.5$ (see their Figure 5). The latter YSO region is overlapped with the quasar region shown in Figure \ref{fig:ccdwise}, therefore the contamination should be taken care of.

AGB stars are evolved stars with low temperatures and high luminosities. They are surrounded by circumstellar envelopes, and IR excess exists in their broad SEDs. According to Figure 5 from \citet{2014ApJ...791..131K}, only a minority of AGB stars actually overlap with Class I, II, and Flat spectrum YSOs (and thus quasars) in the $W1-W2$ versus $W2-W3$ diagram. Therefore we expect the contamination from AGB stars is less than that from YSOs.

The PNe have a series of narrow emission lines as well as IR excess. Known PNe can be later removed from the GPQ candidate sample by cross-matching with the Simbad database \citep{2000A&AS..143....9W}.

In order to remove stellar contaminants such as white dwarfs, M/L/T dwarfs, YSOs, and AGB stars from GPQ candidates, we apply an additional cut based on \gaia\ proper motion, because the proper motion distribution of quasars is different from that of Milky Way stars. Although quasars should have negligible transverse motions, non-zero proper motions of them are measured by \gaia\ due to various effects, such as photocenter variability of quasars \citep[see][and references therein]{2016A&A...589A..71B}. In addition, proper motions with large uncertainties are not reliable. Therefore we need a probabilistic cut instead of a cut on the total proper motion. We define the probability density of zero proper motion ($f_{\mathrm{PM0}}$) of a source based on the bivariate normal distribution of proper motion measurements of the source as:

\begin{equation}
    f_{\mathrm{PM0}}=\frac{1}{2\pi \sigma_x \sigma_y\sqrt{1-\rho^2}}\mathrm{exp}\left\{-\frac{1}{2(1-\rho^2)}\left[ \left(\frac{x}{\sigma_{x}}\right)^{2} - \frac{2\rho xy}{\sigma_{x}\sigma_{y}} + \left(\frac{y}{\sigma_{y}}\right)^{2} \right] \right\}
\end{equation}

\noindent where $x=\mathrm{pmra}$, $y=\mathrm{pmdec}$, and $\rho=\mathrm{pmra\_pmdec\_corr}$ (correlation coefficient between pmra and pmdec) are obtained from \gdr{2} catalog, while $\sigma_{x}$ and $\sigma_{y}$ are the true external proper motion uncertainties calculated with the method suggested by \citet{lindegren2018gaia,lindegren2018gaia2}. The external proper motion uncertainty can be expressed as $\sigma_{\mathrm{ext}}=(k^2\sigma_{i}^2+\sigma_{s}^2)^{\frac{1}{2}}$, where $\sigma_{\mathrm{ext}}$ can be $\sigma_{x}$ or $\sigma_{y}$, $k=1.08$ is a multiplicative factor, $\sigma_{i}$ is the catalog uncertainty (pmra\_error or pmdec\_error), and $\sigma_{s}$ is the systematic error. For bright sources ($G<13$), $\sigma_{s}=0.032~\mathrm{mas/yr}$; for faint sources ($G>13$), $\sigma_{s}=0.066~\mathrm{mas/yr}$. Under the same uncertainty level, sources with smaller proper motions will have higher $f_{\mathrm{PM0}}$ by definition.

\begin{figure}[htb!]
    \centering
    \includegraphics[width=0.5\textwidth]{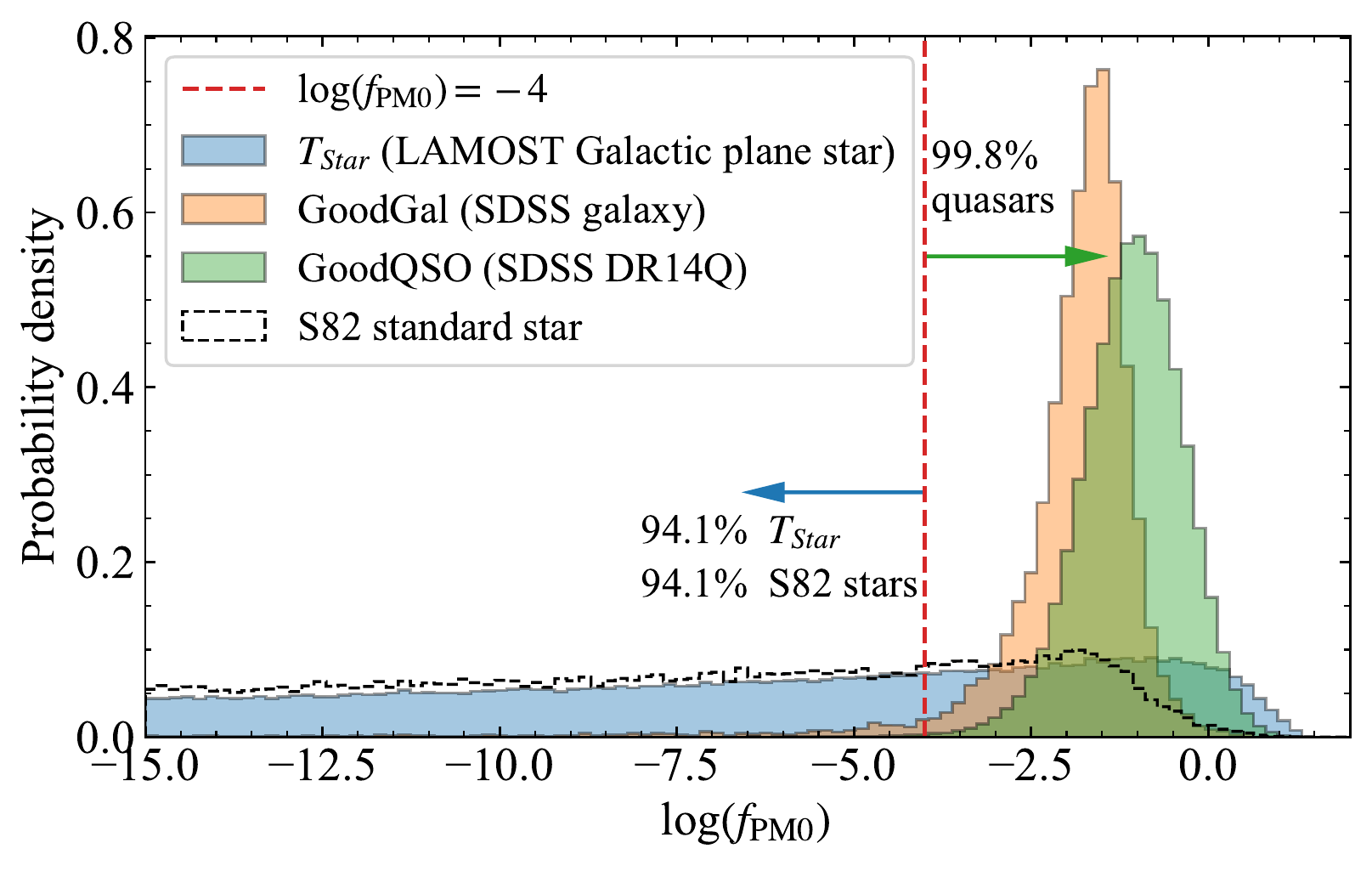}
    \caption{Histograms of $\mathrm{log}(f_{\mathrm{PM0}})$ of $T_{Star}$ (LAMOST Galactic plane star), $GoodGal$ (from SDSS galaxy), $GoodQSO$ (from SDSS DR14Q), and sources from the SDSS Stripe 82 Standard Star Catalog. Because $f_{\mathrm{PM0}}$ is the probability density which can be greater than 1 (the integral of the probability density function over the entire space is equal to 1), $\mathrm{log}(f_{\mathrm{PM0}})$ can have positive values.}
    \label{fig:hist_logpzpm}
\end{figure}

We take the logarithm of $f_{\mathrm{PM0}}$ for better comparison between samples. Figure  \ref{fig:hist_logpzpm} shows distributions of $\mathrm{log}(f_{\mathrm{PM0}})$ of stars, galaxies and quasars used in this study. For stellar samples, in addition to $T_{Star}$ (LAMOST Galactic plane star sample), a subsample of the SDSS Stripe 82 Standard Star Catalog \citep[hereafter S82 star;][]{2007AJ....134..973I} that meets the same constraints in Section \ref{sec:data-ps1} and Section \ref{sec:data-wise} is also included for comparison. We choose a $\mathrm{log}(f_{\mathrm{PM0}})\geq -4$ cut that excludes 94.1\% of both LAMOST Galactic plane stars and S82 stars, while retains 99.8\% of the quasars. Nevertheless, faint stars can be major contaminants even with such strict cut on $\mathrm{log}(f_{\mathrm{PM0}})$. 

We calculate $\mathrm{log}(f_{\mathrm{PM0}})$ for GPQ candidates after cross-matching them with \gdr{2}. For candidates without \gdr{2} proper motion records, we assign a default value of 99 for $\mathrm{log}(f_{\mathrm{PM0}})$. Sources with $\mathrm{log}(f_{\mathrm{PM0}})\geq -4$ are kept as reliable GPQ candidates.

\section{Photometric redshift estimation for GPQ candidates} \label{sec:photoz}

Measuring redshifts is an important step for quasar surveys. For quasar candidates, photometric redshifts (photo-$z$) estimation is a key to follow-up studies. Many different approaches have been proposed for calculating photo-$z$s of quasars, including quasar template fitting \citep[e.g.][]{2001AJ....122.1163B,2004MNRAS.353..654B,2009ApJ...690.1250S}, the empirical color-redshift relation \citep[e.g.][]{2001AJ....122.1151R,2004ApJS..155..243W,2004ChJAA...4...17W,2010MNRAS.406.1583W,2012AJ....144...49W}, machine learning \citep[e.g.][]{2010A&A...523A..14Y,2011MNRAS.418.2165L,2013ApJ...772..140B,2013AJ....146...22Z,2018A&A...611A..97P}, XDQSOz method \citep{2012ApJ...749...41B}, and Skew-QSO method \citep{yang2017quasar}. As the photo-$z$ estimation problem can be well described by the regression problem in machine learning, we also use XGBoost to train the regression model and predict photo-$z$s for our reliable GPQ candidates. 

To build the training set and validation set, we randomly split the de-reddened $GoodQSO$ sample with a ratio of 4:1. Our application set (reliable GPQ candidates) is also de-reddened. The same 13 features as those in Section \ref{sec:class} are used for photo-$z$ regression: $g-r$, $r-i$, $i-z$, $z-y$, $g-W1$, $r-W1$, $i-W1$, $z-W1$, $y-W1$, $W1-W2$, $W2-W3$, $i-i_{\mathrm{Kron}}$ and $z-z_{\mathrm{Kron}}$. The morphological features $i-i_{\mathrm{Kron}}$ and $z-z_{\mathrm{Kron}}$ are included as they may help distinguish quasars at different cosmological distances. To obtain the optimal model, we also tune the parameters with five-fold cross-validations using optuna.

\begin{figure}[htb!]
    \centering
    \includegraphics[width=0.5\textwidth]{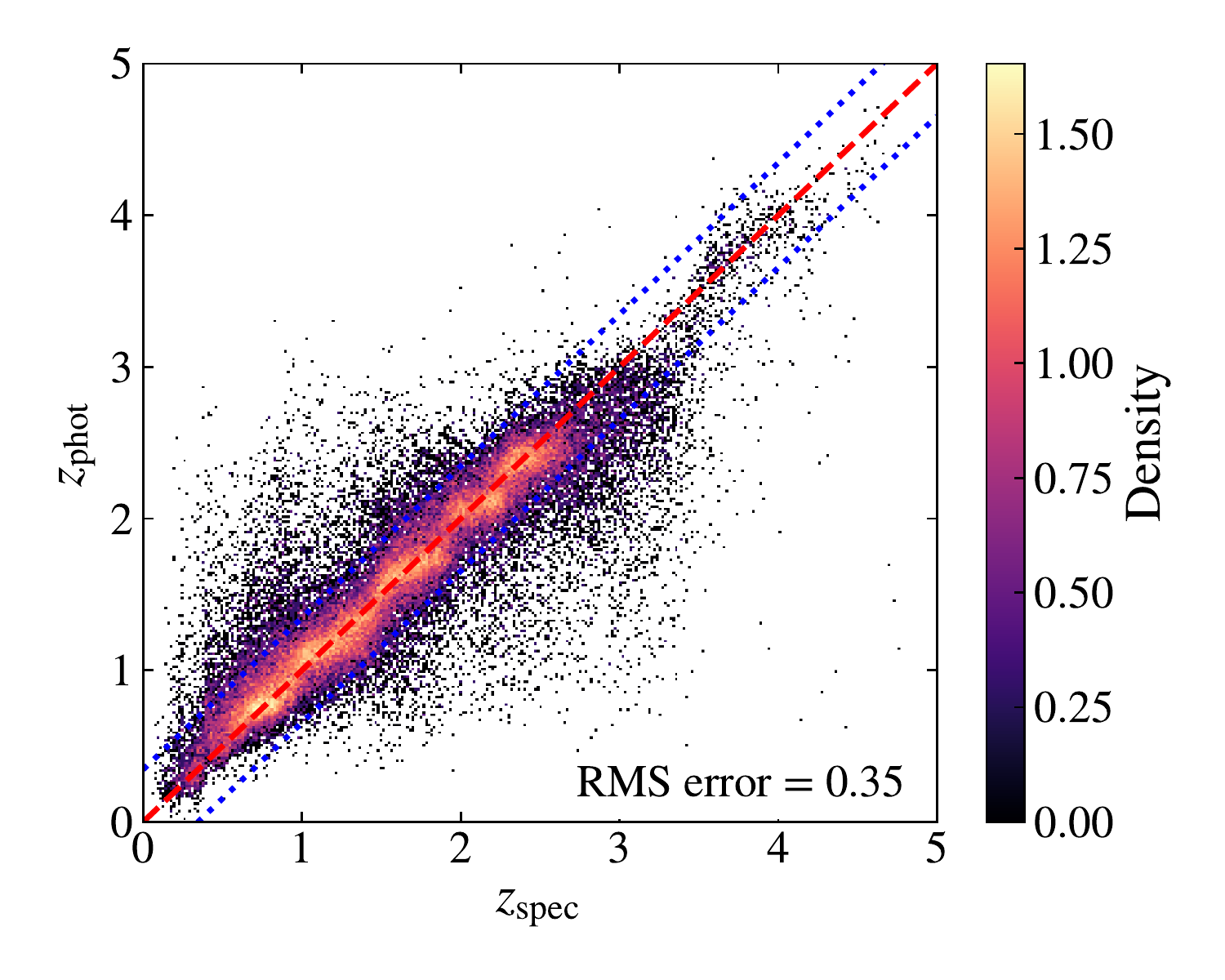}
    \caption{Photometric redshift obtained with XGBoost regression model against spectral redshift of de-reddened validation set with 57,855 quasars. The red dashed line denotes $z_{\mathrm{phot}}=z_{\mathrm{spec}}$ and the blue dotted lines mark the margin within one RMSE from the red dashed line.} 
    \label{fig:photoz_rms}
\end{figure}

\begin{figure*}[htb!]
    \centering
    \includegraphics[width=1\textwidth]{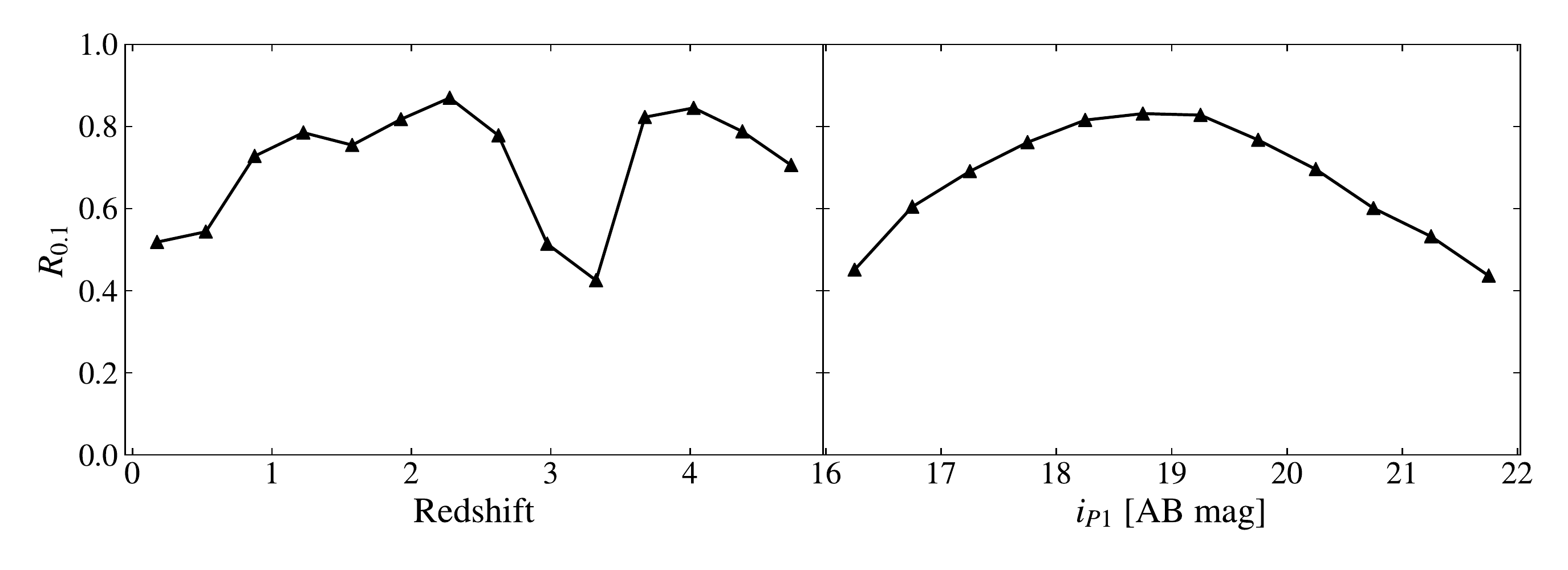}
    \caption{Photo-z accuracy $R_{0.1}$ (the fraction of quasars with $|\Delta z|\leq 0.1$, where $|\Delta z| = |z_{\mathrm{spec}}-z_{\mathrm{phot}}|/(1+z_{\mathrm{spec}})$) as a function of redshift (left panel) and magnitude (de-reddened; right panel).}
    \label{fig:R01_photoz}
\end{figure*}

The performance of the XGBoost photo-$z$ regression model on the test set can be examined in $z_{\mathrm{phot}}$-$z_{\mathrm{spec}}$ (photometric redshift versus spectral redshift) plot (Figure \ref{fig:photoz_rms}) or with two quantities: the root-mean-square error (RMSE) and photo-$z$ accuracy. For a validation set with sample size $n$, the root-mean-square error is $\mathrm{RMSE} = \sqrt{{\sum^{n}_{i=1}(z_{\mathrm{phot}}-z_{\mathrm{spec}})^{2}}/{n}}$. On our validation set with a sample size of 57,855, the RMSE is 0.35. The photo-$z$ accuracy $R_{0.1}$ is defined as the fraction of quasars with $|\Delta z|\leq 0.1$, where $|\Delta z| = |z_{\mathrm{spec}}-z_{\mathrm{phot}}|/(1+z_{\mathrm{spec}})$. Our XGBoost regression model yields a photo-$z$ accuracy of 74\% on the validation set, which is comparable to that of \citet{yang2017quasar} on PS1 and WISE data (79\%). \citet{yang2017quasar} adopted a multivariate Skew-t model and prior probabilities from the quasar luminosity function (QLF) to achieve the high photo-$z$ accuracy. Figure \ref{fig:R01_photoz} shows the photo-$z$ accuracy $R_{0.1}$ as a function of spectral redshift (left panel) and de-reddened $i_{P1}$-band magnitude (right panel) respectively. $R_{0.1}$ has maximum values at $z\approx2.3$ and $z\approx4$, and reaches a minimum at $z\approx 3$. Most $z\approx 3$ quasars have underestimated photometric redshifts (see Figure \ref{fig:photoz_rms}) due to a degeneracy of broad-band photometry in response to quasar SEDs at different redshifts. The strong Ly$\alpha$ emission line enters $g_{P1}$ band at $z\approx2.4$ and moves into $r_{P1}$ band at $z\approx3.5$, which leads to large excess in $g_{P1}$ magnitudes and hence similar PS1 colors of quasars within $2.4\lesssim z\lesssim 3.5$. This kind of degeneracy can be alleviated if SDSS $u$-band data are available to characterize the Lyman limit systems \citep[see Section 4 of][]{yang2017quasar}. The photo-$z$ accuracy is improved at $z\gtrsim 3.5$, because the Lyman limit enters $g_{P1}$ band. $R_{0.1}$ also drops at low redshift ($z<1$), and at both bright and faint ends, because the training sample is biased towards intermediate redshifts and magnitudes.

\section{The GPQ candidate catalog} \label{sec:results}

\subsection{Validation of the GPQ candidates with Simbad, Milliquas, and SDSS DR16Q} \label{subsec:gpq_sb}

With our Transfer Learning Framework and aforementioned additional selection criteria, we obtain a reliable GPQ candidate sample with 161,532 sources from PS1 and AllWISE. We cross-match the GPQ candidates with Simbad database \citep{2000A&AS..143....9W}, and find 2,786 matches. The object types and summary are shown in Table \ref{tab:simbad}. 

We categorize all matched sources to four groups: AGN/QSO, star (including PNe and PN candidates), galaxy, and other-type objects. Among all the matches, 53.98\% (1,504) are recorded as AGN/QSO (including candidates), 8.97\% (250) are recorded as star (including candidates), 4.02\% (112) are recorded as galaxy, and 33.02\% (920) are other-type objects labeled according to the detection properties (e.g. wavelength). Those other-type objects have higher probabilities to be AGN/QSO than stars, as most (728+27) of them are radio sources, and 64 are X-ray sources (see Table \ref{tab:simbad}). For the 4.02\% sources labeled as galaxy, a number of them may also host AGN/QSO as we have applied careful selection criteria to remove possible galaxy contaminants. Among the 250 sources labeled as star, 40 are candidates, and the other 210 are known stars. 101 of the known stars were once selected as QSO candidates using SDSS photometry, and then identified as stars by the 2dF-SDSS LRG and QSO Survey \citep{2009MNRAS.392...19C}. From these analyses, we can conclude that, the purity of our GPQ candidates on the small subset of 2,786 Simbad matches can be as high as $\sim90$\%. The true purity may vary at different locations in the Galactic plane. 

\begin{deluxetable*}{p{0.14\textwidth}l|p{0.16\textwidth}l|p{0.16\textwidth}l|p{0.18\textwidth}l}[htb!]
    \tablenum{4}
    \tablecaption{Matching results of GPQ candidates and Simbad database \label{tab:simbad}}
    \tabletypesize{\scriptsize}
    \tablehead{AGN/QSO & \colhead{Number} & Star & \colhead{Number} & Galaxy & \colhead{Number} & Other types & \colhead{Number}}
    \startdata
    QSO & 1121 & Star & 175 & Galaxy & 106 & Radio source & 728 \\
AGN candidate & 150 & YSO Candidate & 22 & Radio galaxy & 3 & IR source & 77 \\
BL Lac object & 143 & YSO & 13 & Brightest galaxy in a cluster (BCG) & 2 & X-ray source & 64 \\
Seyfert 1 galaxy & 38 & Cataclysmic binary candidate & 7 & Cluster of galaxies & 1 & Centimetric radio source & 27 \\
AGN & 31 & Planetary Nebula (PN) & 5 &  &  & Blue object & 22 \\
Other subclasses & 21 & Other subclasses & 28 &  &  & Far-IR source ($\lambda\geq30\mu \mathrm{m}$) & 2 \\
Total & 1504 & Total & 250 & Total & 112 & Total & 920 \\
Fraction & 53.98\% & Fraction & 8.97\% & Fraction & 4.02\% & Fraction & 33.02\%
    \enddata
\end{deluxetable*}

As another test on stellar contamination, we cross-match our GPQ candidates with LAMOST Galactic plane star sample. This match identifies 29 LAMOST stars, all of which are not recorded in Simbad. Therefore the total number of known stars in the GPQ candidates is 239.

We also examine the fraction of known GPQs that can be recovered with our candidates table. The known GPQ sample is $MLQSUB$ with 1,853 sources, which is retrieved from Milliquas catalog and described in Section \ref{sec:data-mlq}. The $MLQSUB$ is selected with same constraints as those on our application PS1-AllWISE data, to get a consistent analysis result. Cross-matching $MLQSUB$ with our GPQ candidates results in 1,763 matches, meaning that 95.14\% of GPQs from Milliquas can be selected with our methods, under same quality constraints on the photometric data. The recent sixteenth data release of SDSS Quasar Catalog \citep[DR16Q;][]{2020ApJS..250....8L} has a total of 750,414 sources, in which 3,727 sources are located at $|b|<20^{\circ}$. Only 1,320 of these SDSS GPQs meet the photometric quality constraints in Section \ref{sec:data-ps1} and Section \ref{sec:data-wise}. Cross-matching our GPQ candidates with SDSS DR16Q gives 1,292 matches, which corresponds to a recall rate of 97.88\% under the same photometric quality constraints, or a recall rate of 34.67\% to the whole identified sample. The overall completeness of the sample of candidates is mainly limited by the photometric quality constraints.

\subsection{Description of the GPQ candidate catalog}

We remove 239 known stars (see Section \ref{subsec:gpq_sb}) from our GPQ candidate sample. We then match the remaining GPQ candidates by coordinates with TOPCAT internally, and found 347 close pairs within $0.2^{\prime\prime}$. These pairs are very likely duplicated sources, because the PS1 survey can not resolve two sources within an angular distance of $0.2^{\prime\prime}$. The median image quality for PS1 $3\pi$ survey is FWHM = (1.31, 1.19, 1.11, 1.07, 1.02) arcseconds for ($grizy_{\mathrm{P1}}$) \citep{2016arXiv161205242M}. Therefore we only keep one source for each close pair, and obtain the final GPQ candidates sample with \finalgpqnum\ sources. The GPQ candidate catalog is compiled based on this sample, with photometric data from PS1 DR1, AllWISE, and astrometric data from \gdr{2}. The descriptions for the catalog are displayed in Table \ref{tab:gpqc}. 

\startlongtable
\begin{deluxetable*}{p{0.08\textwidth}p{0.08\textwidth}p{0.15\textwidth}p{0.55\textwidth}}
    \tablenum{5}
    \tabletypesize{\footnotesize}
    \tablehead{Column & Units & Label & Explanations}
    \tablecaption{Contents of the GPQ candidate catalog\label{tab:gpqc}}
    \startdata
    1 & --- & Designation & Catalog designation hhmmss.ss+ddmmss.s (J2000) based on Pan-STARRS1 (PS1) coordinates \\
2 & deg & ra & PS1 right ascension in decimal degrees (J2000) (weighted mean) at mean epoch \\
3 & deg & dec & PS1 declination in decimal degrees (J2000) (weighted mean) at mean epoch \\
4 & deg & l & Galactic longitude in decimal degrees \\
5 & deg & b & Galactic latitude in decimal degrees \\
6 & --- & photoz & Photometric redshift predicted with XGBoost regressor \\
7 & --- & p\_star & Probability of the object to be a star, predicted by the first XGBoost classifier, a.k.a. $p_{\mathrm{star}}$ (p\_star+p\_ext=1) \\
8 & --- & p\_ext & Probability of the object to be an extragalactic object, predicted by the first XGBoost classifier, a.k.a. $p_{\mathrm{ext}}$ (p\_star+p\_ext=1) \\
9 & --- & p2\_gal & Probability of the object to be a galaxy, predicted by the second XGBoost classifier, a.k.a. $p_{\mathrm{gal}}$ (p2\_gal+p2\_qso=1) \\
10 & --- & p2\_qso & Probability of the object to be a quasar, predicted by the second XGBoost classifier, a.k.a. $p_{\mathrm{QSO}}$ (p2\_gal+p2\_qso=1) \\
11 & --- & fpm0 & Probability density of zero proper motion ($f_{\mathrm{PM0}}$) of the source \\
12 & --- & log\_fpm0 & The logarithm of fpm0 ($\mathrm{log}(f_{\mathrm{PM0}})$) \\
13 & mag & ebv & Line-of-sight $E(B-V)$ given by the Planck14 dust map \\
14 & --- & PS\_objID & Pan-STARRS1 (PS1) unique object identifier \\
15 & mag & gmag & Mean PSF AB magnitude from PS1 g filter detections \\
16 & mag & e\_gmag & Error in gmag \\
17 & mag & gKmag & Mean Kron AB magnitude from PS1 g filter detections \\
18 & mag & e\_gKmag & Error in gKmag \\
19 & mag & rmag & Mean PSF AB magnitude from PS1 r filter detections \\
20 & mag & e\_rmag & Error in rmag \\
21 & mag & rKmag & Mean Kron AB magnitude from PS1 r filter detections \\
22 & mag & e\_rKmag & Error in rKmag \\
23 & mag & imag & Mean PSF AB magnitude from PS1 i filter detections \\
24 & mag & e\_imag & Error in imag \\
25 & mag & iKmag & Mean Kron AB magnitude from PS1 i filter detections \\
26 & mag & e\_iKmag & Error in iKmag \\
27 & mag & zmag & Mean PSF AB magnitude from PS1 z filter detections \\
28 & mag & e\_zmag & Error in zmag \\
29 & mag & zKmag & Mean Kron AB magnitude from PS1 z filter detections \\
30 & mag & e\_zKmag & Error in zKmag \\
31 & mag & ymag & Mean PSF AB magnitude from PS1 y filter detections \\
32 & mag & e\_ymag & Error in ymag \\
33 & mag & yKmag & Mean Kron AB magnitude from PS1 y filter detections \\
34 & mag & e\_yKmag & Error in yKmag \\
35 & --- & AllWISE\_ID & AllWISE unique source ID \\
36 & mag & W1mag & W1 (Vega) magnitude (3.35$\mu$m) \\
37 & mag & e\_W1mag & Mean error on W1 magnitude \\
38 & mag & W2mag & W2 (Vega) magnitude (4.6$\mu$m) \\
39 & mag & e\_W2mag & Mean error on W2 magnitude \\
40 & mag & W3mag & W3 (Vega) magnitude (11.6$\mu$m) \\
41 & mag & e\_W3mag & Mean error on W3 magnitude \\
42 & mag & W4mag & W4 (Vega) magnitude (22.1$\mu$m) \\
43 & mag & e\_W4mag & Mean error on W4 magnitude \\
44 & mag & Jmag & 2MASS J (Vega) magnitude (1.25$\mu$m) \\
45 & mag & e\_Jmag & Mean error on J magnitude \\
46 & mag & Hmag & 2MASS H (Vega) magnitude (1.65$\mu$m) \\
47 & mag & e\_Hmag & Mean error on H magnitude \\
48 & mag & Kmag & 2MASS Ks (Vega) magnitude (2.17$\mu$m) \\
49 & mag & e\_Kmag & Mean error on Ks magnitude \\
50 & --- & Gaia\_source\_id & \gdr{2} unique source identifier \\
51 & mas & parallax & \gdr{2} parallax \\
52 & mas & parallax\_error & Standard error of parallax \\
53 & mas/yr & pmra & \gdr{2} proper motion in right ascension direction \\
54 & mas/yr & pmra\_error & Standard error of pmra \\
55 & mas/yr & pmdec & \gdr{2} proper motion in declination direction \\
56 & mas/yr & pmdec\_error & Standard error of pmdec \\
57 & --- & pmdec\_pmdec\_corr & Correlation between pmra and pmdec \\
58 & mas/yr & pmra\_error\_ext & True external uncertainty of pmra \\
59 & mas/yr & pmdec\_error\_ext & True external uncertainty of pmdec \\
60 & --- & sb\_main\_id & Main identifier for an object in Simbad database \\
61 & --- & sb\_main\_type & Main object type for an object in Simbad database \\
62 & --- & sb\_redshift & Redshift of an object recorded in Simbad database
    \enddata
    \tablecomments{This table is published in its entirety in the machine-readable format.}
    \end{deluxetable*}

\begin{figure*}[htb!]
    \centering
    \includegraphics[width=1\textwidth]{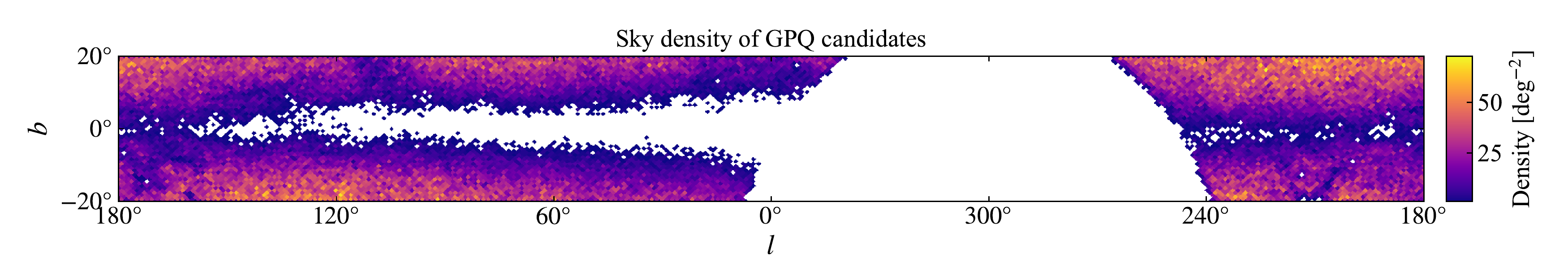}
    \caption{Sky density plot of GPQ candidates in Galactic coordinates.}
    \label{fig:gpqc_sky}
\end{figure*}

The sky density of sources from the GPQ candidate catalog is shown in Figure \ref{fig:gpqc_sky}. In general, the sky distribution of the GPQ candidates is consistent with the prediction in Section \ref{sec:est_dens}. The highest sky density of the candidates is 72.7 $\mathrm{deg}^{-2}$, which is slightly higher than the estimation (66.7 $\mathrm{deg}^{-2}$). The median density is 16.7 $\mathrm{deg}^{-2}$, which is comparable but lower than the estimated value (or the median density of $GoodQSO$). As can be seen from Figure \ref{fig:gpqc_sky}, the sky densities of GPQ candidates at $|b|\lesssim10^{\circ}$ are lower than those of the estimation in Figure \ref{fig:est_dens_gpq} (d), which indicates that the modelling process overestimates the sky density of GPQs at lower Galactic latitudes. The region with $\delta\lesssim-30^{\circ}$ ($0^{\circ} \lesssim~l ~\lesssim~240^{\circ}$) is blank, because it is not covered by the PS1 $3\pi$ survey. 

\begin{figure}[htb!]
    \centering
    \gridline{\fig{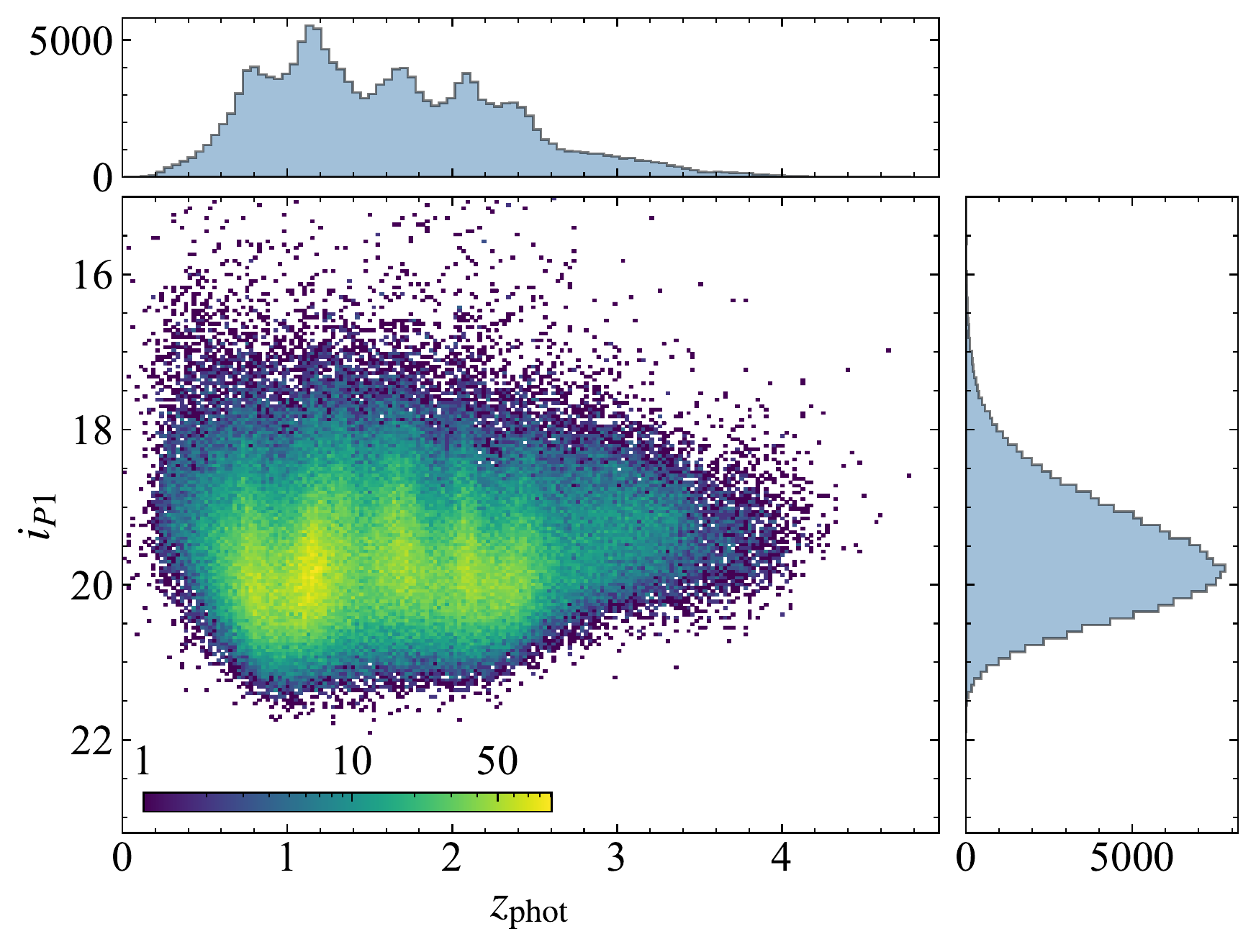}{0.49\textwidth}{(a) GPQ candidates}
              \fig{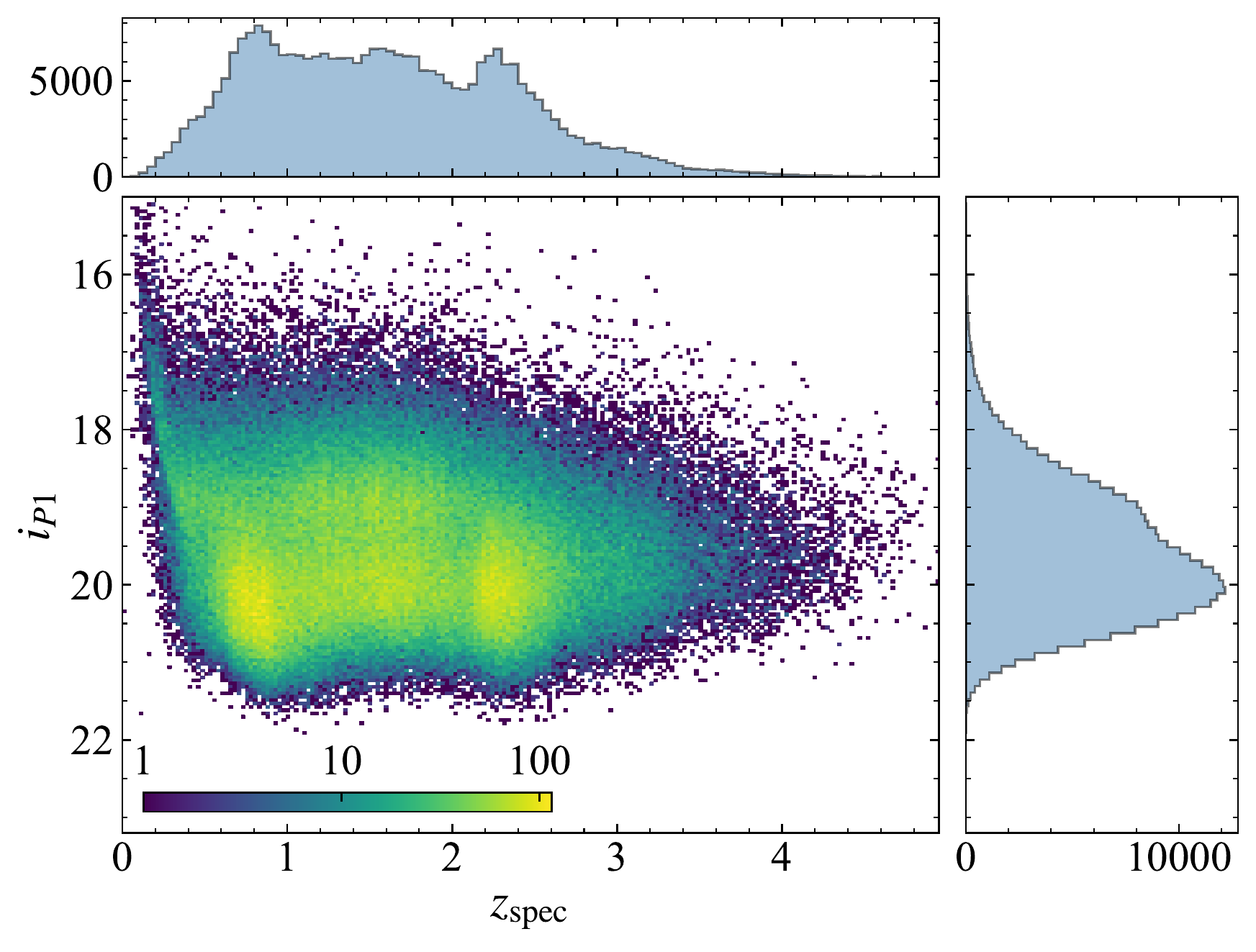}{0.49\textwidth}{(b) GoodQSO (SDSS DR14Q)}}
    \caption{(a) De-reddened $i_{P1}$ magnitude and photometric redshift distribution of GPQ candidates. (b) De-reddened $i_{P1}$ magnitude and redshift distribution of GoodQSO sample. The $i_{P1}$ magnitude is de-reddened according to the \citetalias{abergel2014planck} dust map.}
    \label{fig:gpqc_photoz}
\end{figure}

The distributions of de-reddened $i_{P1}$ magnitudes and photometric redshifts of our GPQ candidates are displayed in Figure \ref{fig:gpqc_photoz} (a). The lowest and highest photometric redshift are $z_{\mathrm{phot}}=0.016$ and $z_{\mathrm{phot}}=4.777$, respectively. Taking into account the uncertainties in photo-$z$ estimations, the actual highest redshift of the GPQs can be up to 5. Five peaks appear in the histogram of photometric redshift (Figure \ref{fig:gpqc_photoz} (a)) at $z_{\mathrm{phot}}\approx$ (0.8, 1.2, 1.7, 2.1, 2.4), which are caused by selection effects and sample bias of the training set. Quasars with these redshifts have higher chances to be selected with PS1 photometry: (i) when $z\approx 0.8$, the Mg \textsc{ii} emission line enters $g_{\mathrm{P1}}$ band; (ii) when $z\approx 1.2$, Mg \textsc{ii} enters $r_{\mathrm{P1}}$ band; (iii) when $z\approx 1.7$, C \textsc{iii]} enters $g_{\mathrm{P1}}$ band, and Mg \textsc{ii} enters $i_{\mathrm{P1}}$ band; (iv) when $z\approx 2.1$, both Si \textsc{iv} and C \textsc{iv} line enter $g_{\mathrm{P1}}$ band, and C \textsc{iii]} enters $r_{\mathrm{P1}}$ band; and (v) when $z\approx 2.4$, Ly$\alpha$ and Si \textsc{iv} enter $g_{\mathrm{P1}}$ band.

The distributions of de-reddened $i_{P1}$ magnitudes and spectroscopic redshifts of $GoodQSO$ sample from SDSS DR14Q are also shown in Figure \ref{fig:gpqc_photoz} (b) for comparison. The $GoodQSO$ sample and the sample of GPQ candidates have similar redshift distributions with some subtle differences. The magnitude distributions are also similar to each other, except that $GoodQSO$ has a larger fraction of bright sources ($i_{P1}<19$) than the GPQ candidates. Such differences in both redshift and magnitude distributions of these two samples are mainly due to their different target selection strategies. Our GPQ candidates are selected from one single parent sample, while SDSS DR14Q includes many quasar samples in various redshift and magnitude ranges \citep[see Section 2.2 of][]{paris2018sloan}.

\begin{figure*}[htb!]
    \gridline{\fig{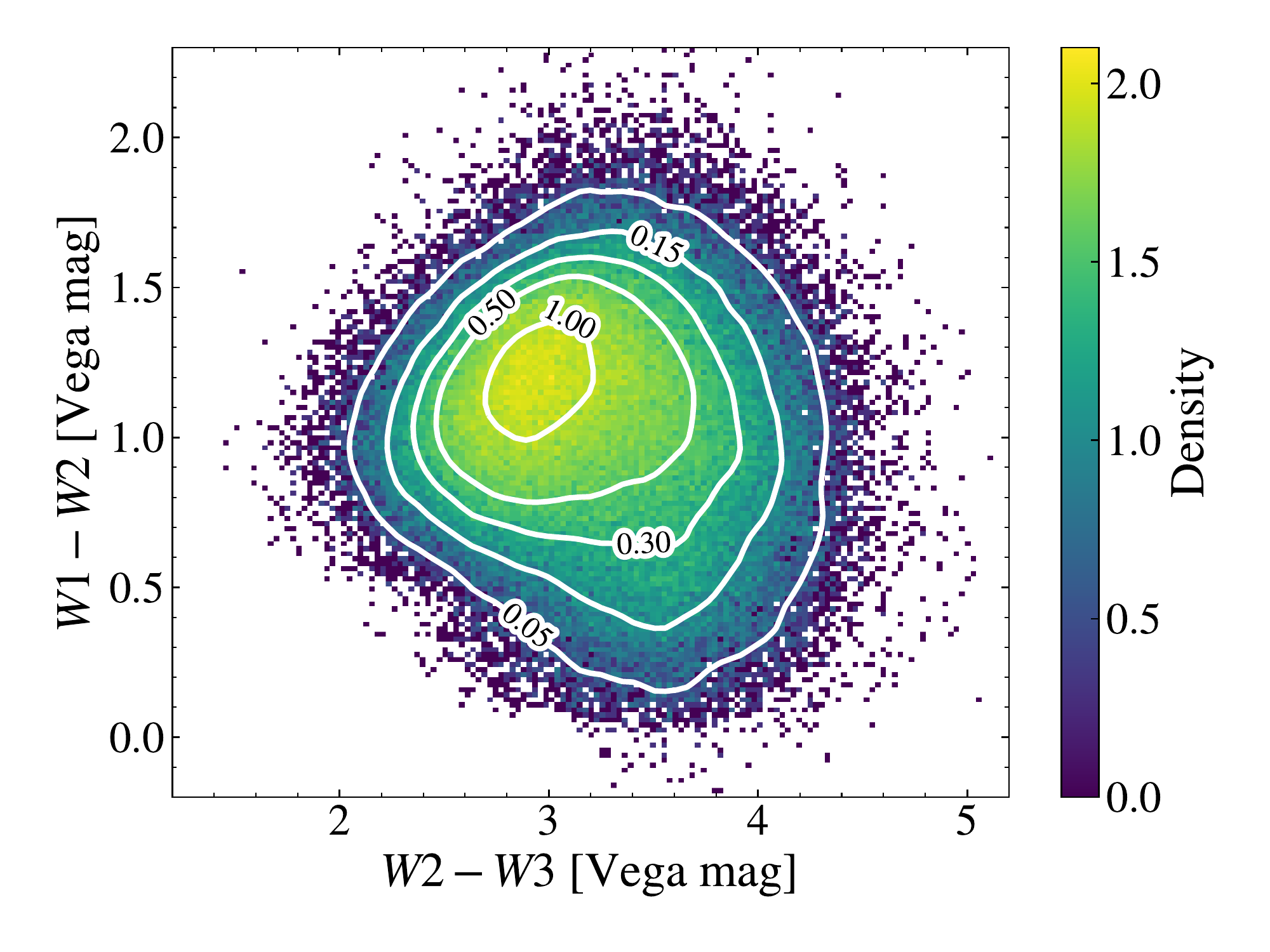}{0.42\textwidth}{(a)}
              \fig{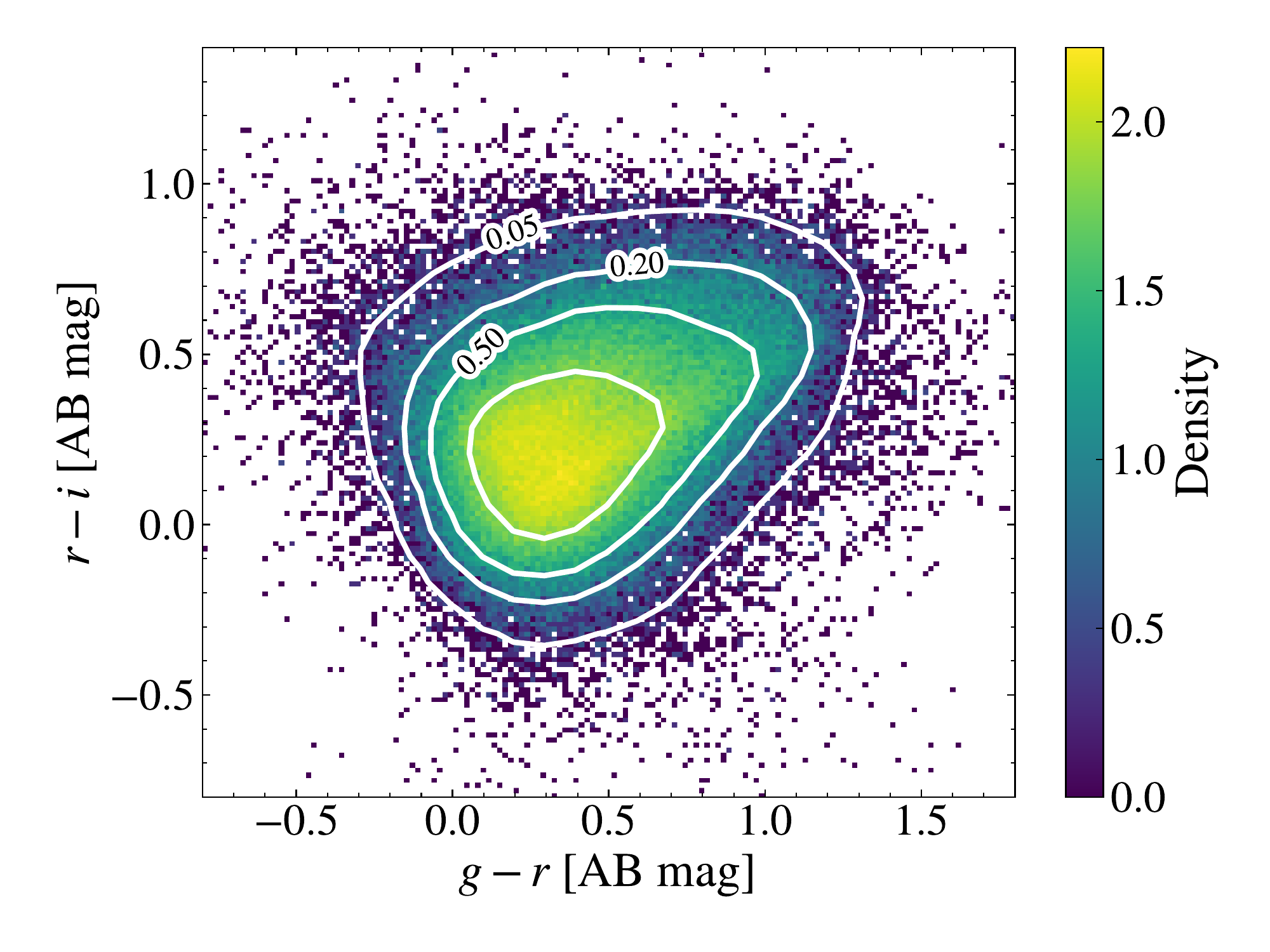}{0.42\textwidth}{(b)}}
    \gridline{\fig{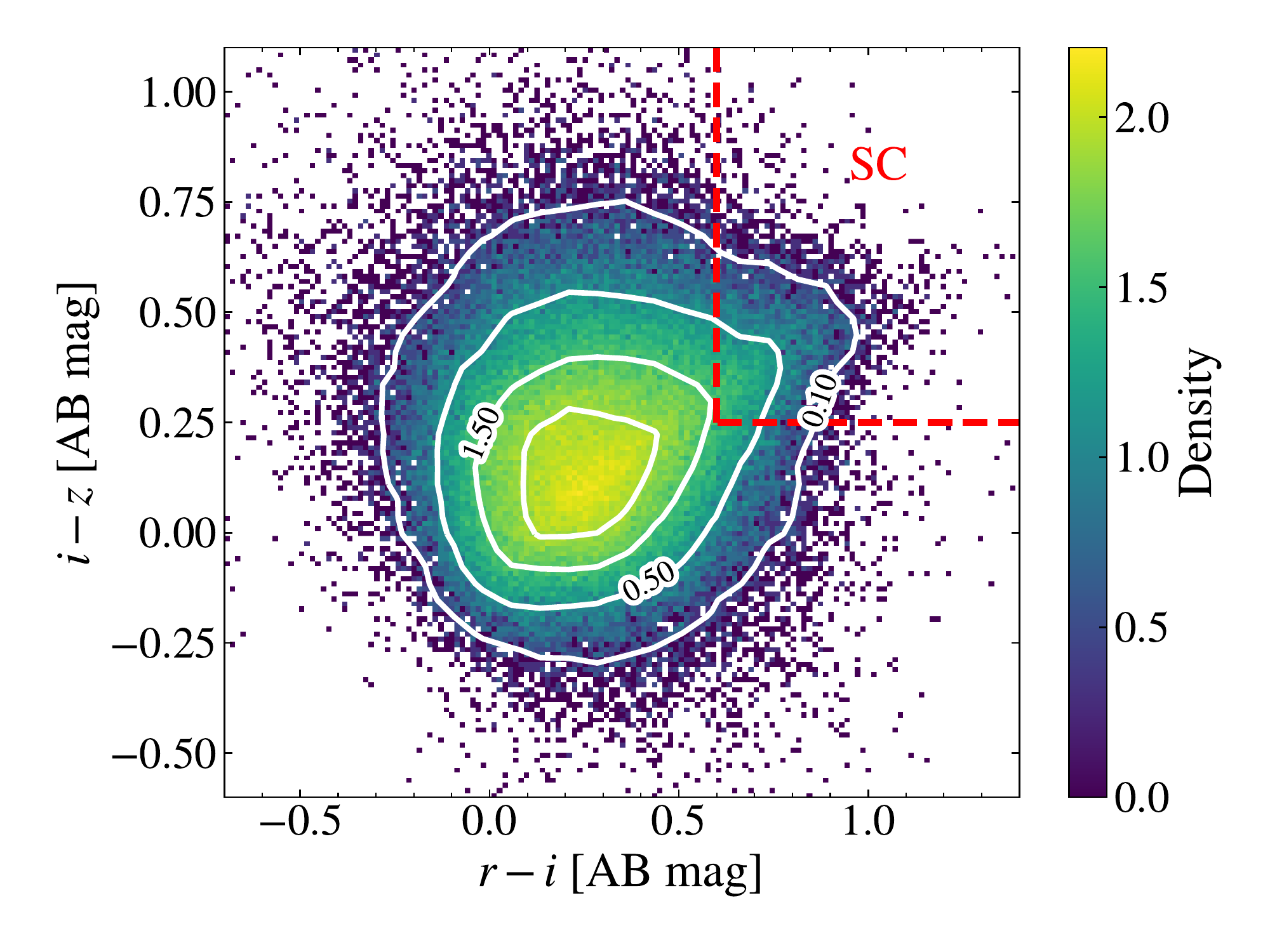}{0.42\textwidth}{(c)}
              \fig{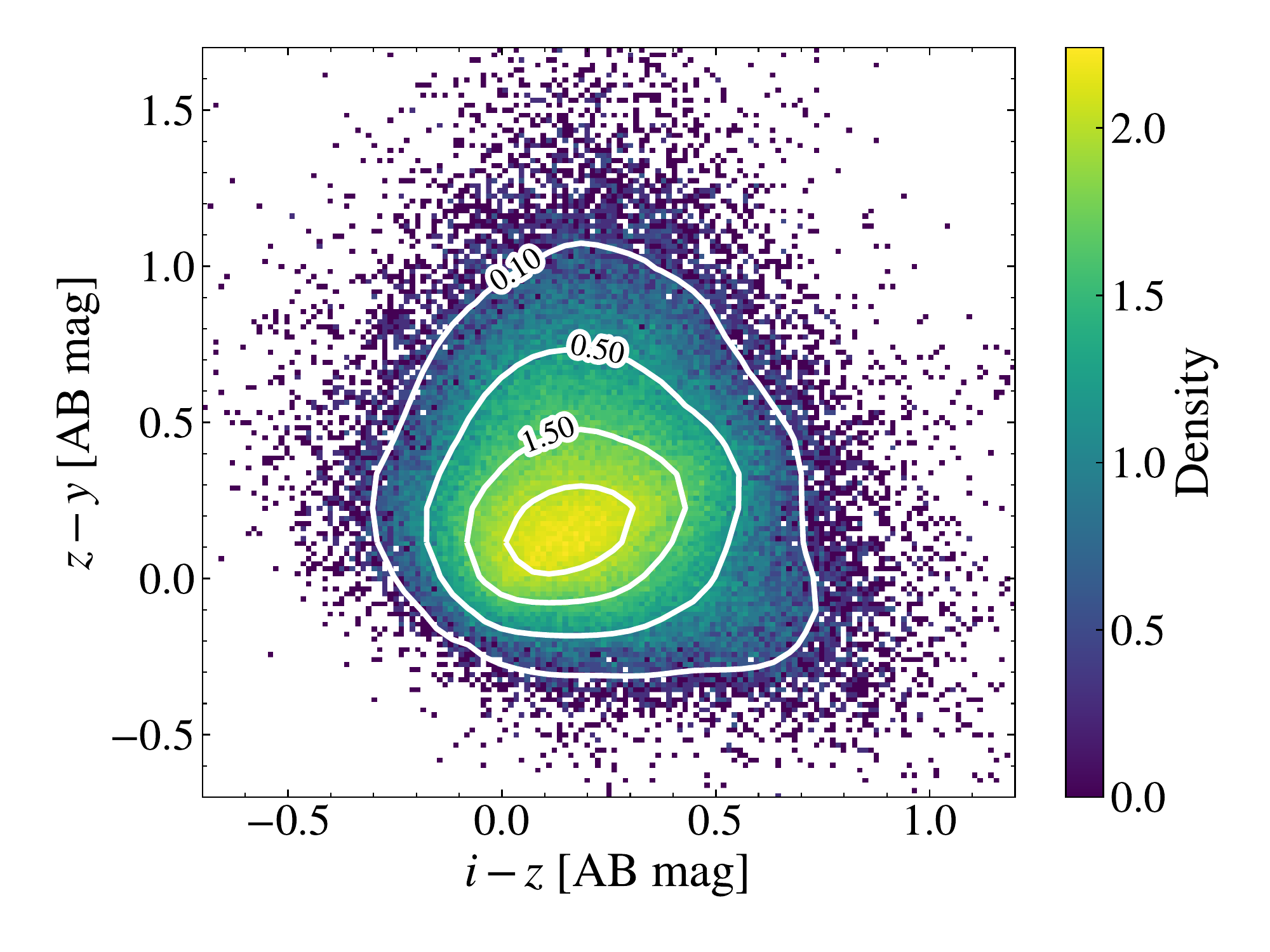}{0.42\textwidth}{(d)}}
    \caption{AllWISE (a) and PS1 (b, c, d) color-color diagrams of sources from the GPQ candidate catalog. Contour lines based on 2d KDE are displayed on the density plots. The red dashed lines and ``SC'' in subplot (c) mark the region which is slightly contaminated by stars ($r-i>0.6\ \&\ i-z>0.25$).}
    \label{fig:ccd_new}
\end{figure*}

The color-color properties of sources from the GPQ candidate catalog are shown in Figure \ref{fig:ccd_new}. In general, GPQ candidates have color-color distributions that are well matched to those of $GoodMockQSO$-PS1 (see Figure \ref{fig:ccds} and \ref{fig:ccdwise}). The unimodal structures seen from both AllWISE and PS1 colors imply a low level of contamination from stars and galaxies. However, contamination from stars can be recognized from the $i-z$ versus $r-i$ diagram, where some sources are concentrated along the stellar locus (see the slightly contaminated region ``SC'' in  Figure \ref{fig:ccd_new} (c)). We apply no cut on the ``SC'' region because it only contains 7,892 sources (4.90\% of whole catalog) and any cut is likely to also remove reddened quasars (see Figure \ref{fig:ccds}).

\section{Summary and conclusions} \label{sec:summary}

We present a Transfer Learning Framework for quasar selection, and its application on finding GPQs. We construct mock samples of quasars and galaxies behind the Galactic plane, by assigning new locations and extinction values to the extinction-corrected high-$b$ SDSS extragalactic sources. We use PS1 limiting magnitudes to select good mock sources, and compare them with high-$b$ sources in color-color spaces. We show that the covariate change of source colors is significant from high-$b$ regions to the Galactic plane. We synthesize training and validation data for machine learning with: (i) good mock samples, (ii) SDSS extragalactic sources which do not have counterparts in the good mock samples, and (iii) real LAMOST Galactic plane star sample.

We apply XGBoost algorithm for machine learning in this study. To help reduce the effects of \textit{class imbalance} and \textit{class-balance change}, we turn the three-class classification task (star, galaxy, and quasar) into two binary classification problems. A total of 13 features are used for the two classification steps: $g-r$, $r-i$, $i-z$, $z-y$, $g-W1$, $r-W1$, $i-W1$, $z-W1$, $y-W1$, $W1-W2$, $W2-W3$, $i-i_{\mathrm{Kron}}$ and $z-z_{\mathrm{Kron}}$. In order to remove star and galaxy contaminants, we use high thresholds of model-predicted probabilities ($p_{\mathrm{EXT}}>0.99\ \&\ p_{\mathrm{QSO}}>0.95$) to select extragalactic and quasar candidates. We perform an additional cut on probability density of zero proper motion ($\mathrm{log}(f_{\mathrm{PM0}})\geq -4$) based on \gdr{2}\ data to further reduce stellar contamination. Using the extinction-corrected SDSS DR14Q sources, we build the photometric redshift estimator with $\mathrm{RMSE}=0.35$ on the validation set. 

Our GPQ candidate sample is validated with Simbad database and Milliquas catalog. The purity of quasars is $\sim 90$\% on the Simbad matches. Under our constraints for good PS1 and AllWISE photometry, 95.14\% of the GPQs in the Milliquas catalog and 97.88\% of the GPQs from SDSS DR16Q can be recalled with our GPQ candidate sample. The photometric quality constraints ensure reliability of the candidates, but at the cost of lower overall completeness of the candidate sample. The sky density of GPQ candidates is consistent with the estimation based on mock GPQ catalog. The median marginal probability of GPQs to the PS1-AllWISE sample with good photometry is $\sim 10^{-3}$ and the lowest marginal probability is $\sim 10^{-4}$. We compile the GPQ candidate catalog after removing known stars in Simbad and LAMOST, and some duplicated sources. The GPQ candidate catalog consists of \finalgpqnum\ sources. In addtion to our machine learning predictions, we include PS1 and AllWISE photometry, as well as \gdr{2}\ astrometry in the table. The GPQ candidate sample has a broad redshift coverage ($0<z\lesssim 5$), indicating that our selection methods can be used on wide redshift ranges. 

Colors of GPQ candidates agree well with those of the mock GPQ catalog, which also indicates high purity of the candidates even though the marginal probability is low. Contamination from stars and galaxies still exists in the GPQ candidate sample, but at a low level. Because most stars in the training sample ($T_{Star}$) are bright ones, identifying and removing faint stars can be challenging for the XGBoost classification model (CLF-1). Using colors instead of magnitudes as features helps to lessen the effects of such training sample bias. The strict $\mathrm{log}(f_{\mathrm{PM0}})\geq -4$ cut can additionally remove most stellar contaminants. Galaxies overlap heavily with quasars on PS1 color-color diagrams, and show similar $\mathrm{log}(f_{\mathrm{PM0}})$ distribution with quasars. The use of AllWISE colors ($W1-W2$ and $W2-W3$) and morphological separators ($i-i_{\mathrm{Kron}}$ and $z-z_{\mathrm{Kron}}$) largely aids the galaxy-quasar classification. For future GPQ candidate selections, we expect to improve the machine learning performance by compiling a Galactic plane star training sample with more stars in the faint end. 

We have been carrying out a series of spectroscopic identification for GPQ candidates since 2018, using optical telescopes including two-meter telescopes based at Lijiang and Xinglong in China and Siding Spring in Australia, and 200-inch Hale Telescope in the US. The success rate of identifying new GPQs is $\sim$90\% in our spectroscopic campaign, which is consistent with the estimated reliability of the GPQ candidate catalog. We have also been exploring the LAMOST spectral data to find new GPQs. All these efforts yield promising results that will be presented in the next paper of this series. 

\acknowledgments

We thank the support from the National Key R\&D Program of China (2016YFA0400703) and the National Science Foundation of China (11533001, 11721303 \& 11927804). We thank the referee very much for constructive and helpful suggestions to improve this paper. We thank Prof. Yanxia Zhang (NAOC) for helping us cross-match PS1 and AllWISE catalogs. We thank Dr. Jinyi Yang and Dr. Feige Wang from Steward Observatory for useful suggestions. YF thanks Dr. Hassen Yesuf (KIAA-PKU), Dr. Bojin Zhuang (Ping An Technology), Hongdong Zheng (PKU) and Dinghuai Zhang (Mila/PKU) for helpful discussions on machine learning. YF thanks Prof. Gregory J. Herczeg, Bitao Wang, Mingyang Zhuang, Yun Zheng, Yuanhang Ning and Niankun Yu from PKU for helping edit the draft.

This publication makes use of data from the Pan-STARRS1 Surveys. The Pan-STARRS1 Surveys (PS1) and the PS1 public science archive have been made possible through contributions by the Institute for Astronomy, the University of Hawaii, the Pan-STARRS Project Office, the Max-Planck Society and its participating institutes, the Max Planck Institute for Astronomy, Heidelberg and the Max Planck Institute for Extraterrestrial Physics, Garching, The Johns Hopkins University, Durham University, the University of Edinburgh, the Queen's University Belfast, the Harvard-Smithsonian Center for Astrophysics, the Las Cumbres Observatory Global Telescope Network Incorporated, the National Central University of Taiwan, the Space Telescope Science Institute, the National Aeronautics and Space Administration under Grant No. NNX08AR22G issued through the Planetary Science Division of the NASA Science Mission Directorate, the National Science Foundation Grant No. AST-1238877, the University of Maryland, Eotvos Lorand University (ELTE), the Los Alamos National Laboratory, and the Gordon and Betty Moore Foundation.

This publication makes use of data products from the Wide-field Infrared Survey Explorer, which is a joint project of the University of California, Los Angeles, and the Jet Propulsion Laboratory/California Institute of Technology, funded by the National Aeronautics and Space Administration.

This work has made use of data from the European Space Agency (ESA) mission {\it Gaia} (\url{https://www.cosmos.esa.int/gaia}), processed by the {\it Gaia} Data Processing and Analysis Consortium (DPAC, \url{https://www.cosmos.esa.int/web/gaia/dpac/consortium}). Funding for the DPAC has been provided by national institutions, in particular the institutions participating in the {\it Gaia} Multilateral Agreement.

The Guoshoujing Telescope (the Large Sky Area Multi-object Fiber Spectroscopic Telescope LAMOST) is a National Major Scientific Project built by the Chinese Academy of Sciences. Funding for the project has been provided by the National Development and Reform Commission. LAMOST is operated and managed by the National Astronomical Observatories, Chinese Academy of Sciences.

Funding for the Sloan Digital Sky Survey IV has been provided by the Alfred P. Sloan Foundation, the U.S. Department of Energy Office of Science, and the Participating Institutions. SDSS-IV acknowledges support and resources from the Center for High-Performance Computing at the University of Utah. The SDSS web site is www.sdss.org. SDSS-IV is managed by the Astrophysical Research Consortium for the 
Participating Institutions of the SDSS Collaboration including the 
Brazilian Participation Group, the Carnegie Institution for Science, 
Carnegie Mellon University, the Chilean Participation Group, the French Participation Group, Harvard-Smithsonian Center for Astrophysics, 
Instituto de Astrof\'isica de Canarias, The Johns Hopkins University, Kavli Institute for the Physics and Mathematics of the Universe (IPMU) / 
University of Tokyo, the Korean Participation Group, Lawrence Berkeley National Laboratory, 
Leibniz Institut f\"ur Astrophysik Potsdam (AIP),  
Max-Planck-Institut f\"ur Astronomie (MPIA Heidelberg), 
Max-Planck-Institut f\"ur Astrophysik (MPA Garching), 
Max-Planck-Institut f\"ur Extraterrestrische Physik (MPE), 
National Astronomical Observatories of China, New Mexico State University, New York University, University of Notre Dame, 
Observat\'ario Nacional / MCTI, The Ohio State University, 
Pennsylvania State University, Shanghai Astronomical Observatory, 
United Kingdom Participation Group,
Universidad Nacional Aut\'onoma de M\'exico, University of Arizona, 
University of Colorado Boulder, University of Oxford, University of Portsmouth, 
University of Utah, University of Virginia, University of Washington, University of Wisconsin, 
Vanderbilt University, and Yale University.

%

\vspace{5mm}


\software{astropy \citep{astropy:2013,astropy:2018},
          dustmaps \citep{2018JOSS....3..695M},
          GNU Parallel \citep{Tange2011a},
          healpy \citep{Zonca2019},
          HEALPix \citep{2005ApJ...622..759G},
          optuna \citep{akiba2019optuna},
          scikit-learn \citep{pedregosa2011scikit},
          TOPCAT \citep{2005ASPC..347...29T},
          XGBoost \citep{chen2016xgboost}.
}

\bibliography{gpqpaperone}

\begin{thebibliography}{}
\expandafter\ifx\csname natexlab\endcsname\relax\def\natexlab#1{#1}\fi
\providecommand{\url}[1]{\href{#1}{#1}}
\providecommand{\dodoi}[1]{doi:~\href{http://doi.org/#1}{\nolinkurl{#1}}}
\providecommand{\doeprint}[1]{\href{http://ascl.net/#1}{\nolinkurl{http://ascl.net/#1}}}
\providecommand{\doarXiv}[1]{\href{https://arxiv.org/abs/#1}{\nolinkurl{https://arxiv.org/abs/#1}}}

\bibitem[{Acero {et~al.}(2015)Acero, Ackermann, Ajello, Albert, Atwood,
  Axelsson, Baldini, Ballet, Barbiellini, Bastieri,
  {et~al.}}]{2015ApJS..218...23A}
Acero, F., Ackermann, M., Ajello, M., {et~al.} 2015, \apjs, 218, 23

\bibitem[{{Aguado} {et~al.}(2019){Aguado}, {Ahumada}, {Almeida}, {Anderson},
  {Andrews}, {Anguiano}, {Aquino Ort{\'\i}z}, {Arag{\'o}n-Salamanca},
  {Argudo-Fern{\'a}ndez}, {Aubert}, \& et~al.}]{2019ApJS..240...23A}
{Aguado}, D.~S., {Ahumada}, R., {Almeida}, A., {et~al.} 2019, \apjs, 240, 23

\bibitem[{Akiba {et~al.}(2019)Akiba, Sano, Yanase, Ohta, \&
  Koyama}]{akiba2019optuna}
Akiba, T., Sano, S., Yanase, T., Ohta, T., \& Koyama, M. 2019, in Proceedings
  of the 25th ACM SIGKDD International Conference on Knowledge Discovery \&
  Data Mining, 2623--2631

\bibitem[{{Arenou} {et~al.}(2018){Arenou}, {Luri}, {Babusiaux}, {Fabricius},
  {Helmi}, {Muraveva}, {Robin}, {Spoto}, {Vallenari}, {Antoja},
  {Cantat-Gaudin}, {Jordi}, {Leclerc}, {Reyl{\'e}}, {Romero-G{\'o}mez}, {Shih},
  {Soria}, \& et~al.}]{Arenou2018}
{Arenou}, F., {Luri}, X., {Babusiaux}, C., {et~al.} 2018, \aap, 616, A17

\bibitem[{Argyriou {et~al.}(2006)Argyriou, Evgeniou, \&
  Pontil}]{argyriou2006multi}
Argyriou, A., Evgeniou, T., \& Pontil, M. 2006, in Proceedings of the 19th
  International Conference on Neural Information Processing Systems, 41–48

\bibitem[{{Assef} {et~al.}(2018){Assef}, {Stern}, {Noirot}, {Jun}, {Cutri}, \&
  {Eisenhardt}}]{Assef2018}
{Assef}, R.~J., {Stern}, D., {Noirot}, G., {et~al.} 2018, \apjs, 234, 23

\bibitem[{{Astropy Collaboration} {et~al.}(2013){Astropy Collaboration},
  {Robitaille}, {Tollerud}, {Greenfield}, {Droettboom}, {Bray}, {Aldcroft},
  {Davis}, {Ginsburg}, {Price-Whelan}, {Kerzendorf}, {Conley}, \&
  et~al.}]{astropy:2013}
{Astropy Collaboration}, {Robitaille}, T.~P., {Tollerud}, E.~J., {et~al.} 2013,
  \aap, 558, A33

\bibitem[{{Babbedge} {et~al.}(2004){Babbedge}, {Rowan-Robinson},
  {Gonzalez-Solares}, {Polletta}, {Berta}, {P{\'e}rez-Fournon}, {Oliver},
  {Salaman}, {Irwin}, \& {Weatherley}}]{2004MNRAS.353..654B}
{Babbedge}, T.~S.~R., {Rowan-Robinson}, M., {Gonzalez-Solares}, E., {et~al.}
  2004, \mnras, 353, 654

\bibitem[{{Bachchan} {et~al.}(2016){Bachchan}, {Hobbs}, \&
  {Lindegren}}]{2016A&A...589A..71B}
{Bachchan}, R.~K., {Hobbs}, D., \& {Lindegren}, L. 2016, \aap, 589, A71

\bibitem[{{Bailer-Jones} {et~al.}(2019){Bailer-Jones}, {Fouesneau}, \&
  {Andrae}}]{2019MNRAS.490.5615B}
{Bailer-Jones}, C. A.~L., {Fouesneau}, M., \& {Andrae}, R. 2019, \mnras, 490,
  5615

\bibitem[{{Becker} {et~al.}(2001){Becker}, {White}, {Gregg},
  {Laurent-Muehleisen}, {Brotherton}, {Impey}, {Chaffee}, {Richards},
  {Helfand}, {Lacy}, {Courbin}, \& {Proctor}}]{2001ApJS..135..227B}
{Becker}, R.~H., {White}, R.~L., {Gregg}, M.~D., {et~al.} 2001, \apjs, 135, 227

\bibitem[{{Ben Bekhti} {et~al.}(2008){Ben Bekhti}, {Richter}, {Westmeier}, \&
  {Murphy}}]{2008A&A...487..583B}
{Ben Bekhti}, N., {Richter}, P., {Westmeier}, T., \& {Murphy}, M.~T. 2008,
  \aap, 487, 583

\bibitem[{{Ben Bekhti} {et~al.}(2012){Ben Bekhti}, {Winkel}, {Richter}, {Kerp},
  {Klein}, \& {Murphy}}]{2012A&A...542A.110B}
{Ben Bekhti}, N., {Winkel}, B., {Richter}, P., {et~al.} 2012, \aap, 542, A110

\bibitem[{Bishop(2006)}]{bishop2006pattern}
Bishop, C.~M. 2006, Pattern recognition and machine learning (springer)

\bibitem[{{Blanton} {et~al.}(2017){Blanton}, {Bershady}, {Abolfathi},
  {Albareti}, {Allende Prieto}, {Almeida}, {Alonso-Garc{\'\i}a}, {Anders},
  {Anderson}, {Andrews}, {Aquino-Ort{\'\i}z}, {Arag{\'o}n-Salamanca},
  {Argudo-Fern{\'a}ndez}, {Armengaud}, {Aubourg}, {Avila-Reese}, {Badenes}, \&
  et~al.}]{2017AJ....154...28B}
{Blanton}, M.~R., {Bershady}, M.~A., {Abolfathi}, B., {et~al.} 2017, \aj, 154,
  28

\bibitem[{Blitzer {et~al.}(2006)Blitzer, McDonald, \&
  Pereira}]{blitzer2006domain}
Blitzer, J., McDonald, R., \& Pereira, F. 2006, in Proceedings of the 2006
  Conference on Empirical Methods in Natural Language Processing, 120–128

\bibitem[{{Bovy} {et~al.}(2011){Bovy}, {Hennawi}, {Hogg}, {Myers},
  {Kirkpatrick}, {Schlegel}, {Ross}, {Sheldon}, {McGreer}, {Schneider}, \&
  {Weaver}}]{2011ApJ...729..141B}
{Bovy}, J., {Hennawi}, J.~F., {Hogg}, D.~W., {et~al.} 2011, \apj, 729, 141

\bibitem[{{Bovy} {et~al.}(2012){Bovy}, {Myers}, {Hennawi}, {Hogg}, {McMahon},
  {Schiminovich}, {Sheldon}, {Brinkmann}, {Schneider}, \&
  {Weaver}}]{2012ApJ...749...41B}
{Bovy}, J., {Myers}, A.~D., {Hennawi}, J.~F., {et~al.} 2012, \apj, 749, 41

\bibitem[{Boyd {et~al.}(2013)Boyd, Eng, \& Page}]{boyd2013area}
Boyd, K., Eng, K.~H., \& Page, C.~D. 2013, in Joint European conference on
  machine learning and knowledge discovery in databases, Springer, 451--466

\bibitem[{{Brescia} {et~al.}(2013){Brescia}, {Cavuoti}, {D'Abrusco}, {Longo},
  \& {Mercurio}}]{2013ApJ...772..140B}
{Brescia}, M., {Cavuoti}, S., {D'Abrusco}, R., {Longo}, G., \& {Mercurio}, A.
  2013, \apj, 772, 140

\bibitem[{{Budav{\'a}ri} {et~al.}(2001){Budav{\'a}ri}, {Csabai}, {Szalay},
  {Connolly}, {Szokoly}, {Vand en Berk}, \& {Richards}}]{2001AJ....122.1163B}
{Budav{\'a}ri}, T., {Csabai}, I., {Szalay}, A. e.~S., {et~al.} 2001, \aj, 122,
  1163

\bibitem[{{Chambers} {et~al.}(2016){Chambers}, {Magnier}, {Metcalfe},
  {Flewelling}, {Huber}, {Waters}, {Denneau}, {Draper}, {Farrow}, {Finkbeiner},
  {Holmberg}, {Koppenhoefer}, {Price}, {Saglia}, {Schlafly}, {Smartt},
  {Sweeney}, {Wainscoat}, {Burgett}, {Grav}, {Heasley}, {Hodapp}, {Jedicke},
  {Kaiser}, \& et~al.}]{Chambers2016}
{Chambers}, K.~C., {Magnier}, E.~A., {Metcalfe}, N., {et~al.} 2016, arXiv
  e-prints, arXiv:1612.05560

\bibitem[{Chen \& Guestrin(2016)}]{chen2016xgboost}
Chen, T., \& Guestrin, C. 2016, in Proceedings of the 22nd acm sigkdd
  international conference on knowledge discovery and data mining, 785--794

\bibitem[{Chicco \& Jurman(2020)}]{chicco2020advantages}
Chicco, D., \& Jurman, G. 2020, BMC genomics, 21, 6

\bibitem[{{Chiu} {et~al.}(2006){Chiu}, {Fan}, {Leggett}, {Golimowski}, {Zheng},
  {Geballe}, {Schneider}, \& {Brinkmann}}]{2006AJ....131.2722C}
{Chiu}, K., {Fan}, X., {Leggett}, S.~K., {et~al.} 2006, \aj, 131, 2722

\bibitem[{Condon {et~al.}(1998)Condon, Cotton, Greisen, Yin, Perley, Taylor, \&
  Broderick}]{condon1998nrao}
Condon, J.~J., Cotton, W., Greisen, E., {et~al.} 1998, \aj, 115, 1693

\bibitem[{{Croom} {et~al.}(2009){Croom}, {Richards}, {Shanks}, {Boyle},
  {Sharp}, {Bland-Hawthorn}, {Bridges}, {Brunner}, {Cannon}, {Carson}, {Chiu},
  {Colless}, {Couch}, {De Propris}, {Drinkwater}, {Edge}, {Fine}, {Loveday},
  {Miller}, {Myers}, {Nichol}, {Outram}, {Pimbblet}, {Roseboom}, {Ross},
  {Schneider}, {Smith}, {Stoughton}, {Strauss}, \&
  {Wake}}]{2009MNRAS.392...19C}
{Croom}, S.~M., {Richards}, G.~T., {Shanks}, T., {et~al.} 2009, \mnras, 392, 19

\bibitem[{{Cui} {et~al.}(2012){Cui}, {Zhao}, {Chu}, {Li}, {Li}, {Zhang}, {Su},
  {Yao}, {Wang}, {Xing}, {Li}, {Zhu}, {Wang}, {Gu}, {Luo}, {Xu}, {Zhang},
  {Liu}, {Zhang}, {Yang}, {Cao}, {Chen}, {Chen}, {Chen}, {Chen}, {Chu}, {Feng},
  {Gong}, {Hou}, {Hu}, {Hu}, {Hu}, {Jia}, {Jiang}, {Jiang}, {Jiang}, {Jin},
  {Li}, {Li}, {Li}, {Liu}, {Liu}, {Lu}, {Mao}, {Men}, {Qi}, {Qi}, {Shi},
  {Tang}, {Tao}, {Wang}, {Wang}, {Wang}, {Wang}, {Wang}, {Wang}, {Wang},
  {Wang}, {Wang}, {Wang}, {Wang}, {Wang}, {Xu}, {Xu}, {Yang}, {Yu}, {Yuan},
  {Yuan}, {Zhai}, {Zhang}, {Zhang}, {Zhang}, {Zhao}, {Zhou}, {Zhou}, {Zhu}, \&
  {Zou}}]{2012RAA....12.1197C}
{Cui}, X.-Q., {Zhao}, Y.-H., {Chu}, Y.-Q., {et~al.} 2012, Research in Astronomy
  and Astrophysics, 12, 1197

\bibitem[{Davis \& Goadrich(2006)}]{davis2006relationship}
Davis, J., \& Goadrich, M. 2006, in Proceedings of the 23rd international
  conference on Machine learning, 233--240

\bibitem[{{Deng} {et~al.}(2012){Deng}, {Newberg}, {Liu}, {Carlin}, {Beers},
  {Chen}, {Chen}, {Christlieb}, {Grillmair}, {Guhathakurta}, {Han}, {Hou},
  {Lee}, {L{\'e}pine}, {Li}, {Liu}, {Pan}, {Sellwood}, {Wang}, {Wang}, {Yang},
  {Yanny}, {Zhang}, {Zhang}, {Zheng}, \& {Zhu}}]{2012RAA....12..735D}
{Deng}, L.-C., {Newberg}, H.~J., {Liu}, C., {et~al.} 2012, Research in
  Astronomy and Astrophysics, 12, 735

\bibitem[{{Dobrzycki} {et~al.}(2003){Dobrzycki}, {Macri}, {Stanek}, \&
  {Groot}}]{2003AJ....125.1330D}
{Dobrzycki}, A., {Macri}, L.~M., {Stanek}, K.~Z., \& {Groot}, P.~J. 2003, \aj,
  125, 1330

\bibitem[{Du~Plessis \& Sugiyama(2014)}]{du2014semi}
Du~Plessis, M.~C., \& Sugiyama, M. 2014, Neural Networks, 50, 110

\bibitem[{{Fan} {et~al.}(2001){Fan}, {Narayanan}, {Lupton}, {Strauss}, {Knapp},
  {Becker}, {White}, {Pentericci}, {Leggett}, {Haiman}, {Gunn}, {Ivezi{\'c}},
  {Schneider}, {Anderson}, {Brinkmann}, {Bahcall}, {Connolly}, {Csabai},
  {oridoi}, {Fukugita}, {Geballe}, {Grebel}, {Harbeck}, {Hennessy}, {Lamb},
  {Miknaitis}, {Munn}, {Nichol}, {Okamura}, {Pier}, {Prada}, {Richards},
  {Szalay}, \& {York}}]{2001AJ....122.2833F}
{Fan}, X., {Narayanan}, V.~K., {Lupton}, R.~H., {et~al.} 2001, \aj, 122, 2833

\bibitem[{{Farrow} {et~al.}(2014){Farrow}, {Cole}, {Metcalfe}, {Draper},
  {Norberg}, {Foucaud}, {Burgett}, {Chambers}, {Kaiser}, {Kudritzki},
  {Magnier}, {Price}, {Tonry}, \& {Waters}}]{2014MNRAS.437..748F}
{Farrow}, D.~J., {Cole}, S., {Metcalfe}, N., {et~al.} 2014, \mnras, 437, 748

\bibitem[{{Fischer} {et~al.}(2019){Fischer}, {Rigby}, {Mahler}, {Gladders},
  {Sharon}, {Florian}, {Kraemer}, {Bayliss}, {Dahle}, {Barrientos}, {Lopez},
  {Tejos}, {Johnson}, \& {Wuyts}}]{2019ApJ...875..102F}
{Fischer}, T.~C., {Rigby}, J.~R., {Mahler}, G., {et~al.} 2019, \apj, 875, 102

\bibitem[{{Flesch}(2015)}]{Flesch2015}
{Flesch}, E.~W. 2015, \pasa, 32, e010

\bibitem[{{Flesch}(2019)}]{2019arXiv191205614F}
---. 2019, arXiv e-prints, arXiv:1912.05614

\bibitem[{Friedman {et~al.}(2000)Friedman, Hastie, Tibshirani,
  {et~al.}}]{friedman2000additive}
Friedman, J., Hastie, T., Tibshirani, R., {et~al.} 2000, The annals of
  statistics, 28, 337

\bibitem[{Friedman(2001)}]{friedman2001greedy}
Friedman, J.~H. 2001, Annals of statistics, 1189

\bibitem[{{Gaia Collaboration} {et~al.}(2016){Gaia Collaboration}, Prusti,
  De~Bruijne, Brown, Vallenari, Babusiaux, Bailer-Jones, Bastian, Biermann,
  Evans, Eyer, {et~al.}}]{prusti2016gaia}
{Gaia Collaboration}, Prusti, T., De~Bruijne, J., {et~al.} 2016, \aap, 595, A1

\bibitem[{{Gaia Collaboration} {et~al.}(2018{\natexlab{a}}){Gaia
  Collaboration}, {Mignard}, {Klioner}, {Lindegren}, {Hern{\'a}ndez},
  {Bastian}, {Bombrun}, {Hobbs}, {Lammers}, {Michalik}, {Ramos-Lerate},
  {Biermann}, {Fern{\'a}ndez-Hern{\'a}ndez}, {Geyer}, \&
  et~al.}]{mignard2018gaia}
{Gaia Collaboration}, {Mignard}, F., {Klioner}, S.~A., {et~al.}
  2018{\natexlab{a}}, \aap, 616, A14

\bibitem[{{Gaia Collaboration} {et~al.}(2018{\natexlab{b}}){Gaia
  Collaboration}, {Brown}, {Vallenari}, {Prusti}, {de Bruijne}, {Babusiaux},
  {Bailer-Jones}, {Biermann}, {Evans}, {Eyer}, {Jansen}, {Jordi}, {Klioner},
  {Lammers}, {Lindegren}, {Luri}, {Mignard}, \& et~al.}]{brown2018gaia}
{Gaia Collaboration}, {Brown}, A.~G.~A., {Vallenari}, A., {et~al.}
  2018{\natexlab{b}}, \aap, 616, A1

\bibitem[{{Glikman} {et~al.}(2006){Glikman}, {Helfand}, \&
  {White}}]{2006ApJ...640..579G}
{Glikman}, E., {Helfand}, D.~J., \& {White}, R.~L. 2006, \apj, 640, 579

\bibitem[{Gong {et~al.}(2019)Gong, Zhong, \& Hu}]{gong2019diversity}
Gong, Z., Zhong, P., \& Hu, W. 2019, IEEE Access, 7, 64323

\bibitem[{{G{\'o}rski} {et~al.}(2005){G{\'o}rski}, {Hivon}, {Banday},
  {Wandelt}, {Hansen}, {Reinecke}, \& {Bartelmann}}]{2005ApJ...622..759G}
{G{\'o}rski}, K.~M., {Hivon}, E., {Banday}, A.~J., {et~al.} 2005, \apj, 622,
  759

\bibitem[{{Grazian} {et~al.}(2000){Grazian}, {Cristiani}, {D'Odorico},
  {Omizzolo}, \& {Pizzella}}]{2000AJ....119.2540G}
{Grazian}, A., {Cristiani}, S., {D'Odorico}, V., {Omizzolo}, A., \& {Pizzella},
  A. 2000, \aj, 119, 2540

\bibitem[{{Green}(2018)}]{2018JOSS....3..695M}
{Green}, G. 2018, The Journal of Open Source Software, 3, 695

\bibitem[{{Green} {et~al.}(2019){Green}, {Schlafly}, {Zucker}, {Speagle}, \&
  {Finkbeiner}}]{2019ApJ...887...93G}
{Green}, G.~M., {Schlafly}, E., {Zucker}, C., {Speagle}, J.~S., \&
  {Finkbeiner}, D. 2019, \apj, 887, 93

\bibitem[{{Green} {et~al.}(2018){Green}, {Schlafly}, {Finkbeiner}, {Rix},
  {Martin}, {Burgett}, {Draper}, {Flewelling}, {Hodapp}, {Kaiser}, {Kudritzki},
  {Magnier}, {Metcalfe}, {Tonry}, {Wainscoat}, \&
  {Waters}}]{2018MNRAS.478..651G}
{Green}, G.~M., {Schlafly}, E.~F., {Finkbeiner}, D., {et~al.} 2018, \mnras,
  478, 651

\bibitem[{{Green} {et~al.}(1986){Green}, {Schmidt}, \&
  {Liebert}}]{1986ApJS...61..305G}
{Green}, R.~F., {Schmidt}, M., \& {Liebert}, J. 1986, \apjs, 61, 305

\bibitem[{{Gregg} {et~al.}(1996){Gregg}, {Becker}, {White}, {Helfand},
  {McMahon}, \& {Hook}}]{1996AJ....112..407G}
{Gregg}, M.~D., {Becker}, R.~H., {White}, R.~L., {et~al.} 1996, \aj, 112, 407

\bibitem[{Hastie {et~al.}(2009)Hastie, Tibshirani, \&
  Friedman}]{hastie2009elements}
Hastie, T., Tibshirani, R., \& Friedman, J. 2009, The elements of statistical
  learning: data mining, inference, and prediction (Springer Science \&
  Business Media)

\bibitem[{Hern{\'a}n-Caballero {et~al.}(2016)Hern{\'a}n-Caballero,
  Hatziminaoglou, Alonso-Herrero, \& Mateos}]{hernan2016near}
Hern{\'a}n-Caballero, A., Hatziminaoglou, E., Alonso-Herrero, A., \& Mateos, S.
  2016, \mnras, 463, 2064

\bibitem[{Huo {et~al.}(2010)Huo, Liu, Yuan, Zhang, Zhao, Chen, Bai, Zhang,
  Zhang, Garc{\'\i}a-Benito, {et~al.}}]{huo2010new}
Huo, Z.-Y., Liu, X.-W., Yuan, H.-B., {et~al.} 2010, Research in Astronomy and
  Astrophysics, 10, 612

\bibitem[{Huo {et~al.}(2013)Huo, Liu, Xiang, Yuan, Huang, Zhang, Yan, Bai,
  Chen, Chen, {et~al.}}]{huo2013lamost}
Huo, Z.-Y., Liu, X.-W., Xiang, M.-S., {et~al.} 2013, \aj, 145, 159

\bibitem[{Huo {et~al.}(2015)Huo, {Liu}, {Xiang}, {Shi}, {Yuan}, {Huang},
  {Zhang}, {Hou}, {Wang}, \& {Yang}}]{huo2015lamost}
Huo, Z.-Y., {Liu}, X.-W., {Xiang}, M.-S., {et~al.} 2015, Research in Astronomy
  and Astrophysics, 15, 1438

\bibitem[{{Im} {et~al.}(2007){Im}, {Lee}, {Cho}, {Choi}, {Ko}, \&
  {Song}}]{Im2007}
{Im}, M., {Lee}, I., {Cho}, Y., {et~al.} 2007, \apj, 664, 64

\bibitem[{{Ivezi{\'c}} {et~al.}(2007){Ivezi{\'c}}, {Smith}, {Miknaitis}, {Lin},
  {Tucker}, {Lupton}, {Gunn}, {Knapp}, {Strauss}, {Sesar}, {Doi}, {Tanaka},
  {Fukugita}, {Holtzman}, {Kent}, {Yanny}, {Schlegel}, {Finkbeiner},
  {Padmanabhan}, {Rockosi}, {Juri{\'c}}, {Bond}, {Lee}, {Stoughton}, {Jester},
  {Harris}, {Harding}, {Morrison}, {Brinkmann}, {Schneider}, \&
  {York}}]{2007AJ....134..973I}
{Ivezi{\'c}}, {\v{Z}}., {Smith}, J.~A., {Miknaitis}, G., {et~al.} 2007, \aj,
  134, 973

\bibitem[{{Jin} {et~al.}(2019){Jin}, {Zhang}, {Zhang}, {Zhao}, {Wu}, \&
  {Fan}}]{2019MNRAS.485.4539J}
{Jin}, X., {Zhang}, Y., {Zhang}, J., {et~al.} 2019, \mnras, 485, 4539

\bibitem[{{Kirkpatrick} {et~al.}(1997){Kirkpatrick}, {Henry}, \&
  {Irwin}}]{1997AJ....113.1421K}
{Kirkpatrick}, J.~D., {Henry}, T.~J., \& {Irwin}, M.~J. 1997, \aj, 113, 1421

\bibitem[{{Koenig} \& {Leisawitz}(2014)}]{2014ApJ...791..131K}
{Koenig}, X.~P., \& {Leisawitz}, D.~T. 2014, \apj, 791, 131

\bibitem[{{Koz{\l}owski} \& {Kochanek}(2009)}]{2009ApJ...701..508K}
{Koz{\l}owski}, S., \& {Kochanek}, C.~S. 2009, \apj, 701, 508

\bibitem[{{Kron}(1980)}]{1980ApJS...43..305K}
{Kron}, R.~G. 1980, \apjs, 43, 305

\bibitem[{{Lacy} {et~al.}(2004){Lacy}, {Storrie-Lombardi}, {Sajina},
  {Appleton}, {Armus}, {Chapman}, {Choi}, {Fadda}, {Fang}, {Frayer},
  {Heinrichsen}, {Helou}, {Im}, {Marleau}, {Masci}, {Shupe}, {Soifer},
  {Surace}, {Teplitz}, {Wilson}, \& {Yan}}]{2004ApJS..154..166L}
{Lacy}, M., {Storrie-Lombardi}, L.~J., {Sajina}, A., {et~al.} 2004, \apjs, 154,
  166

\bibitem[{Lada(1987)}]{lada1987star}
Lada, C.~J. 1987, in Symposium-International astronomical union, Vol. 115,
  Cambridge University Press, 1--18

\bibitem[{{Laurino} {et~al.}(2011){Laurino}, {D'Abrusco}, {Longo}, \&
  {Riccio}}]{2011MNRAS.418.2165L}
{Laurino}, O., {D'Abrusco}, R., {Longo}, G., \& {Riccio}, G. 2011, \mnras, 418,
  2165

\bibitem[{{Lindegren} {et~al.}(2018{\natexlab{a}}){Lindegren}, {Hern{\'a}ndez},
  {Bombrun}, {Klioner}, {Bastian}, {Ramos-Lerate}, {de Torres},
  {Steidelm{\"u}ller}, {Stephenson}, {Hobbs}, {Lammers}, {Biermann}, \&
  et~al.}]{lindegren2018gaia}
{Lindegren}, L., {Hern{\'a}ndez}, J., {Bombrun}, A., {et~al.}
  2018{\natexlab{a}}, \aap, 616, A2

\bibitem[{{Lindegren} {et~al.}(2018{\natexlab{b}}){Lindegren}, {Hern{\'a}ndez},
  {Bombrun}, {Klioner}, {Bastian}, {Ramos-Lerate}, {de Torres},
  {Steidelm{\"u}ller}, {Stephenson}, {Hobbs}, {Lammers}, {Biermann}, \&
  et~al.}]{lindegren2018gaia2}
{Lindegren}, L., {Hern{\'a}ndez}, J., {Bombrun}, A., {et~al.}
  2018{\natexlab{b}}, in IAU 30 GA - Division A: Fundamental Astronomy

\bibitem[{{Luo} {et~al.}(2012){Luo}, {Zhang}, {Zhao}, {Zhao}, {Cui}, {Li},
  {Chu}, {Shi}, {Wang}, {Zhang}, {Bai}, {Chen}, {Wang}, {Guo}, {Chen}, {Du},
  {Kong}, {Lei}, {Li}, {Song}, {Wu}, {Zhang}, {Zhou}, {Zuo}, {Du}, {He}, {Hou},
  {Dong}, {Li}, {Li}, {Li}, {Song}, {Tian}, {Wang}, {Wu}, {Yang}, {Yuan},
  {Cao}, {Chen}, {Chen}, {Chen}, {Chu}, {Feng}, {Gong}, {Gu}, {Hou}, {Huo},
  {Hu}, {Hu}, {Hu}, {Jia}, {Jiang}, {Jiang}, {Jiang}, {Jin}, {Li}, {Li}, {Li},
  {Li}, {Li}, {Liu}, {Liu}, {Liu}, {Lu}, {Lu}, {Luo}, {Mao}, {Men}, {Ni}, {Qi},
  {Qi}, {Shi}, {Su}, {Sun}, {Su}, {Tang}, {Tao}, {Tu}, {Wang}, {Wang}, {Wang},
  {Wang}, {Wang}, {Wang}, {Wang}, {Wang}, {Wang}, {Wang}, {Wang}, {Wang},
  {Wang}, {Wang}, {Wei}, {Xue}, {Xing}, {Xu}, {Xu}, {Xu}, {Yang}, {Yang},
  {Yao}, {Yu}, {Yuan}, {Zhai}, {Zhang}, {Zhang}, {Zhang}, {Zhang}, {Zhang},
  {Zhang}, {Zhao}, {Zhou}, {Zhu}, {Zhu}, \& {Zou}}]{2012RAA....12.1243L}
{Luo}, A.~L., {Zhang}, H.-T., {Zhao}, Y.-H., {et~al.} 2012, Research in
  Astronomy and Astrophysics, 12, 1243

\bibitem[{Luo {et~al.}(2015)Luo, Zhao, Zhao, Deng, Liu, Jing, Wang, Zhang, Shi,
  Cui, {et~al.}}]{2015RAA....15.1095L}
Luo, A.-L., Zhao, Y.-H., Zhao, G., {et~al.} 2015, Research in Astronomy and
  Astrophysics, 15, 1095

\bibitem[{{Lyke} {et~al.}(2020){Lyke}, {Higley}, {McLane}, {Schurhammer},
  {Myers}, {Ross}, {Dawson}, {Chabanier}, {Martini}, {Busca}, {Mas des
  Bourboux}, {Salvato}, {Streblyanska}, {Zarrouk}, {Burtin}, {Anderson},
  {Bautista}, {Bizyaev}, {Brandt}, {Brinkmann}, {Brownstein}, {Comparat},
  {Green}, {de la Macorra}, {Mu{\~n}oz Guti{\'e}rrez}, {Hou}, {Newman},
  {Palanque-Delabrouille}, {P{\^a}ris}, {Percival}, {Petitjean}, {Rich},
  {Rossi}, {Schneider}, {Smith}, {Vivek}, \& {Weaver}}]{2020ApJS..250....8L}
{Lyke}, B.~W., {Higley}, A.~N., {McLane}, J.~N., {et~al.} 2020, \apjs, 250, 8

\bibitem[{{Magnier} {et~al.}(2016){Magnier}, {Schlafly}, {Finkbeiner}, {Tonry},
  {Goldman}, {R{\"o}ser}, {Schilbach}, {Chambers}, {Flewelling}, {Huber},
  {Price}, {Sweeney}, {Waters}, {Denneau}, {Draper}, {Hodapp}, {Jedicke},
  {Kudritzki}, {Metcalfe}, {Stubbs}, \& {Wainscoast}}]{2016arXiv161205242M}
{Magnier}, E.~A., {Schlafly}, E.~F., {Finkbeiner}, D.~P., {et~al.} 2016, arXiv
  e-prints, arXiv:1612.05242

\bibitem[{Mainzer {et~al.}(2011)Mainzer, Bauer, Grav, Masiero, Cutri, Dailey,
  Eisenhardt, McMillan, Wright, Walker, {et~al.}}]{mainzer2011preliminary}
Mainzer, A., Bauer, J., Grav, T., {et~al.} 2011, \apj, 731, 53

\bibitem[{{Mateos} {et~al.}(2012){Mateos}, {Alonso-Herrero}, {Carrera},
  {Blain}, {Watson}, {Barcons}, {Braito}, {Severgnini}, {Donley}, \&
  {Stern}}]{2012MNRAS.426.3271M}
{Mateos}, S., {Alonso-Herrero}, A., {Carrera}, F.~J., {et~al.} 2012, \mnras,
  426, 3271

\bibitem[{{Mirabal} {et~al.}(2016){Mirabal}, {Charles}, {Ferrara}, {Gonthier},
  {Harding}, {S{\'a}nchez-Conde}, \& {Thompson}}]{2016ApJ...825...69M}
{Mirabal}, N., {Charles}, E., {Ferrara}, E.~C., {et~al.} 2016, \apj, 825, 69

\bibitem[{{Palanque-Delabrouille} {et~al.}(2011){Palanque-Delabrouille},
  {Yeche}, {Myers}, {Petitjean}, {Ross}, {Sheldon}, {Aubourg}, {Delubac}, {Le
  Goff}, {P{\^a}ris}, {Rich}, {Dawson}, {Schneider}, \&
  {Weaver}}]{2011A&A...530A.122P}
{Palanque-Delabrouille}, N., {Yeche}, C., {Myers}, A.~D., {et~al.} 2011, \aap,
  530, A122

\bibitem[{Pan \& Yang(2009)}]{pan2009survey}
Pan, S.~J., \& Yang, Q. 2009, IEEE Transactions on knowledge and data
  engineering, 22, 1345

\bibitem[{P{\^a}ris {et~al.}(2018)P{\^a}ris, Petitjean, Aubourg, Myers,
  Streblyanska, Lyke, Anderson, Armengaud, Bautista, Blanton,
  {et~al.}}]{paris2018sloan}
P{\^a}ris, I., Petitjean, P., Aubourg, {\'E}., {et~al.} 2018, \aap, 613, A51

\bibitem[{{Pasquet-Itam} \& {Pasquet}(2018)}]{2018A&A...611A..97P}
{Pasquet-Itam}, J., \& {Pasquet}, J. 2018, \aap, 611, A97

\bibitem[{Pedregosa {et~al.}(2011)Pedregosa, Varoquaux, Gramfort, Michel,
  Thirion, Grisel, Blondel, Prettenhofer, Weiss, Dubourg,
  {et~al.}}]{pedregosa2011scikit}
Pedregosa, F., Varoquaux, G., Gramfort, A., {et~al.} 2011, the Journal of
  machine Learning research, 12, 2825

\bibitem[{{Planck Collaboration} {et~al.}(2014){Planck Collaboration},
  {Abergel}, {Ade}, {Aghanim}, {Alves}, {Aniano}, {Armitage-Caplan}, {Arnaud},
  {Ashdown}, {Atrio-Barandela}, \& et~al.}]{abergel2014planck}
{Planck Collaboration}, {Abergel}, A., {Ade}, P.~A.~R., {et~al.} 2014, \aap,
  571, A11

\bibitem[{{Pounds}(1979)}]{1979RSPSA.366..375P}
{Pounds}, K.~A. 1979, Proceedings of the Royal Society of London Series A, 366,
  375

\bibitem[{{Price-Whelan} {et~al.}(2018){Price-Whelan}, {Sip{\H{o}}cz},
  {G{\"u}nther}, {Lim}, {Crawford}, {Conseil}, {Shupe}, {Craig}, {Dencheva},
  {Ginsburg}, {VanderPlas}, {Bradley}, {P{\'e}rez-Su{\'a}rez}, \&
  et~al.}]{astropy:2018}
{Price-Whelan}, A.~M., {Sip{\H{o}}cz}, B.~M., {G{\"u}nther}, H.~M., {et~al.}
  2018, \aj, 156, 123

\bibitem[{Provost {et~al.}(1998)Provost, Fawcett, \& Kohavi}]{provost1998case}
Provost, F., Fawcett, T., \& Kohavi, R. 1998, in ICML Conference

\bibitem[{Quionero-Candela {et~al.}(2009)Quionero-Candela, Sugiyama,
  Schwaighofer, \& Lawrence}]{quionero2009dataset}
Quionero-Candela, J., Sugiyama, M., Schwaighofer, A., \& Lawrence, N.~D. 2009,
  Dataset shift in machine learning (The MIT Press)

\bibitem[{{Richards} {et~al.}(2001){Richards}, {Weinstein}, {Schneider}, {Fan},
  {Strauss}, {Vanden Berk}, {Annis}, \& {Burles}}]{2001AJ....122.1151R}
{Richards}, G.~T., {Weinstein}, M.~A., {Schneider}, D.~P., {et~al.} 2001, \aj,
  122, 1151

\bibitem[{Richards {et~al.}(2002)Richards, Fan, Newberg, Strauss, Berk,
  Schneider, Yanny, Boucher, Burles, Frieman,
  {et~al.}}]{richards2002spectroscopic}
Richards, G.~T., Fan, X., Newberg, H.~J., {et~al.} 2002, \aj, 123, 2945

\bibitem[{{Richards} {et~al.}(2004){Richards}, {Nichol}, {Gray}, {Brunner},
  {Lupton}, {Vand en Berk}, {Chong}, {Weinstein}, {Schneider}, {Anderson},
  {Munn}, {Harris}, {Strauss}, {Fan}, {Gunn}, {Ivezi{\'c}}, {York},
  {Brinkmann}, \& {Moore}}]{2004ApJS..155..257R}
{Richards}, G.~T., {Nichol}, R.~C., {Gray}, A.~G., {et~al.} 2004, \apjs, 155,
  257

\bibitem[{Saerens {et~al.}(2002)Saerens, Latinne, \&
  Decaestecker}]{saerens2002adjusting}
Saerens, M., Latinne, P., \& Decaestecker, C. 2002, Neural computation, 14, 21

\bibitem[{Saito \& Rehmsmeier(2015)}]{saito2015precision}
Saito, T., \& Rehmsmeier, M. 2015, PloS one, 10

\bibitem[{{Salvato} {et~al.}(2009){Salvato}, {Hasinger}, {Ilbert}, {Zamorani},
  {Brusa}, {Scoville}, {Rau}, {Capak}, {Arnouts}, {Aussel}, {Bolzonella},
  {Buongiorno}, {Cappelluti}, \& {Caputi}}]{2009ApJ...690.1250S}
{Salvato}, M., {Hasinger}, G., {Ilbert}, O., {et~al.} 2009, \apj, 690, 1250

\bibitem[{{Sandage}(1965)}]{1965ApJ...141.1560S}
{Sandage}, A. 1965, \apj, 141, 1560

\bibitem[{Savage {et~al.}(1993)Savage, Lu, Bahcall, Bergeron, Boksenberg,
  Hartig, Jannuzi, Kirhakos, Lockman, Sargent, {et~al.}}]{savage1993hubble}
Savage, B.~D., Lu, L., Bahcall, J.~N., {et~al.} 1993, \apj, 413, 116

\bibitem[{Savage {et~al.}(2000)Savage, Wakker, Jannuzi, Bahcall, Bergeron,
  Boksenberg, Hartig, Kirhakos, Murphy, Sargent, {et~al.}}]{savage2000hubble}
Savage, B.~D., Wakker, B., Jannuzi, B.~T., {et~al.} 2000, \apjs, 129, 563

\bibitem[{{Schlafly} {et~al.}(2016){Schlafly}, {Meisner}, {Stutz},
  {Kainulainen}, {Peek}, {Tchernyshyov}, {Rix}, {Finkbeiner}, {Covey}, {Green},
  {Bell}, {Burgett}, {Chambers}, {Draper}, {Flewelling}, {Hodapp}, {Kaiser},
  {Magnier}, {Martin}, {Metcalfe}, {Wainscoat}, \&
  {Waters}}]{2016ApJ...821...78S}
{Schlafly}, E.~F., {Meisner}, A.~M., {Stutz}, A.~M., {et~al.} 2016, \apj, 821,
  78

\bibitem[{{Schmidt}(1963)}]{schmidt19633c273}
{Schmidt}, M. 1963, \nat, 197, 1040

\bibitem[{{Secrest} {et~al.}(2015){Secrest}, {Dudik}, {Dorland}, {Zacharias},
  {Makarov}, {Fey}, {Frouard}, \& {Finch}}]{Secrest2015}
{Secrest}, N.~J., {Dudik}, R.~P., {Dorland}, B.~N., {et~al.} 2015, \apjs, 221,
  12

\bibitem[{{Shen} {et~al.}(2016){Shen}, {Argudo-Fern{\'a}ndez}, {Chen}, {Chen},
  {Feng}, {Hou}, {Hou}, {Jiang}, {Jing}, {Kong}, {Luo}, {Luo}, {Shao}, {Wang},
  {Wang}, {Wang}, {Wu}, {Wu}, {Yang}, {Yang}, {Yuan}, {Yuan}, {Zhang}, {Zhang},
  \& {Zhang}}]{2016RAA....16...43S}
{Shen}, S.-Y., {Argudo-Fern{\'a}ndez}, M., {Chen}, L., {et~al.} 2016, Research
  in Astronomy and Astrophysics, 16, 43

\bibitem[{Shimodaira(2000)}]{shimodaira2000improving}
Shimodaira, H. 2000, Journal of statistical planning and inference, 90, 227

\bibitem[{Skrutskie {et~al.}(2006)Skrutskie, Cutri, Stiening, Weinberg,
  Schneider, Carpenter, Beichman, Capps, Chester, Elias,
  {et~al.}}]{skrutskie2006two}
Skrutskie, M., Cutri, R., Stiening, R., {et~al.} 2006, \aj, 131, 1163

\bibitem[{{Stern} {et~al.}(2005){Stern}, {Eisenhardt}, {Gorjian}, {Kochanek},
  {Caldwell}, {Eisenstein}, {Brodwin}, {Brown}, {Cool}, {Dey}, {Green},
  {Jannuzi}, {Murray}, {Pahre}, \& {Willner}}]{2005ApJ...631..163S}
{Stern}, D., {Eisenhardt}, P., {Gorjian}, V., {et~al.} 2005, \apj, 631, 163

\bibitem[{{Stern} {et~al.}(2012){Stern}, {Assef}, {Benford}, {Blain}, {Cutri},
  {Dey}, {Eisenhardt}, {Griffith}, {Jarrett}, {Lake}, {Masci}, {Petty},
  {Stanford}, {Tsai}, {Wright}, {Yan}, {Harrison}, \&
  {Madsen}}]{2012ApJ...753...30S}
{Stern}, D., {Assef}, R.~J., {Benford}, D.~J., {et~al.} 2012, \apj, 753, 30

\bibitem[{{Strauss} {et~al.}(2002){Strauss}, {Weinberg}, {Lupton}, {Narayanan},
  {Annis}, {Bernardi}, {Blanton}, {Burles}, {Connolly}, {Dalcanton}, {oridoi},
  {Eisenstein}, \& et~al.}]{2002AJ....124.1810S}
{Strauss}, M.~A., {Weinberg}, D.~H., {Lupton}, R.~H., {et~al.} 2002, \aj, 124,
  1810

\bibitem[{{Su} \& {Cui}(2004)}]{2004ChJAA...4....1S}
{Su}, D.-Q., \& {Cui}, X.-Q. 2004, \cjaa, 4, 1, \dodoi{10.1088/1009-9271/4/1/1}

\bibitem[{Sugiyama \& Kawanabe(2012)}]{sugiyama2012machine}
Sugiyama, M., \& Kawanabe, M. 2012, Machine learning in non-stationary
  environments: Introduction to covariate shift adaptation (MIT press)

\bibitem[{Tange(2011)}]{Tange2011a}
Tange, O. 2011, ;login: The USENIX Magazine, 36, 42

\bibitem[{{Taylor}(2005)}]{2005ASPC..347...29T}
{Taylor}, M.~B. 2005, in Astronomical Society of the Pacific Conference Series,
  Vol. 347, Astronomical Data Analysis Software and Systems XIV, ed.
  P.~{Shopbell}, M.~{Britton}, \& R.~{Ebert}, 29

\bibitem[{Vanden~Berk {et~al.}(2001)Vanden~Berk, Richards, Bauer, Strauss,
  Schneider, Heckman, York, Hall, Fan, Knapp, {et~al.}}]{berk2001composite}
Vanden~Berk, D.~E., Richards, G.~T., Bauer, A., {et~al.} 2001, \aj, 122, 549

\bibitem[{Vapnik(2013)}]{vapnik2013nature}
Vapnik, V. 2013, The nature of statistical learning theory (Springer science \&
  business media)

\bibitem[{{Vennes} {et~al.}(2002){Vennes}, {Smith}, {Boyle}, {Croom}, {Kawka},
  {Shanks}, {Miller}, \& {Loaring}}]{2002MNRAS.335..673V}
{Vennes}, S., {Smith}, R.~J., {Boyle}, B.~J., {et~al.} 2002, \mnras, 335, 673

\bibitem[{{Wang} \& {Chen}(2019)}]{wang2019optical}
{Wang}, S., \& {Chen}, X. 2019, \apj, 877, 116

\bibitem[{Wang {et~al.}(1996)Wang, Su, Chu, Cui, \& Wang}]{wang1996special}
Wang, S.-g., Su, D.-q., Chu, Y.-q., Cui, X., \& Wang, Y.-n. 1996, Applied
  Optics, 35, 5155

\bibitem[{{Weinstein} {et~al.}(2004){Weinstein}, {Richards}, {Schneider},
  {Younger}, {Strauss}, {Hall}, {Budav{\'a}ri}, {Gunn}, {York}, \&
  {Brinkmann}}]{2004ApJS..155..243W}
{Weinstein}, M.~A., {Richards}, G.~T., {Schneider}, D.~P., {et~al.} 2004,
  \apjs, 155, 243

\bibitem[{{Wenger} {et~al.}(2000){Wenger}, {Ochsenbein}, {Egret}, {Dubois},
  {Bonnarel}, {Borde}, {Genova}, {Jasniewicz}, {Lalo{\"e}}, {Lesteven}, \&
  {Monier}}]{2000A&AS..143....9W}
{Wenger}, M., {Ochsenbein}, F., {Egret}, D., {et~al.} 2000, \aaps, 143, 9

\bibitem[{{Westmeier}(2018)}]{2018MNRAS.474..289W}
{Westmeier}, T. 2018, \mnras, 474, 289

\bibitem[{{White} {et~al.}(2000){White}, {Becker}, {Gregg},
  {Laurent-Muehleisen}, {Brotherton}, {Impey}, {Petry}, {Foltz}, {Chaffee},
  {Richards}, {Oegerle}, {Helfand}, {McMahon}, \&
  {Cabanela}}]{2000ApJS..126..133W}
{White}, R.~L., {Becker}, R.~H., {Gregg}, M.~D., {et~al.} 2000, \apjs, 126, 133

\bibitem[{Wright {et~al.}(2010)Wright, Eisenhardt, Mainzer, Ressler, Cutri,
  Jarrett, Kirkpatrick, Padgett, McMillan, Skrutskie, \&
  et~al.}]{wright2010wide}
Wright, E.~L., Eisenhardt, P.~R., Mainzer, A.~K., {et~al.} 2010, \aj, 140, 1868

\bibitem[{{Wu} {et~al.}(2012){Wu}, {Hao}, {Jia}, {Zhang}, \&
  {Peng}}]{2012AJ....144...49W}
{Wu}, X.-B., {Hao}, G., {Jia}, Z., {Zhang}, Y., \& {Peng}, N. 2012, \aj, 144,
  49

\bibitem[{{Wu} \& {Jia}(2010)}]{2010MNRAS.406.1583W}
{Wu}, X.-B., \& {Jia}, Z. 2010, \mnras, 406, 1583

\bibitem[{{Wu} {et~al.}(2004){Wu}, {Zhang}, \& {Zhou}}]{2004ChJAA...4...17W}
{Wu}, X.-B., {Zhang}, W., \& {Zhou}, X. 2004, \cjaa, 4, 17

\bibitem[{Yan {et~al.}(2013)Yan, Donoso, Tsai, Stern, Assef, Eisenhardt, Blain,
  Cutri, Jarrett, Stanford, {et~al.}}]{yan2013characterizing}
Yan, L., Donoso, E., Tsai, C.-W., {et~al.} 2013, \aj, 145, 55

\bibitem[{Yang {et~al.}(2017)Yang, Wu, Fan, Jiang, McGreer, Green, Yang,
  Schindler, Wang, Zuo, {et~al.}}]{yang2017quasar}
Yang, Q., Wu, X.-B., Fan, X., {et~al.} 2017, \aj, 154, 269

\bibitem[{{Yao} {et~al.}(2019){Yao}, {Wu}, {Ai}, {Yang}, {Yang}, {Dong},
  {Joshi}, {Wang}, {Feng}, {Fu}, {Hou}, {Luo}, {Kong}, {Liu}, {Zhao}, {Zhang},
  {Yuan}, \& {Shen}}]{2019ApJS..240....6Y}
{Yao}, S., {Wu}, X.-B., {Ai}, Y.~L., {et~al.} 2019, \apjs, 240, 6

\bibitem[{{Y{\`e}che} {et~al.}(2010){Y{\`e}che}, {Petitjean}, {Rich},
  {Aubourg}, {Busca}, {Hamilton}, {Le Goff}, {Paris}, {Peirani}, {Pichon},
  {Rollinde}, \& {Vargas-Maga{\~n}a}}]{2010A&A...523A..14Y}
{Y{\`e}che}, C., {Petitjean}, P., {Rich}, J., {et~al.} 2010, \aap, 523, A14

\bibitem[{{Yi} {et~al.}(2019){Yi}, {Chen}, {Pan}, {Yue}, {Lu}, {Li}, \&
  {Luo}}]{2019ApJ...887..241Y}
{Yi}, Z., {Chen}, Z., {Pan}, J., {et~al.} 2019, \apj, 887, 241

\bibitem[{{York} {et~al.}(2000){York}, {Adelman}, {Anderson}, {Anderson},
  {Annis}, {Bahcall}, {Bakken}, {Barkhouser}, {Bastian}, {Berman}, {Boroski},
  {Bracker}, {Briegel}, {Briggs}, {Brinkmann}, {Brunner}, {Burles}, {Carey},
  {Carr}, {Castander}, {Chen}, {Colestock}, \& et~al.}]{2000AJ....120.1579Y}
{York}, D.~G., {Adelman}, J., {Anderson}, Jr., J.~E., {et~al.} 2000, \aj, 120,
  1579

\bibitem[{Yuan {et~al.}(2015)Yuan, Liu, Huo, Xiang, Huang, Chen, Zhang, Sun,
  Wang, Zhang, {et~al.}}]{yuan2015lamost}
Yuan, H.-B., Liu, X.-W., Huo, Z.-Y., {et~al.} 2015, \mnras, 448, 855

\bibitem[{{Zhang} {et~al.}(2013){Zhang}, {Ma}, {Peng}, {Zhao}, \&
  {Wu}}]{2013AJ....146...22Z}
{Zhang}, Y., {Ma}, H., {Peng}, N., {Zhao}, Y., \& {Wu}, X.-b. 2013, \aj, 146,
  22

\bibitem[{{Zhao} {et~al.}(2012){Zhao}, {Zhao}, {Chu}, {Jing}, \&
  {Deng}}]{2012RAA....12..723Z}
{Zhao}, G., {Zhao}, Y.-H., {Chu}, Y.-Q., {Jing}, Y.-P., \& {Deng}, L.-C. 2012,
  Research in Astronomy and Astrophysics, 12, 723

\bibitem[{Zonca {et~al.}(2019)Zonca, Singer, Lenz, Reinecke, Rosset, Hivon, \&
  Gorski}]{Zonca2019}
Zonca, A., Singer, L., Lenz, D., {et~al.} 2019, Journal of Open Source
  Software, 4, 1298

\end{thebibliography}



\end{document}